\documentclass[submission,copyright,creativecommons]{eptcs}
\usepackage[utf8]{inputenc}

\usepackage{xcolor}
\usepackage{pgf}
\usepackage{pgfarrows}

\providecommand{\event}{TTC 2011} 

\title{Solving the TTC 2011 Compiler Optimization Task with \metatools}

\author{Markus Lepper
\institute{\texttt{<}semantics\texttt{/>} GmbH, Berlin, Germany}
\email{post@markuslepper.eu}
\and
Baltasar Tranc{\'o}n y Widemann
\institute{Universit\"at Bayreuth, Germany}
\email{Baltasar.Trancon@uni-bayreuth.de}
}
\def\titlerunning{Compiler Optimization with \metatools}
\def\authorrunning{M. Lepper, B. Tranc{\'o}n}
\begin{document}

\newcommand{\metatools}{\ensuremath{{}^\textsf{meta\_}}\textsf{tools}}
\maketitle

\newcommand{\tdom}{\texttt{tdom}}
\newcommand{\umod}{\texttt{umod}}

\newcommand{\xt}[1]{``\texttt{#1}''}

\newcommand{\java}{\textsf{JAVA}}
\newcommand{\icmt}{\textsf{ICMT}}
\newcommand{\icmtz}{\textsf{ICMT 2011}}
\newcommand{\tools}{\textsf{Tools}}
\newcommand{\toolse}{\textsf{Tools Europe}}
\newcommand{\toolsez}{\textsf{Tools Europe 2011}}
\newcommand{\TTC}{\textsf{Transformation Tool Contest}}
\newcommand{\TTCz}{\textsf{Transformation Tool Contest 2011}}
\newcommand{\ttcz}{\textsf{TTC 2011}}

\newcommand{\first}[1]{\emph{#1}}

\let\xACOLsave\addcontentsline
\def\addcontentsline#1#2#3{\relax}%

\begin{abstract}
The authors' \metatools{} are a collection of tools for generic
programming. This includes generating Java sources from mathematically well-founded
specifications, as well as the creation of strictly typed document object
models for XML encoded texts.
In this context, almost every computer-internal structure is 
treated as a ``model'', and every computation is a kind of model transformation.

This concept differs significantly from ``classical model transformation'' executed
by specialized tools and languages.
Therefore it seemed promising  to the organizers of the 
\TTCz{} in Zürich, as well as to the authors, to apply {\metatools}
to one of the challenges, namely to the ``compiler optimization task''.
This is a report on the resulting experiences.
\end{abstract}


\section{Principles of \metatools{} and Context of this Report}

The authors' view on model transformation \cite{lt2011} 
is in some aspects a ``dual'' approach to the construction of dedicated
model transformation tools:
It implies that \emph{every} computer-internal data structure can 
(and should!) be seen as a model, and every computation is a kind of
model transformation.
Consequently, models are realized as type definitions, invariants
and processing code in the hosting computer
language: In a first step, 
source code is generated automatically 
from a mathematical
description. In a second step, the programmer is free to handle the model objects
by combining the generated API and the corresponding runtime libraries
with his/her familiar coding techniques.
During the last decade the authors' coding activities in this area have
been collected into  \metatools{} \cite{mt2011}, a tool set 
for \java{} source code generation. In this context, even source code
generation itself is realized as a transformation between models which adhere
to different meta-models.

Obviously this view differs significantly from dedicated model
transformation languages. Both approaches have their merits.
It may be valuable to compare the experiences, it may be possible
to learn from each other, and maybe both approaches turn out to
be two sides of a bridge which slowly grow and will meet somewhere
in the middle.

Regrettably, the authors did not get to know about the ``\TTCz'' \cite{Ttc2011}
until it had started.
Nevertheless, the organizers invited them to present their approach and 
the authors afterwards developed a solution to the ``compiler optimization task''
from \ttcz{} using \metatools. The results are presented in this paper.

\subsection{Solved Tasks}
The source text and a runnable demonstration of this solution 
is found at \cite{ourShare}.
It can read, visualize and optimize the files 
\texttt{min.gxl}, 
\texttt{const.gxl}, 
\texttt{zero.gxl}, 
and \texttt{testcase.gxl}, which are included as 
copies from the task's original test data.
The other files employ ``memory operations'', which are not yet supported
by the importing code.

\section{Stacking Models for Solving the Compiler Optimization Task}

\subsection{Import and Export Pipeline}

The ``Compiler Optimization Task'' is presented in
\cite{compileroptimizationcase}, 
containing an informal description and the necessary
test data.
The latter is given as an intermediate graph following the  
``Firm'' syntax \cite{TrBoLi-99}.
This in turn is encoded in the ``GXL'' format, 
a standardized XML format for exchanging graph-like data \cite{GxlHome}.

Given this setting, it soon turned out that a 
\emph{stack of models} is an adequate approach.
Not only due to the fact that 
two programmers were involved who created code for 
GXL and Firm independently,
but also for clean, modular and correct implementation.
Figure~\ref{fig_cot_pipeline} shows the transformation pipeline.

Two kinds of model transformations are involved:
First, identical information is transformed from one representation
to another. Then the optimization task itself is performed as a transformation
on the best-suited representation. Afterwards the first transformation chain
is reversed.

\newcommand{\citeGxlDtd}{\cite[/gxl-1.0.dtd]{GxlHome}}

\newcommand{\citeGxlDtdCommented}{\cite[/dtd/gxl-1.0.html]{GxlHome}}

Directly 
from the DTD  at \citeGxlDtd{}, the \tdom{} generator from \metatools{} generated a 
type-safe document object model, in form of source
text for \java{} classes.
Only a small driver DTD (see appendix~\ref{file_gxl_driver}) had to be 
prepended, for  manually identifying some common attributes and abstractions of 
common content models. These will be translated to abstract super classes.
Appendix~\ref{txt_call_tdom_creation} shows the code snippet for
creating a model instance from a SAX event stream.


\definecolor{lightblue}{rgb}{0.8,0.8,1.0}
\definecolor{lightyellow}{rgb}{0.99,0.99,0.3}

\newcommand{\xarrowRead}[2]{
\begin{pgfscope}
\pgfsetendarrow{\pgfarrowlargepointed{0.3cm}}
\pgfsetdash{{0.15cm}{0.1cm}}{0cm}
\pgfline{\pgfbackoff{0.5cm}{#1}{#2}}{\pgfbackoff{0.5cm}{#2}{#1}}
\end{pgfscope}
}

\newcommand{\xarrowCreate}[2]{
\begin{pgfscope}
\pgfsetendarrow{\pgfarrowlargepointed{0.3cm}}
\pgfline{\pgfbackoff{0.5cm}{#1}{#2}}{\pgfbackoff{0.5cm}{#2}{#1}}
\end{pgfscope}
}

\newdimen\xdimy
\newdimen\xdimx

\newcommand{\xarrowInheritXX}[2]{
\begin{pgfscope}
\pgftranslateto{#1}
\pgfextracty{\xdimy}{#2}
\pgfextractx{\xdimx}{#2}
\pgfsetyvec{#2}
\pgfsetxvec{\pgfpoint{\xdimy}{-1.0\xdimx}}
\pgfmoveto{\pgfxy(-0.8,0)}
\pgflineto{\pgfxy(0,0.8)}
\pgflineto{\pgfxy(0.8,0)}
\pgfclosepath
\pgfstroke
\end{pgfscope}
}

\newcommand{\xarrowInheritX}[3]{
\pgfline{\pgfbackoff{0.5cm}{#1}{#2}}{#3}
\xarrowInheritXX{#3}{\pgfdiff{#3}{#2}} 
}

\newcommand{\xarrowInherit}[2]{
\xarrowInheritX{#1}{\pgfbackoff{0.5cm}{#2}{#1}}{\pgfbackoff{0.8cm}{#2}{#1}}
}

\newcommand{\xarrowData}[2]{
\begin{pgfscope}
\pgfsetlinewidth{4pt}
\color{black!20}
\pgfsetendarrow{\pgfarrowtriangle{10pt}}
\pgfline{\pgfbackoff{0.5cm}{#1}{#2}}{\pgfbackoff{0.7cm}{#2}{#1}}
\end{pgfscope}
}

\newcommand{\xarrowDataX}[4]{
\begin{pgfscope}
\pgfsetlinewidth{4pt}
\color{black!20}
\pgfmoveto{#1}
\pgfcurveto{#2}{#3}{#4}
\pgfsetendarrow{\pgfarrowtriangle{10pt}}
\pgfstroke
\end{pgfscope}
}

\newcommand{\xcode}[5]{
\begin{pgfscope}
\pgfsetlinewidth{#5}
\color{#4}
\pgfputat{#1}{\pgfrect[fill]{\pgfxy(-0.80,-0.4)}{\pgfxy(1.6,0.8)}}
\color{#3}
\pgfputat{#1}{\pgfrect[stroke]{\pgfxy(-0.80,-0.4)}{\pgfxy(1.6,0.8)}}
\pgfputat{#1}{\pgfbox[center,center]{#2}}
\end{pgfscope}
}

\newcommand{\xcodeX}[5]{
\begin{pgfscope}
\pgfsetlinewidth{#5}
\color{#4}
\pgfputat{#1}{\pgfrect[fill]{\pgfxy(-0.75,-0.4)}{\pgfxy(1.5,0.8)}}
\color{#3}
\pgfputat{#1}{\pgfrect[stroke]{\pgfxy(-0.75,-0.4)}{\pgfxy(1.5,0.8)}}
\pgfputat{#1}{\pgfbox[center,center]{#2}}
\end{pgfscope}
}

\newcommand{\xmodel}[4]{
\begin{pgfscope}
\pgfsetlinewidth{1pt}
\color{#4}
\pgfcircle[fill]{#1}{0.5cm}
\color{#3}
\pgfcircle[stroke]{#1}{0.5cm}
\pgfputat{#1}{\pgfbox[center,center]{#2}}
\end{pgfscope}
}

\newcommand{\xcurveA}{
\pgfcurveto{\pgfxy(0.0,0.0)}{\pgfxy(0.1,0.2)}{\pgfxy(0.3,0.2)}{\pgfxy(0.4,0.0)}}
\newcommand{\xcurveV}{
\pgfcurveto{\pgfxy(0.0,0.0)}{\pgfxy(0.1,-0.2)}{\pgfxy(0.3,-0.2)}{\pgfxy(0.4,0.0)}}

\newcommand{\xpfull}{
\pgfmoveto{\pgfxy(-0.4,-0.4)}
\pgflineto{\pgfxy(-0.4,0.4)}
\pgfcurveto{\pgfxy(-0.3,0.6)}{\pgfxy(-0.1,0.6)}{\pgfxy(0.0,0.4)}
\pgfcurveto{\pgfxy(0.1,0.2)}{\pgfxy(0.3,0.2)}{\pgfxy(0.4,0.4)}
\pgflineto{\pgfxy(0.4,-0.4)}
\pgfcurveto{\pgfxy(0.3,-0.6)}{\pgfxy(0.1,-0.6)}{\pgfxy(0.0,-0.4)}
\pgfcurveto{\pgfxy(-0.1,-0.2)}{\pgfxy(-0.3,-0.2)}{\pgfxy(-0.4,-0.4)}
\pgfclosepath
}

\newcommand{\xpage}[4]{
\begin{pgfscope}
\pgfsetlinewidth{2pt}
\pgfputat{#1}{\xpfull\color{#4}\pgffill}
\pgfputat{#1}{\xpfull\color{#3}\pgfstroke}
\pgfputat{#1}{\pgfputat{\pgfxy(0,0.8)}{\pgfbox[center,center]{#2}}}
\end{pgfscope}
}

\newcommand{\xpageGiven}[2]{\xpage{#1}{#2}{black}{white}}
\newcommand{\xpageHandwritten}[2]{\xpage{#1}{#2}{black}{lightblue}}
\newcommand{\xpageUserInput}[2]{\xpage{#1}{#2}{black}{lightyellow}}
\newcommand{\xpageUserOutput}[2]{\xpage{#1}{#2}{black}{green}}

\newcommand{\xcodeHandwritten}[2]{\xcode{#1}{#2}{blue}{white}{2pt}}
\newcommand{\xcodeHandwrittenX}[2]{\xcodeX{#1}{#2}{blue}{white}{2pt}}
\newcommand{\xcodeGenerated}[2]{\xcode{#1}{#2}{black}{white}{1pt}}
\newcommand{\xcodeExisting}[2]{\xcode{#1}{#2}{black}{white}{1pt}}

\newcommand{\xmodelGenerated}[2]{\xmodel{#1}{#2}{black}{white}}


\begin{figure}[t]
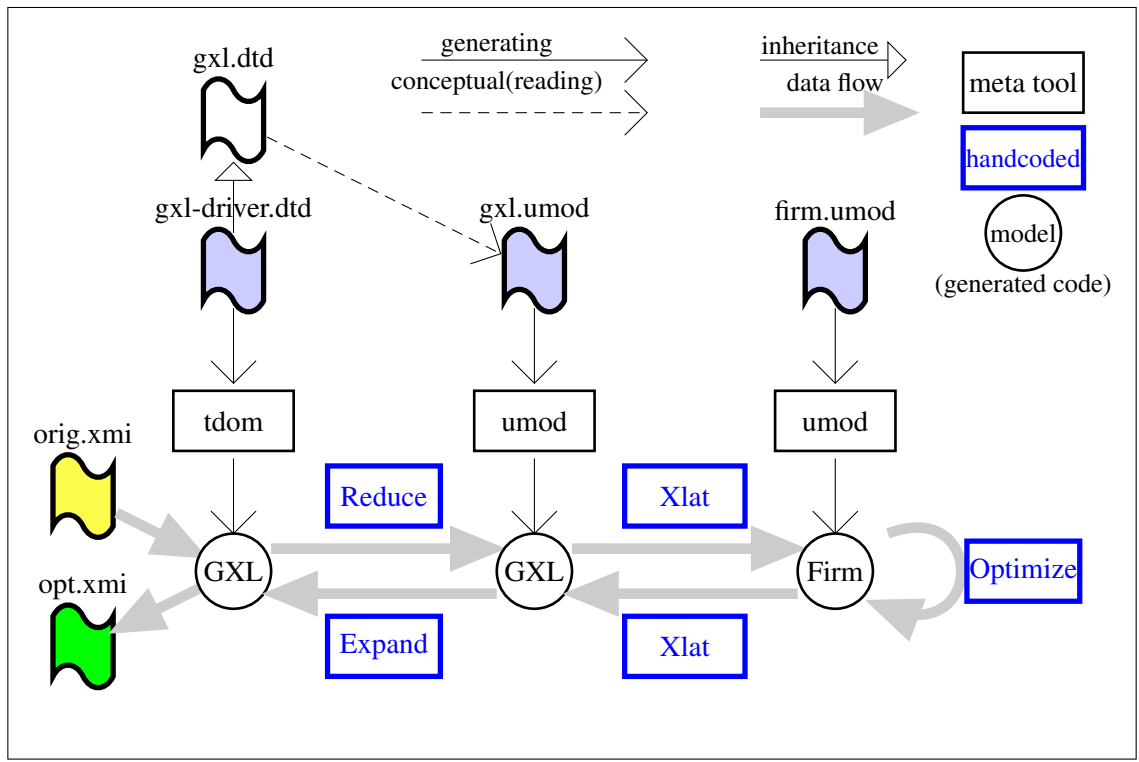

\center

\begin{pgfpicture}{0cm}{-1.5cm}{15cm}{9cm}
\pgfrect[stroke]{\pgfxy(0,-1.5)}{\pgfxy(15,10)}

\xcodeExisting{\pgfxy(13.5,7.5)}{meta tool}
\xcodeHandwritten{\pgfxy(13.5,6.5)}{\small handcoded}
\xmodelGenerated{\pgfxy(13.5,5.5)}{\small model}
\pgfputat{\pgfxy(13.5,4.8)}{\pgfbox[center,center]{\small(generated code)}}

\pgfputat{\pgfxy(6.5,8.0)}{\pgfbox[center,center]{\small generating}}
\xarrowCreate{\pgfxy(5,7.8)}{\pgfxy(9,7.8)}
\pgfputat{\pgfxy(6.5,7.5)}{\pgfbox[center,center]{\small conceptual(reading)}}
\xarrowRead{\pgfxy(5,7.1)}{\pgfxy(9,7.1)}
\pgfputat{\pgfxy(10.8,8.0)}{\pgfbox[center,center]{\small inheritance}}
\xarrowInherit{\pgfxy(9.5,7.8)}{\pgfxy(12.5,7.8)}
\pgfputat{\pgfxy(11,7.5)}{\pgfbox[center,center]{\small data flow}}
\xarrowData{\pgfxy(9.5,7.1)}{\pgfxy(12.5,7.1)}

\xpageUserInput{\pgfxy(1,2)}{orig.xmi}
\xpageUserOutput{\pgfxy(1,0)}{opt.xmi}
\xarrowData{\pgfxy(1,2)}{\pgfxy(3,1)}
\xarrowData{\pgfxy(3,1)}{\pgfxy(1,0)}

\xpageGiven{\pgfxy(3,7)}{gxl.dtd}
\xarrowInherit{\pgfxy(3,5)}{\pgfxy(3,7)}
\xpageHandwritten{\pgfxy(3,5)}{gxl-driver.dtd}
\xarrowCreate{\pgfxy(3,5)}{\pgfxy(3,3)}
\xcodeExisting{\pgfxy(3,3)}{tdom}
\xarrowCreate{\pgfxy(3,3)}{\pgfxy(3,1)}
\xmodelGenerated{\pgfxy(3,1)}{GXL}

\xcodeHandwrittenX{\pgfxy(5,2)}{Reduce}
\xcodeHandwrittenX{\pgfxy(5,0)}{Expand}
\xarrowData{\pgfxy(3,1.3)}{\pgfxy(7,1.3)}
\xarrowData{\pgfxy(7,0.7)}{\pgfxy(3,0.7)}

\xpageHandwritten{\pgfxy(7,5)}{gxl.umod}
\xarrowRead{\pgfxy(3,7)}{\pgfxy(7,5)}
\xarrowCreate{\pgfxy(7,5)}{\pgfxy(7,3)}
\xcodeExisting{\pgfxy(7,3)}{umod}
\xarrowCreate{\pgfxy(7,3)}{\pgfxy(7,1)}
\xmodelGenerated{\pgfxy(7,1)}{GXL}

\xcodeHandwritten{\pgfxy(9,2)}{Xlat}
\xcodeHandwritten{\pgfxy(9,0)}{Xlat}
\xarrowData{\pgfxy(7,1.3)}{\pgfxy(11,1.3)}
\xarrowData{\pgfxy(11,0.7)}{\pgfxy(7,0.7)}

\xpageHandwritten{\pgfxy(11,5)}{firm.umod}
\xarrowCreate{\pgfxy(11,5)}{\pgfxy(11,3)}
\xcodeExisting{\pgfxy(11,3)}{umod}
\xarrowCreate{\pgfxy(11,3)}{\pgfxy(11,1)}
\xmodelGenerated{\pgfxy(11,1)}{Firm}


\xarrowDataX{\pgfxy(11.7,1.5)}{\pgfxy(13,2.0)}{\pgfxy(13,0.0)}{\pgfxy(11.7,0.5)}
\xcodeHandwrittenX{\pgfxy(13.5,1)}{Optimize}


\end{pgfpicture}

\caption{The processing pipeline for solving the ``Compiler Optimization Task'' 
with \metatools{}\label{fig_cot_pipeline}}
\end{figure}

Then this model is transformed into a equivalent \umod{}  model,
(see next section for explanation and appendix~\ref{file_gxl_umod} for the source).
The inheritance relation which spans the \umod{} model
resemble that from the ``XML~Schema'' version of the GXL definition,
see \cite[xmlschema/xmlschema.html]{GxlHome}, for natural reasons.
The decoders and encoders between both models of GXL 
are written manually, and derived from the automatically generated base visitors.
The source is in appendices~\ref{file_gxl_decode_java} and \ref{file_gxl_encode_java}.
The structure of this code 
clearly expresses 
the inheritance relation of the visited model.

%
%
%
%
%

\subsection{The \umod{} Model of Firm}

Finally, the data to process is transformed from the GXL  model
into a \umod{} model for representing Firm structures.
\umod{} is one of the central tools in \metatools{} for defining computer-internal data models, \cite{mt2011}, \cite{lt2011}. 
It takes a mathematical description of a model,
i.e.\ of classes, inheritance relation, types and field structures,
and creates an API which allows to create and process \java{} objects
which represent the model elements in a type-safe way.
Esp., it supports the free composition of non-syntactic
data types like sets, sequences and maps, and eliminates the ubiquitous threat
of the \texttt{null} value.
The generated code includes configurable, general purpose
\emph{visitors} and \emph{rewriters},
serving as base classes for user code, as well as dedicated subclasses for
visualization and serialization.
By extending this generated code, user-defined code which
realizes the intended semantics becomes very compact, 
cf.~appendices~\ref{file_checker_java} and \ref{file_constantfolding_java}.
%
%
The \umod{} source is contained in appendix~\ref{file_firm_umod}.
Its design closely follows \cite[table~1]{FirmArticle}  and 
\cite[table~2.1]{TrBoLi-99}.

A program graph in Firm is a kind of ``inverted'' program control flow
graph: Every node points to its predecessors which have to be evaluated
in advance. 
This structure corresponds to those of pure expressions
and mathematical terms.
Consequently, in the \umod{} version each program graph is identified and
represented \emph{solely by its end node}, since from this every 
other node and  block is reachable.

Nodes and edges are \emph{typed}, from a small fixed set of 
\first{node types} and \first{edge types}. 
The latter are called ``modes'' for historical reasons.
Every node class defines for its instances the allowed combinations 
of types for the incoming and the outgoing edges.

While the GXL model contains the Firm edges as reified, the \umod{} version
reduces them to unidirectional references. 
As a consequence, all backward information necessary during processing
must (and always can!) be locally memorized. 
This elegant and efficient 
technique has already been proposed by \cite[section~2.2.2.8]{TrBoLi-99}.
It is only possible because in general 
no role information or sequential order is imposed on \emph{incoming} edges.

The drawings in \cite{TrBoLi-99} introduce ``sockets'', which impose
a partitioning of the incoming and on the outgoing side of each node.
These sockets are required only on the conceptual
level, not in the implementation,
because this partitioning  is uniquely determined by the \emph{type}
of the connected edges.
This type, in turn, is uniquely determined by the ``role'' of the definition of the 
object's  field which contains the reference which realizes the edge.

Whenever a definition line in \cite[table~1]{FirmArticle}  or 
\cite[table~2.1]{TrBoLi-99} mentions two different letters on the right
side of the arrow, then an instance of this node type can be the end point 
of edges of two different edge types. In the \umod{} model, edge types
and node roles are realized by \java{}  ``interfaces''. 
On the conceptual level,
e.g.\ each node of type ``Start'' can be the target of a 
``data flow edge'' and of a ``control flow edge''.
So the corresponding \umod{} class 
implements the \texttt{Numeric} and the \texttt{ControlFlow}
interface, and a reference to an instance object of this class
can be held by a field of \texttt{Numeric} and of  \texttt{ControlFlow} type.

The only node definitions which break this concept are those in which 
the same edge type appears more than once on the right side of the arrow, 
representing output of the same type, but with
different roles. This can be seen as 
a \emph{sequential order} on incoming edges.
In this case, the proposal from \cite[section~2.2.2.8]{TrBoLi-99} is followed:
The node is the target of many edges, but not directly
from a consuming node, but from a special \first{projection node},
inserted to select one component of the incoming tuple.
%
In the \umod{} model there exist two such node classes, \texttt{Proj\_X}
and \texttt{Proj\_N}, for control flow and for data flow.


The sequential order of the \emph{outgoing} edges is realized most
naturally by the distinction
of the fields which contain the references, and  by special container classes
like lists and maps for containing more than one outgoing edge.

\subsection{Consistency Checks on the Firm Model}

Both the textual description of the task in
\cite[section~3 ``Getting started. Verifier'']{Coptim_data} 
and the Firm documentation in 
\cite[section~2.1.3 ``Further restrictions to Firm Graphs'']{TrBoLi-99}, 
suggest to check certain consistency properties.
In our approach, some of these are already guaranteed ``by construction'',
i.e.\ by the algorithm  which translates from the GXL into the Firm \umod{}
model, see appendix~\ref{file_gxl2firm_java}.
Further properties are checked explicitly be a dedicated
\texttt{Checker} class, see appendix~\ref{file_checker_java}.
Others are guaranteed implicitly, by the structure of the \umod{} model.
Table~\ref{tabChecks} lists these properties (just by a few keywords
for the reader familiar with both documents) and indicates which part of
our solution does cover them.

The \texttt{Checker} class is derived from the generated visitor class,
but indirectly, because we have to deal with cycles explicitly.
We assume that \emph{the only cycles} in the
graph are made by blocks and their final ``Jump/Cond''  nodes.
The class \texttt{VisitBlocksOnce} cuts these cycles. It will
be compiled as a static inner class in \texttt{Firm.java}, since
it is contained as a \java{} escape directly in the \umod{} source,
see appendix~\ref{file_firm_umod}.


\begin{table}[t]

\begin{tabular}{|p{0.7\textwidth}*4{|c}|}
\hline
\multicolumn{5}{|l|}{
\textit{The property below is currently \ldots}}\\

\cline{5-5}

\multicolumn{4}{|r|}{
\textit{\ldots checked on the fly by the translation 
algorithm in}
               \texttt{Gxl2Firm} =} &
\multicolumn{1}{|c|}{T}
\\
\cline{1-5}

\multicolumn{3}{|r|}{\textit{\ldots checked explicitly in }
              \texttt{Checker} =} &
\multicolumn{1}{|c|}{C}&
\multicolumn{1}{l}{}\\
\cline{1-4}

\multicolumn{2}{|r|}{\textit{\ldots guaranteed implicitly, by 
the properties of}
                            \texttt{Firm.umod} =}&
\multicolumn{1}{|c|}{i}&
\multicolumn{2}{l}{}\\
\cline{1-3}

\multicolumn{1}{|r|}{\textit{\ldots currently still unchecked, 
but urgently required!} = }&
\multicolumn{1}{|c|}{?}&
\multicolumn{3}{l}{}\\

\hline
\hline
only one start node                 &  &  &  &T \\
\hline
start node only node in block       &  &  &  &  \\
\hline
start block no predecs              &  &  &  &  \\
\hline
only one end node                   &  &i &  &T \\
\hline
end node only node in block         &  &  &  &  \\
\hline
end block has no successors         &  &  &C &  \\
\hline
all blocks reach end node            &  &i &  &  \\
\hline
start node reaches all blocks       &? &  &  &  \\
\hline
inter-block graphs are acyclic      &? &  &  &  \\
\hline
phi node predecs $\equiv$ block predecs
                                    &  &  &C &  \\
\hline
inter-block edge requires corresponding control flow edge
                                    &? &  &  &  \\
\hline
dto., w.r.t.\ phi nodes and predecessor numbering 
                                    &? &  &  &  \\
\hline
\hline
edge from node to block is on position ``-1''
                                    &  &  &  &T \\
\hline
all constants contained in start block
                                    &  &  &  &  \\
\hline
\hline
each block (w/o end block) has one control flow node 
                                    &  &  &C &  \\
\hline

\end{tabular}
\caption{Consistency checks and their implementation\label{tabChecks}.
}
\end{table}

\subsection{Optimizing Transformations on the Firm Model}

The code of the class \texttt{ConstantFolding}, see
appendix~\ref{file_constantfolding_java}, 
performs simple constant folding in arithmetic operations,
and the resulting control flow simplification.
Most of the transformation-performing classes are 
derived from the generated \first{rewriter} code, a framework for copy-on-write updating of model data.
Similar to a visitor, this generated class contains an 
generated \texttt{action(C)} method for every model element class. 
This method creates a clone of the visited object, and calls
the method \texttt{rewriteFields(C clone)}.
The \emph{generated version} of \texttt{rewriteFields()} first 
calls \texttt{rewriteFields((D) clone)}, when \texttt{D} is the superclass of \texttt{C}.
Then it initiates the rewriting process recursively on
each field \texttt{f} defined on the level of \texttt{C} by calling
\texttt{rewrite(clone.get\_f())}.
When this returns, 
(1) it  compares the result of rewriting with  the original, for detecting
changes, (2) stores the result into
the field of the clone, and (3) re-adjusts its own result value,
which at the beginning pointed to the original, to
refer to the clone, whenever such a change of a field value
has been detected.

The generated methods are now partly overridden by user defined, specialized
methods:

The method \texttt{rewriteFields(Binary)} will be called 
on each binary numeric node in the model.
Like the generated version, it first 
calls the rewrite process bottom-up on all sub-objects referred to.
Additionally it 
\emph{substitutes} 
the result by a new constant, where
appropriate.
%
%

User-defined \texttt{action(Proj\_X)} looks whether the selecting
value of the \texttt{Cond} is a constant after rewriting. 
In this case it substitutes a constant \texttt{Jmp} or an empty list, 
reflecting whether the index number of the \texttt{Proj\_X} is equal to 
the selecting constant.
References to control flow nodes can only occur in the 
\texttt{.predecs} map of a block. In the context of such an aggregate,
a single reference
can be re-written to a multiplicity of references, or to the empty list.
This decrease in alternatives is in turn checked by 
\texttt{action(Phi)}, which may replace the visited object
by its single surviving input.

\section{Conclusion}

We have presented a solution for the ``Compiler Optimization Task'' 
from the ``\TTCz'' by generic programming, i.e.\ by automated generation
of model source code, as supported by the \metatools{} toolkit.
During the last decade, 
generated source code of this kind has been employed in different
software projects, serving as the starting point for very different styles of
of \java{} programming.
This shows that the mental model of ``coding as model transformation'' is 
adequate for developing general purpose mid-scale software architectures.

The model definitions in appendices~\ref{file_gxl_umod} and 
\ref{file_firm_umod} show the compactness of model definitions in our framework,
compared to the generated API doc at \cite[ttc2011/apidoc]{ourShare}.

The sources in appendices~\ref{file_gxl_decode_java}, \ref{file_gxl_encode_java},
\ref{file_gxl2firm_java}, and 
\ref{file_firm2gxl_java} show how the structure of the model
organizes the transformation code in a natural and readable way.

The code in appendices~\ref{file_checker_java} and 
\ref{file_constantfolding_java} shows how an intended transformation is
broken down into simple steps, and then combined in a pipeline.

The main \emph{advantages}
of this kind of model code generation are, in our experience:

\begin{enumerate}
\item
After the initial step, the generation of sources by \metatools, no
further dedicated tools are required, and no additional programming
language front-end has to be learned.
\item
Nearly everything (beside some primitive run-time functions) is visible to the
programmer, nothing mystic happens ``behind the scene''.
\item
Many properties of a correct model instance are checked automatically
by the generated constructors and setter methods, or mapped to the
type system of the hosting language, see table~\ref{tabChecks} above.
\item
Visitor and rewriter based code is seamlessly integrated
with all other handwritten code: Both kinds of code can control, call
and parameterize each other, freely and recursively.
For this, only the native constructs from the hosting language are required.
\end{enumerate}

The special problems and \emph{disadvantages}:
\begin{enumerate}
\item
The first disadvantage is caused by the same fact as the preceding advantage:
Due to the integration of generated and hand-written code,
the consistency and integrity 
of the model's internal state variables can only be guaranteed
in the limits imposed by the hosting language. E.g., in case of \java, putting 
both kinds of code into the same ``package'' would allow the programmer
to clutter anything, with unforeseeable effects.
\item
The design of the model definition is crucial: All benefits
mentioned above are only fully enjoyable with an adequate
design of the  \texttt{.umod} source. 
E.g., in case of \texttt{Firm.umod}: 
whether the different numeric 
data types shall be mapped to subclasses of `\texttt{Numeric}'',
--- whether all three interface types are really required,
--- whether one or more ``\texttt{Proj\_<>}'' nodes are sensible,
--- all these questions require some experience, or even some experiments.
\item
The visitor and rewriter API has to be understood and the protocol has
to be respected by the programmer. Otherwise effects can occur which seem
``mystical'' indeed.
\item
Due to the freedom of calling any code,
inferring properties of the transformation system
as a whole, like termination, completeness, confluence, may become
infeasible, or at least more restricted than in case of 
a dedicated model transformation tool.
But the authors' work on visitor optimization \cite{lt2011} showed that
even in the open environment of an imperative hosting language
some analysis is possible.
\end{enumerate}


\bibliographystyle{eptcs}
\bibliography{lepper_trancon_ttc2011}

\appendix

\def\contentsname{Contents of Appendices}
\tableofcontents{}

\let\addcontentsline\xACOLsave

\newcounter{xlinecounter}
\newcommand{\xlinecounterreset}{\setcounter{xlinecounter}0}
\newcommand{\xline}{%
{\tiny\arabic{xlinecounter}\stepcounter{xlinecounter}\ldots\ldots}}

\section{The GXL DTD Driver File}
\label{file_gxl_driver}
\xlinecounterreset{}
\bgroup\footnotesize
\xline\verb~<?tdom default private?>~\\[-1.0ex]
\xline\verb~<?tdom public gxl?>~\\[-1.0ex]
\xline\verb~~\\[-1.0ex]
\xline\verb~<?tdom attribute ~\\[-1.0ex]
\xline\verb~        xlink:type (simple) #FIXED "simple"~\\[-1.0ex]
\xline\verb~        xlink:href CDATA #REQUIRED~\\[-1.0ex]
\xline\verb~        id ID #IMPLIED~\\[-1.0ex]
\xline\verb~        isdirected (true | false) #IMPLIED~\\[-1.0ex]
\xline\verb~ ?>~\\[-1.0ex]
\xline\verb~~\\[-1.0ex]
\xline\verb~<!ENTITY % gxl.dtd SYSTEM "../gxl-1.0.dtd">~\\[-1.0ex]
\xline\verb~%gxl.dtd;~\\[-1.0ex]
\xline\verb~~\\[-1.0ex]
\xline\verb~<!ENTITY % anyval "(%val;)">~\\[-1.0ex]
\xline\verb~<?tdom abstract-entity anyval ?>~\\[-1.0ex]
\xline\verb~<?tdom abstract part (node | edge | rel) ?>~\\[-1.0ex]

\egroup

\section{Code which initiates the creation of the \tdom{} model
from some SAX event stream}
\label{txt_call_tdom_creation}
\xlinecounterreset{}
\bgroup\footnotesize
\begin{verbatim}
import org.xml.sax.InputSource ;
import eu.bandm.tools.xantlrtdom.TdomReader ; 
import eu.bandm.ttc2011.case2.tdom.Document_gxl ; 
import eu.bandm.ttc2011.case2.tdom.DTD ; 
...
  try{
    final Document_gxl allGxl 
     = TdomReader.parseXmlFile
      ( new InputSource(inputfilename),
        Document_gxl.class,
        DTD.dtd,
        /*debug=*/false);
   } catch (TdomException ex){ ...} 
\end{verbatim}
\egroup


\section{The GXL Model, as source to \umod{}}
\label{file_gxl_umod}
\xlinecounterreset{}
\bgroup\footnotesize
\xline\verb~// http://www.gupro.de/GXL/dtd/gxl-1.0.html~\\[-1.0ex]
\xline\verb~~\\[-1.0ex]
\xline\verb~MODEL GxlModel  =~\\[-1.0ex]
\xline\verb~// MODEL GXL = geht nicht unter cygwin/ntfs wg. Konflikt GXL.java Gxl.java~\\[-1.0ex]
\xline\verb~~\\[-1.0ex]
\xline\verb~ENUM Edgemode =  directed, undirected, defaultdirected, defaultundirected~\\[-1.0ex]
\xline\verb~ENUM Direction =  in, out, none~\\[-1.0ex]
\xline\verb~~\\[-1.0ex]
\xline\verb~VISITOR 0 SinglePass  ; ~\\[-1.0ex]
\xline\verb~~\\[-1.0ex]
\xline\verb~EXT Location = eu.bandm.tools.message.Location~\\[-1.0ex]
\xline\verb~    <eu.bandm.tools.message.XMLDocumentIdentifier>~\\[-1.0ex]
\xline\verb~~\\[-1.0ex]
\xline\verb~TOPLEVEL CLASS~\\[-1.0ex]
\xline\verb~~\\[-1.0ex]
\xline\verb~GxlObject ABSTRACT~\\[-1.0ex]
\xline\verb~        location OPT Location~\\[-1.0ex]
\xline\verb~~\\[-1.0ex]
\xline\verb~| Gxl~\\[-1.0ex]
\xline\verb~        graphs SEQ Graph                !       V 0/0 ;~\\[-1.0ex]
\xline\verb~~\\[-1.0ex]
\xline\verb~| Type~\\[-1.0ex]
\xline\verb~        href string                     ! C 0/0 ;~\\[-1.0ex]
\xline\verb~~\\[-1.0ex]
\xline\verb~| Attributed ABSTRACT~\\[-1.0ex]
\xline\verb~        attrs SEQ Attr                  !       V 0/0 ;~\\[-1.0ex]
\xline\verb~| | Typed ABSTRACT~\\[-1.0ex]
\xline\verb~        type OPT Type                   !       V 0/1 ;~\\[-1.0ex]
\xline\verb~| | | Graph~\\[-1.0ex]
\xline\verb~        id string                       ! C 0/0 ;~\\[-1.0ex]
\xline\verb~        parts SEQ Part                  !       V 0/2 ;~\\[-1.0ex]
\xline\verb~        role OPT string~\\[-1.0ex]
\xline\verb~        edgeids bool = "false"~\\[-1.0ex]
\xline\verb~        hypergraph bool = "false"~\\[-1.0ex]
\xline\verb~        edgemode Edgemode = "Edgemode.directed"~\\[-1.0ex]
\xline\verb~| | | Part ABSTRACT~\\[-1.0ex]
\xline\verb~        id ABSTRACT GETTER OPT string~\\[-1.0ex]
\xline\verb~        graphs SEQ Graph                !       V 0/2 ;~\\[-1.0ex]
\xline\verb~| | | | Node~\\[-1.0ex]
\xline\verb~        id string                       ! C 0/0 ;~\\[-1.0ex]
\xline\verb~| | | | Edgy ABSTRACT~\\[-1.0ex]
\xline\verb~        id OPT string~\\[-1.0ex]
\xline\verb~        isdirected OPT bool~\\[-1.0ex]
\xline\verb~| | | | | Edge~\\[-1.0ex]
\xline\verb~        from Part                       ! C 0/0 ;~\\[-1.0ex]
\xline\verb~        to Part                         ! C 0/1 ;~\\[-1.0ex]
\xline\verb~        fromorder OPT int~\\[-1.0ex]
\xline\verb~        toorder OPT int~\\[-1.0ex]
\xline\verb~| | | | | Rel~\\[-1.0ex]
\xline\verb~        relends SEQ Relend              !       V 0/3 ;~\\[-1.0ex]
\xline\verb~| | Relend~\\[-1.0ex]
\xline\verb~        target Part                     ! C 0/0 ;~\\[-1.0ex]
\xline\verb~        role OPT string~\\[-1.0ex]
\xline\verb~        direction OPT Direction~\\[-1.0ex]
\xline\verb~        startorder OPT int~\\[-1.0ex]
\xline\verb~        endorder OPT int~\\[-1.0ex]
\xline\verb~| | Attr~\\[-1.0ex]
\xline\verb~        id OPT string~\\[-1.0ex]
\xline\verb~        name string                     ! C 0/0 ;~\\[-1.0ex]
\xline\verb~        kind OPT string~\\[-1.0ex]
\xline\verb~        val Val                         ! C 0/1 V 0/1 ;~\\[-1.0ex]
\xline\verb~~\\[-1.0ex]
\xline\verb~| Val ABSTRACT~\\[-1.0ex]
\xline\verb~| | Locator~\\[-1.0ex]
\xline\verb~        href string                     ! C 0/0 ;~\\[-1.0ex]
\xline\verb~| | Bool~\\[-1.0ex]
\xline\verb~        value bool                      ! C 0/0 ;~\\[-1.0ex]
\xline\verb~| | Int~\\[-1.0ex]
\xline\verb~        value int                       ! C 0/0 ;~\\[-1.0ex]
\xline\verb~| | Float~\\[-1.0ex]
\xline\verb~        value float                     ! C 0/0 ;~\\[-1.0ex]
\xline\verb~| | String~\\[-1.0ex]
\xline\verb~        value string                    ! C 0/0 ;~\\[-1.0ex]
\xline\verb~| | Enum~\\[-1.0ex]
\xline\verb~        value string                    ! C 0/0 ;~\\[-1.0ex]
\xline\verb~| | Aggregate ABSTRACT~\\[-1.0ex]
\xline\verb~        elems SEQ Val                   !       V 0/0 ;~\\[-1.0ex]
\xline\verb~| | | Seq~\\[-1.0ex]
\xline\verb~| | | Set~\\[-1.0ex]
\xline\verb~| | | Bag~\\[-1.0ex]
\xline\verb~| | | Tup~\\[-1.0ex]
\xline\verb~~\\[-1.0ex]
\xline\verb~END MODEL~\\[-1.0ex]

\egroup

\section{Decoding from the \tdom{} model into the \umod{} model of GXL}
\label{file_gxl_decode_java}
\bgroup\footnotesize
\xlinecounterreset{}
\xline\verb~package eu.bandm.ttc2011.case2.gxlcodec ;~\\[-1.0ex]
\xline\verb~~\\[-1.0ex]
\xline\verb~import eu.bandm.ttc2011.case2.tdom.* ;~\\[-1.0ex]
\xline\verb~~\\[-1.0ex]
\xline\verb~import eu.bandm.tools.tdom.runtime.* ;~\\[-1.0ex]
\xline\verb~import eu.bandm.tools.graph.GraphModels ;~\\[-1.0ex]
\xline\verb~import eu.bandm.tools.graph.CycleException ;~\\[-1.0ex]
\xline\verb~import eu.bandm.tools.message.MessageReceiver ;~\\[-1.0ex]
\xline\verb~import eu.bandm.tools.message.SimpleMessage ;~\\[-1.0ex]
\xline\verb~import eu.bandm.tools.message.Location ;~\\[-1.0ex]
\xline\verb~import eu.bandm.tools.message.XMLDocumentIdentifier ;~\\[-1.0ex]
\xline\verb~import eu.bandm.tools.message.MessageThrower ;~\\[-1.0ex]
\xline\verb~~\\[-1.0ex]
\xline\verb~import java.util.* ;~\\[-1.0ex]
\xline\verb~~\\[-1.0ex]
\xline\verb~~\\[-1.0ex]
\xline\verb~public class GxlDecoder {~\\[-1.0ex]
\xline\verb~~\\[-1.0ex]
\xline\verb~  private MessageReceiver<? super SimpleMessage<XMLDocumentIdentifier>> msg =~\\[-1.0ex]
\xline\verb~    new MessageThrower<SimpleMessage>() ;~\\[-1.0ex]
\xline\verb~~\\[-1.0ex]
\xline\verb~  public void setMessageReceiver(MessageReceiver<? super SimpleMessage<XMLDocumentIdentifier>> msg) {~\\[-1.0ex]
\xline\verb~    this.msg = msg ;~\\[-1.0ex]
\xline\verb~  }~\\[-1.0ex]
\xline\verb~~\\[-1.0ex]
\xline\verb~  protected void error(Location<XMLDocumentIdentifier> location,~\\[-1.0ex]
\xline\verb~                       java.lang.String text) {~\\[-1.0ex]
\xline\verb~    msg.receive(SimpleMessage.error(location, text)) ;~\\[-1.0ex]
\xline\verb~  }~\\[-1.0ex]
\xline\verb~~\\[-1.0ex]
\xline\verb~  protected void hint(Location<XMLDocumentIdentifier> location,~\\[-1.0ex]
\xline\verb~                      java.lang.String text) {~\\[-1.0ex]
\xline\verb~    msg.receive(SimpleMessage.hint(location, text)) ;~\\[-1.0ex]
\xline\verb~  }~\\[-1.0ex]
\xline\verb~~\\[-1.0ex]
\xline\verb~  @SuppressWarnings("unchecked")~\\[-1.0ex]
\xline\verb~  public Gxl decode(Document_gxl d) {~\\[-1.0ex]
\xline\verb~    final Gxl result = new Gxl() ;~\\[-1.0ex]
\xline\verb~    try {~\\[-1.0ex]
\xline\verb~      final GxlGraphElementModel gem = new GxlGraphElementModel(d) ;~\\[-1.0ex]
\xline\verb~      new Visitor() {~\\[-1.0ex]
\xline\verb~        @Override public void visit(Element_graph e) {~\\[-1.0ex]
\xline\verb~          result.get_graphs().add(new GraphDecoder(gem).decode(e)) ;~\\[-1.0ex]
\xline\verb~        }~\\[-1.0ex]
\xline\verb~      }.visit(d) ;~\\[-1.0ex]
\xline\verb~      result.set_location(d.getDocumentElement().getLocation()) ;~\\[-1.0ex]
\xline\verb~      return result ;~\\[-1.0ex]
\xline\verb~    }~\\[-1.0ex]
\xline\verb~    catch (HomonymousIdException e) {~\\[-1.0ex]
\xline\verb~      error(e.getElement2().getLocation(),~\\[-1.0ex]
\xline\verb~            "illegal homonymous element ID: " + e.getId()) ;~\\[-1.0ex]
\xline\verb~      if (e.getElement().getLocation() != null)~\\[-1.0ex]
\xline\verb~        hint(e.getElement().getLocation(), "previous use here") ;~\\[-1.0ex]
\xline\verb~    }~\\[-1.0ex]
\xline\verb~    catch (SynonymousIdException e) {~\\[-1.0ex]
\xline\verb~      error(e.getElement().getLocation(),~\\[-1.0ex]
\xline\verb~            "illegal synonymous element ID: " + e.getId2()) ;~\\[-1.0ex]
\xline\verb~      hint(e.getElement().getLocation(), "previous ID: " + e.getId()) ;~\\[-1.0ex]
\xline\verb~    }~\\[-1.0ex]
\xline\verb~    return null ;~\\[-1.0ex]
\xline\verb~  }~\\[-1.0ex]
\xline\verb~~\\[-1.0ex]
\xline\verb~~\\[-1.0ex]
\xline\verb~  private abstract class AttributedDecoder<R extends Attributed>~\\[-1.0ex]
\xline\verb~    extends Visitor {~\\[-1.0ex]
\xline\verb~~\\[-1.0ex]
\xline\verb~    protected R result ;~\\[-1.0ex]
\xline\verb~~\\[-1.0ex]
\xline\verb~    @Override public void visit(Element_attr e) {~\\[-1.0ex]
\xline\verb~      result.get_attrs().add(new AttrDecoder().decode(e)) ;~\\[-1.0ex]
\xline\verb~    }~\\[-1.0ex]
\xline\verb~~\\[-1.0ex]
\xline\verb~    protected final void visitNoMatch(Element_attr e) {~\\[-1.0ex]
\xline\verb~      super.visit(e) ;~\\[-1.0ex]
\xline\verb~    }~\\[-1.0ex]
\xline\verb~~\\[-1.0ex]
\xline\verb~  }~\\[-1.0ex]
\xline\verb~~\\[-1.0ex]
\xline\verb~  private abstract class TypedDecoder<R extends Typed> extends AttributedDecoder<R> {~\\[-1.0ex]
\xline\verb~~\\[-1.0ex]
\xline\verb~    @Override public void visit(Element_type e) {~\\[-1.0ex]
\xline\verb~      result.set_type(new Type(e.getAttr_xlink_href().getValue())) ;~\\[-1.0ex]
\xline\verb~    }~\\[-1.0ex]
\xline\verb~~\\[-1.0ex]
\xline\verb~  }~\\[-1.0ex]
\xline\verb~~\\[-1.0ex]
\xline\verb~  private class GraphDecoder extends TypedDecoder<Graph> {~\\[-1.0ex]
\xline\verb~~\\[-1.0ex]
\xline\verb~    final GxlGraphElementModel gem ;~\\[-1.0ex]
\xline\verb~    final Map<java.lang.String, Part> parts = new HashMap<java.lang.String, Part>() ;~\\[-1.0ex]
\xline\verb~~\\[-1.0ex]
\xline\verb~    GraphDecoder(GxlGraphElementModel gem) {~\\[-1.0ex]
\xline\verb~      this.gem = gem ;~\\[-1.0ex]
\xline\verb~    }~\\[-1.0ex]
\xline\verb~~\\[-1.0ex]
\xline\verb~    Part getPart(Element e, java.lang.String id) {~\\[-1.0ex]
\xline\verb~      if (parts.containsKey(id))~\\[-1.0ex]
\xline\verb~        return parts.get(id) ;~\\[-1.0ex]
\xline\verb~      else {~\\[-1.0ex]
\xline\verb~        error(e.getLocation(), "undefined IDREF: " + id) ;~\\[-1.0ex]
\xline\verb~        return null ;~\\[-1.0ex]
\xline\verb~      }~\\[-1.0ex]
\xline\verb~    }~\\[-1.0ex]
\xline\verb~~\\[-1.0ex]
\xline\verb~    public Graph decode(Element_graph e) {~\\[-1.0ex]
\xline\verb~      result = new Graph(e.getId()) ;~\\[-1.0ex]
\xline\verb~      visit(e) ;~\\[-1.0ex]
\xline\verb~      final Pass2 pass2 = new Pass2() ;~\\[-1.0ex]
\xline\verb~      try {~\\[-1.0ex]
\xline\verb~        for (Element_part p : GraphModels.postorder(gem.narrow(e)))~\\[-1.0ex]
\xline\verb~          pass2.visit(p) ;~\\[-1.0ex]
\xline\verb~      }~\\[-1.0ex]
\xline\verb~      catch (CycleException ex) {~\\[-1.0ex]
\xline\verb~        final Element_part p = (Element_part)ex.getNode() ;~\\[-1.0ex]
\xline\verb~        error(p.getLocation(), "cyclic cross-reference") ;~\\[-1.0ex]
\xline\verb~        // FIXME: find all members?~\\[-1.0ex]
\xline\verb~        return null ;~\\[-1.0ex]
\xline\verb~      }~\\[-1.0ex]
\xline\verb~      result.set_location(e.getLocation()) ;~\\[-1.0ex]
\xline\verb~      return result ;~\\[-1.0ex]
\xline\verb~    }~\\[-1.0ex]
\xline\verb~~\\[-1.0ex]
\xline\verb~    @Override public void visit(Element_graph.Attr_role a) {~\\[-1.0ex]
\xline\verb~      result.set_role(a.getValue()) ;~\\[-1.0ex]
\xline\verb~    }~\\[-1.0ex]
\xline\verb~~\\[-1.0ex]
\xline\verb~    @Override public void visit(Element_graph.Attr_edgeids a) {~\\[-1.0ex]
\xline\verb~      switch (a.getValue()) {~\\[-1.0ex]
\xline\verb~      case Value_true:~\\[-1.0ex]
\xline\verb~        result.set_edgeids(true) ; break ;~\\[-1.0ex]
\xline\verb~      case Value_false:~\\[-1.0ex]
\xline\verb~        result.set_edgeids(false) ; break ;~\\[-1.0ex]
\xline\verb~      }~\\[-1.0ex]
\xline\verb~    }~\\[-1.0ex]
\xline\verb~~\\[-1.0ex]
\xline\verb~    @Override public void visit(Element_graph.Attr_hypergraph a) {~\\[-1.0ex]
\xline\verb~      switch (a.getValue()) {~\\[-1.0ex]
\xline\verb~      case Value_true:~\\[-1.0ex]
\xline\verb~        result.set_edgeids(true) ; break ;~\\[-1.0ex]
\xline\verb~      case Value_false:~\\[-1.0ex]
\xline\verb~        result.set_edgeids(false) ; break ;~\\[-1.0ex]
\xline\verb~      }~\\[-1.0ex]
\xline\verb~    }~\\[-1.0ex]
\xline\verb~~\\[-1.0ex]
\xline\verb~    @Override public void visit(Element_graph.Attr_edgemode a) {~\\[-1.0ex]
\xline\verb~      switch (a.getValue()) {~\\[-1.0ex]
\xline\verb~      case Value_directed:~\\[-1.0ex]
\xline\verb~        result.set_edgemode(Edgemode.directed) ; break ;~\\[-1.0ex]
\xline\verb~      case Value_undirected:~\\[-1.0ex]
\xline\verb~        result.set_edgemode(Edgemode.undirected) ; break ;~\\[-1.0ex]
\xline\verb~      case Value_defaultdirected:~\\[-1.0ex]
\xline\verb~        result.set_edgemode(Edgemode.defaultdirected) ; break ;~\\[-1.0ex]
\xline\verb~      case Value_defaultundirected:~\\[-1.0ex]
\xline\verb~        result.set_edgemode(Edgemode.defaultundirected) ; break ;~\\[-1.0ex]
\xline\verb~      }~\\[-1.0ex]
\xline\verb~    }~\\[-1.0ex]
\xline\verb~~\\[-1.0ex]
\xline\verb~~\\[-1.0ex]
\xline\verb~    @Override public void visit(Element_node e) {}~\\[-1.0ex]
\xline\verb~    @Override public void visit(Element_edge e) {}~\\[-1.0ex]
\xline\verb~    @Override public void visit(Element_rel e) {}~\\[-1.0ex]
\xline\verb~~\\[-1.0ex]
\xline\verb~    private class Pass2 extends Visitor {~\\[-1.0ex]
\xline\verb~~\\[-1.0ex]
\xline\verb~      @Override public void visit(Element_node e) {~\\[-1.0ex]
\xline\verb~        result.get_parts().add(register(new NodeDecoder().decode(e))) ;~\\[-1.0ex]
\xline\verb~      }~\\[-1.0ex]
\xline\verb~~\\[-1.0ex]
\xline\verb~      @Override public void visit(Element_edge e) {~\\[-1.0ex]
\xline\verb~        result.get_parts().add(register(new EdgeDecoder().decode(e))) ;~\\[-1.0ex]
\xline\verb~      }~\\[-1.0ex]
\xline\verb~~\\[-1.0ex]
\xline\verb~      @Override public void visit(Element_rel e) {~\\[-1.0ex]
\xline\verb~        result.get_parts().add(register(new RelDecoder().decode(e))) ;~\\[-1.0ex]
\xline\verb~      }~\\[-1.0ex]
\xline\verb~~\\[-1.0ex]
\xline\verb~      private Part register(Part x) {~\\[-1.0ex]
\xline\verb~        if (x.get_id() != null)~\\[-1.0ex]
\xline\verb~          parts.put(x.get_id(), x) ;~\\[-1.0ex]
\xline\verb~        return x ;~\\[-1.0ex]
\xline\verb~      }~\\[-1.0ex]
\xline\verb~~\\[-1.0ex]
\xline\verb~    }~\\[-1.0ex]
\xline\verb~~\\[-1.0ex]
\xline\verb~~\\[-1.0ex]
\xline\verb~    private abstract class PartDecoder<R extends Part> extends TypedDecoder<R> {~\\[-1.0ex]
\xline\verb~~\\[-1.0ex]
\xline\verb~      @Override public void visit(Element_graph e) {~\\[-1.0ex]
\xline\verb~        result.get_graphs().add(new GraphDecoder(gem).decode(e)) ;~\\[-1.0ex]
\xline\verb~      }~\\[-1.0ex]
\xline\verb~~\\[-1.0ex]
\xline\verb~    }~\\[-1.0ex]
\xline\verb~~\\[-1.0ex]
\xline\verb~    private abstract class EdgyDecoder<R extends Edgy> extends PartDecoder<R> {~\\[-1.0ex]
\xline\verb~~\\[-1.0ex]
\xline\verb~      Boolean isdirected ;~\\[-1.0ex]
\xline\verb~~\\[-1.0ex]
\xline\verb~      @Override public void visit(Attr_isdirected a) {~\\[-1.0ex]
\xline\verb~        if (a.getValue() != null)~\\[-1.0ex]
\xline\verb~          switch (a.getValue()) {~\\[-1.0ex]
\xline\verb~          case Value_true:~\\[-1.0ex]
\xline\verb~            isdirected = true ; break ;~\\[-1.0ex]
\xline\verb~          case Value_false:~\\[-1.0ex]
\xline\verb~            isdirected = false ; break ;~\\[-1.0ex]
\xline\verb~          }~\\[-1.0ex]
\xline\verb~      }~\\[-1.0ex]
\xline\verb~~\\[-1.0ex]
\xline\verb~      @Override public void visit(Attr_id a) {~\\[-1.0ex]
\xline\verb~        result.set_id(a.getValue()) ;~\\[-1.0ex]
\xline\verb~      }~\\[-1.0ex]
\xline\verb~~\\[-1.0ex]
\xline\verb~    }~\\[-1.0ex]
\xline\verb~~\\[-1.0ex]
\xline\verb~~\\[-1.0ex]
\xline\verb~    private class NodeDecoder extends PartDecoder<Node> {~\\[-1.0ex]
\xline\verb~~\\[-1.0ex]
\xline\verb~      public Node decode(Element_node e) {~\\[-1.0ex]
\xline\verb~        result = new Node(e.getId()) ;~\\[-1.0ex]
\xline\verb~        visit(e) ;~\\[-1.0ex]
\xline\verb~        result.set_location(e.getLocation()) ;~\\[-1.0ex]
\xline\verb~        return result ;~\\[-1.0ex]
\xline\verb~      }~\\[-1.0ex]
\xline\verb~~\\[-1.0ex]
\xline\verb~    }~\\[-1.0ex]
\xline\verb~~\\[-1.0ex]
\xline\verb~~\\[-1.0ex]
\xline\verb~    private class EdgeDecoder extends EdgyDecoder<Edge> {~\\[-1.0ex]
\xline\verb~~\\[-1.0ex]
\xline\verb~      private Location<XMLDocumentIdentifier> location ;~\\[-1.0ex]
\xline\verb~~\\[-1.0ex]
\xline\verb~      public Edge decode(Element_edge e) {~\\[-1.0ex]
\xline\verb~        final Part from = getPart(e, e.getAttr_from().getValue()) ;~\\[-1.0ex]
\xline\verb~        final Part to = getPart(e, e.getAttr_to().getValue()) ;~\\[-1.0ex]
\xline\verb~        result = new Edge(from, to) ;~\\[-1.0ex]
\xline\verb~        location = e.getLocation() ;~\\[-1.0ex]
\xline\verb~        visit(e) ;~\\[-1.0ex]
\xline\verb~        result.set_location(e.getLocation()) ;~\\[-1.0ex]
\xline\verb~        return result ;~\\[-1.0ex]
\xline\verb~      }~\\[-1.0ex]
\xline\verb~~\\[-1.0ex]
\xline\verb~      @Override public void visit(Element_edge.Attr_fromorder a) {~\\[-1.0ex]
\xline\verb~        try {~\\[-1.0ex]
\xline\verb~          if (a.getValue() != null)~\\[-1.0ex]
\xline\verb~            result.set_fromorder(Integer.parseInt(a.getValue())) ;~\\[-1.0ex]
\xline\verb~        }~\\[-1.0ex]
\xline\verb~        catch (NumberFormatException ex) {~\\[-1.0ex]
\xline\verb~          error(location, "illegal integer: " + a.getValue()) ;~\\[-1.0ex]
\xline\verb~        }~\\[-1.0ex]
\xline\verb~      }~\\[-1.0ex]
\xline\verb~~\\[-1.0ex]
\xline\verb~      @Override public void visit(Element_edge.Attr_toorder a) {~\\[-1.0ex]
\xline\verb~        try {~\\[-1.0ex]
\xline\verb~          if (a.getValue() != null)~\\[-1.0ex]
\xline\verb~            result.set_toorder(Integer.parseInt(a.getValue())) ;~\\[-1.0ex]
\xline\verb~        }~\\[-1.0ex]
\xline\verb~        catch (NumberFormatException ex) {~\\[-1.0ex]
\xline\verb~          error(location, "illegal integer: " + a.getValue()) ;~\\[-1.0ex]
\xline\verb~        }~\\[-1.0ex]
\xline\verb~      }~\\[-1.0ex]
\xline\verb~~\\[-1.0ex]
\xline\verb~    }~\\[-1.0ex]
\xline\verb~~\\[-1.0ex]
\xline\verb~~\\[-1.0ex]
\xline\verb~    private class RelDecoder extends EdgyDecoder<Rel> {~\\[-1.0ex]
\xline\verb~~\\[-1.0ex]
\xline\verb~      public Rel decode(Element_rel e) {~\\[-1.0ex]
\xline\verb~        result = new Rel() ;~\\[-1.0ex]
\xline\verb~        visit(e) ;~\\[-1.0ex]
\xline\verb~        result.set_location(e.getLocation()) ;~\\[-1.0ex]
\xline\verb~        return result ;~\\[-1.0ex]
\xline\verb~      }~\\[-1.0ex]
\xline\verb~~\\[-1.0ex]
\xline\verb~      @Override public void visit(Attr_isdirected a) {~\\[-1.0ex]
\xline\verb~        if (a.getValue() != null)~\\[-1.0ex]
\xline\verb~          switch (a.getValue()) {~\\[-1.0ex]
\xline\verb~          case Value_true:~\\[-1.0ex]
\xline\verb~            result.set_isdirected(true) ; break ;~\\[-1.0ex]
\xline\verb~          case Value_false:~\\[-1.0ex]
\xline\verb~            result.set_isdirected(false) ; break ;~\\[-1.0ex]
\xline\verb~          }~\\[-1.0ex]
\xline\verb~      }~\\[-1.0ex]
\xline\verb~~\\[-1.0ex]
\xline\verb~      @Override public void visit(Element_relend e) {~\\[-1.0ex]
\xline\verb~        result.get_relends().add(new RelendDecoder().decode(e)) ;~\\[-1.0ex]
\xline\verb~      }~\\[-1.0ex]
\xline\verb~~\\[-1.0ex]
\xline\verb~    }~\\[-1.0ex]
\xline\verb~~\\[-1.0ex]
\xline\verb~    private class RelendDecoder extends AttributedDecoder<Relend> {~\\[-1.0ex]
\xline\verb~~\\[-1.0ex]
\xline\verb~      private Location<XMLDocumentIdentifier> location ;~\\[-1.0ex]
\xline\verb~~\\[-1.0ex]
\xline\verb~      public Relend decode(Element_relend e) {~\\[-1.0ex]
\xline\verb~        final Part target = getPart(e, e.getAttr_target().getValue()) ;~\\[-1.0ex]
\xline\verb~        result = new Relend(target) ;~\\[-1.0ex]
\xline\verb~        location = e.getLocation() ;~\\[-1.0ex]
\xline\verb~        visit(e) ;~\\[-1.0ex]
\xline\verb~        result.set_location(e.getLocation()) ;~\\[-1.0ex]
\xline\verb~        return result ;~\\[-1.0ex]
\xline\verb~      }~\\[-1.0ex]
\xline\verb~~\\[-1.0ex]
\xline\verb~      @Override public void visit(Element_relend.Attr_role a) {~\\[-1.0ex]
\xline\verb~        result.set_role(a.getValue()) ;~\\[-1.0ex]
\xline\verb~      }~\\[-1.0ex]
\xline\verb~~\\[-1.0ex]
\xline\verb~      @Override public void visit(Element_relend.Attr_direction a) {~\\[-1.0ex]
\xline\verb~        if (a.getValue() != null)~\\[-1.0ex]
\xline\verb~          switch (a.getValue()) {~\\[-1.0ex]
\xline\verb~          case Value_in:~\\[-1.0ex]
\xline\verb~            result.set_direction(Direction.in) ; break ;~\\[-1.0ex]
\xline\verb~          case Value_out:~\\[-1.0ex]
\xline\verb~            result.set_direction(Direction.out) ; break ;~\\[-1.0ex]
\xline\verb~          case Value_none:~\\[-1.0ex]
\xline\verb~            result.set_direction(Direction.none) ; break ;~\\[-1.0ex]
\xline\verb~          }~\\[-1.0ex]
\xline\verb~      }~\\[-1.0ex]
\xline\verb~~\\[-1.0ex]
\xline\verb~      @Override public void visit(Element_relend.Attr_startorder a) {~\\[-1.0ex]
\xline\verb~        try {~\\[-1.0ex]
\xline\verb~          if (a.getValue() != null)~\\[-1.0ex]
\xline\verb~            result.set_startorder(Integer.parseInt(a.getValue())) ;~\\[-1.0ex]
\xline\verb~        }~\\[-1.0ex]
\xline\verb~        catch (NumberFormatException ex) {~\\[-1.0ex]
\xline\verb~          error(location, "illegal integer: " + a.getValue()) ;~\\[-1.0ex]
\xline\verb~        }~\\[-1.0ex]
\xline\verb~      }~\\[-1.0ex]
\xline\verb~~\\[-1.0ex]
\xline\verb~      @Override public void visit(Element_relend.Attr_endorder a) {~\\[-1.0ex]
\xline\verb~        try {~\\[-1.0ex]
\xline\verb~          if (a.getValue() != null)~\\[-1.0ex]
\xline\verb~            result.set_endorder(Integer.parseInt(a.getValue())) ;~\\[-1.0ex]
\xline\verb~        }~\\[-1.0ex]
\xline\verb~        catch (NumberFormatException ex) {~\\[-1.0ex]
\xline\verb~          error(location, "illegal integer: " + a.getValue()) ;~\\[-1.0ex]
\xline\verb~        }~\\[-1.0ex]
\xline\verb~      }~\\[-1.0ex]
\xline\verb~~\\[-1.0ex]
\xline\verb~    }~\\[-1.0ex]
\xline\verb~~\\[-1.0ex]
\xline\verb~  }~\\[-1.0ex]
\xline\verb~~\\[-1.0ex]
\xline\verb~  private class AttrDecoder extends AttributedDecoder<Attr> {~\\[-1.0ex]
\xline\verb~~\\[-1.0ex]
\xline\verb~    public Attr decode(Element_attr e) {~\\[-1.0ex]
\xline\verb~      result = new Attr(e.getAttr_name().getValue(),~\\[-1.0ex]
\xline\verb~                        new ValDecoder().decode(e.getElem_1_anyval())) ;~\\[-1.0ex]
\xline\verb~      visitNoMatch(e) ;~\\[-1.0ex]
\xline\verb~      result.set_location(e.getLocation()) ;~\\[-1.0ex]
\xline\verb~      return result ;~\\[-1.0ex]
\xline\verb~    }~\\[-1.0ex]
\xline\verb~~\\[-1.0ex]
\xline\verb~    @Override public void visit(Attr_id a) {~\\[-1.0ex]
\xline\verb~      result.set_id(a.getValue()) ;~\\[-1.0ex]
\xline\verb~    }~\\[-1.0ex]
\xline\verb~~\\[-1.0ex]
\xline\verb~    @Override public void visit(Element_attr.Attr_kind a) {~\\[-1.0ex]
\xline\verb~      result.set_kind(a.getValue()) ;~\\[-1.0ex]
\xline\verb~    }~\\[-1.0ex]
\xline\verb~~\\[-1.0ex]
\xline\verb~  }~\\[-1.0ex]
\xline\verb~~\\[-1.0ex]
\xline\verb~  private class ValDecoder extends Visitor {~\\[-1.0ex]
\xline\verb~~\\[-1.0ex]
\xline\verb~    private Val val ;~\\[-1.0ex]
\xline\verb~~\\[-1.0ex]
\xline\verb~    public Val decode(Element_anyval e) {~\\[-1.0ex]
\xline\verb~      visit((Element)e) ;~\\[-1.0ex]
\xline\verb~      val.set_location(e.getLocation()) ;~\\[-1.0ex]
\xline\verb~      return val ;~\\[-1.0ex]
\xline\verb~    }~\\[-1.0ex]
\xline\verb~~\\[-1.0ex]
\xline\verb~    @Override public void visit(Element_locator e) {~\\[-1.0ex]
\xline\verb~      val = new Locator(e.getAttr_xlink_href().getValue()) ;~\\[-1.0ex]
\xline\verb~    }~\\[-1.0ex]
\xline\verb~~\\[-1.0ex]
\xline\verb~    @Override public void visit(Element_bool e) {~\\[-1.0ex]
\xline\verb~      final java.lang.String text = e.getPCData() ;~\\[-1.0ex]
\xline\verb~      if (text.equals("true"))~\\[-1.0ex]
\xline\verb~        val = new Bool(true) ;~\\[-1.0ex]
\xline\verb~      else if (text.equals("false"))~\\[-1.0ex]
\xline\verb~        val = new Bool(false) ;~\\[-1.0ex]
\xline\verb~      else~\\[-1.0ex]
\xline\verb~        error(e.getLocation(), "illegal boolean: " + text) ;~\\[-1.0ex]
\xline\verb~    }~\\[-1.0ex]
\xline\verb~~\\[-1.0ex]
\xline\verb~    @Override public void visit(Element_int e) {~\\[-1.0ex]
\xline\verb~      try {~\\[-1.0ex]
\xline\verb~        val = new Int(Integer.parseInt(e.getPCData())) ;~\\[-1.0ex]
\xline\verb~      }~\\[-1.0ex]
\xline\verb~      catch (NumberFormatException ex) {~\\[-1.0ex]
\xline\verb~        error(e.getLocation(), "illegal integer: " + e.getPCData()) ;~\\[-1.0ex]
\xline\verb~      }~\\[-1.0ex]
\xline\verb~    }~\\[-1.0ex]
\xline\verb~~\\[-1.0ex]
\xline\verb~    @Override public void visit(Element_float e) {~\\[-1.0ex]
\xline\verb~      try {~\\[-1.0ex]
\xline\verb~        val = new Float(Double.parseDouble(e.getPCData())) ;~\\[-1.0ex]
\xline\verb~      }~\\[-1.0ex]
\xline\verb~      catch (NumberFormatException ex) {~\\[-1.0ex]
\xline\verb~        error(e.getLocation(), "illegal float: " + e.getPCData()) ;~\\[-1.0ex]
\xline\verb~      }~\\[-1.0ex]
\xline\verb~    }~\\[-1.0ex]
\xline\verb~~\\[-1.0ex]
\xline\verb~    @Override public void visit(Element_string e) {~\\[-1.0ex]
\xline\verb~      val = new String(e.getPCData()) ;~\\[-1.0ex]
\xline\verb~    }~\\[-1.0ex]
\xline\verb~~\\[-1.0ex]
\xline\verb~    @Override public void visit(Element_enum e) {~\\[-1.0ex]
\xline\verb~      val = new Enum(e.getPCData()) ;~\\[-1.0ex]
\xline\verb~    }~\\[-1.0ex]
\xline\verb~~\\[-1.0ex]
\xline\verb~    @Override public void visit(Element_set e) {~\\[-1.0ex]
\xline\verb~      val = subvals(new Set(), e.getElems_1_anyval()) ;~\\[-1.0ex]
\xline\verb~    }~\\[-1.0ex]
\xline\verb~~\\[-1.0ex]
\xline\verb~    @Override public void visit(Element_seq e) {~\\[-1.0ex]
\xline\verb~      val = subvals(new Seq(), e.getElems_1_anyval()) ;~\\[-1.0ex]
\xline\verb~    }~\\[-1.0ex]
\xline\verb~~\\[-1.0ex]
\xline\verb~    @Override public void visit(Element_bag e) {~\\[-1.0ex]
\xline\verb~      val = subvals(new Bag(), e.getElems_1_anyval()) ;~\\[-1.0ex]
\xline\verb~    }~\\[-1.0ex]
\xline\verb~~\\[-1.0ex]
\xline\verb~    @Override public void visit(Element_tup e) {~\\[-1.0ex]
\xline\verb~      val = subvals(new Tup(), e.getElems_1_anyval()) ;~\\[-1.0ex]
\xline\verb~    }~\\[-1.0ex]
\xline\verb~~\\[-1.0ex]
\xline\verb~    private Aggregate subvals(Aggregate parent, Element_anyval... es) {~\\[-1.0ex]
\xline\verb~      for (Element_anyval e : es) parent.get_elems().add(decode(e)) ;~\\[-1.0ex]
\xline\verb~      return parent ;~\\[-1.0ex]
\xline\verb~    }~\\[-1.0ex]
\xline\verb~~\\[-1.0ex]
\xline\verb~  }~\\[-1.0ex]
\xline\verb~~\\[-1.0ex]
\xline\verb~}~\\[-1.0ex]

\egroup

\section{Encoding from the \umod{} model into the \tdom{} model of GXL}
\label{file_gxl_encode_java}
\bgroup\footnotesize
\xlinecounterreset{}
\xline\verb~package eu.bandm.ttc2011.case2.gxlcodec ;~\\[-1.0ex]
\xline\verb~~\\[-1.0ex]
\xline\verb~import eu.bandm.ttc2011.case2.tdom.* ;~\\[-1.0ex]
\xline\verb~import eu.bandm.tools.tdom.runtime.* ;~\\[-1.0ex]
\xline\verb~~\\[-1.0ex]
\xline\verb~import java.util.* ;~\\[-1.0ex]
\xline\verb~~\\[-1.0ex]
\xline\verb~~\\[-1.0ex]
\xline\verb~public class GxlEncoder {~\\[-1.0ex]
\xline\verb~~\\[-1.0ex]
\xline\verb~  private static final Element_graph[] GRAPHS = new Element_graph[0] ;~\\[-1.0ex]
\xline\verb~  private static final Element_attr[] ATTRS = new Element_attr[0] ;~\\[-1.0ex]
\xline\verb~  private static final Element_part[] PARTS = new Element_part[0] ;~\\[-1.0ex]
\xline\verb~  private static final Element_relend[] RELENDS = new Element_relend[0] ;~\\[-1.0ex]
\xline\verb~~\\[-1.0ex]
\xline\verb~  public Document_gxl encode(Gxl x) {~\\[-1.0ex]
\xline\verb~    final Collection<Element_graph> graphs = new ArrayList<Element_graph>() ;~\\[-1.0ex]
\xline\verb~    new SinglePass() {~\\[-1.0ex]
\xline\verb~      @Override protected void action(Graph x) {~\\[-1.0ex]
\xline\verb~        graphs.add(new GraphEncoder().encode(x)) ;~\\[-1.0ex]
\xline\verb~      }~\\[-1.0ex]
\xline\verb~    }.match(x) ;~\\[-1.0ex]
\xline\verb~    final Element_gxl e = new Element_gxl(graphs.toArray(GRAPHS)) ;~\\[-1.0ex]
\xline\verb~    return new Document_gxl(e) ;~\\[-1.0ex]
\xline\verb~  }~\\[-1.0ex]
\xline\verb~~\\[-1.0ex]
\xline\verb~~\\[-1.0ex]
\xline\verb~  private static abstract class AttributedEncoder extends SinglePass {~\\[-1.0ex]
\xline\verb~~\\[-1.0ex]
\xline\verb~    final Collection<Element_attr> attrs = new ArrayList<Element_attr>() ;~\\[-1.0ex]
\xline\verb~~\\[-1.0ex]
\xline\verb~    @Override protected void action(Attr x) {~\\[-1.0ex]
\xline\verb~      attrs.add(new AttrEncoder().encode(x)) ;~\\[-1.0ex]
\xline\verb~    }~\\[-1.0ex]
\xline\verb~~\\[-1.0ex]
\xline\verb~    protected final void actionNoMatch(Attr x) {~\\[-1.0ex]
\xline\verb~      super.action(x) ;~\\[-1.0ex]
\xline\verb~    }~\\[-1.0ex]
\xline\verb~~\\[-1.0ex]
\xline\verb~  }~\\[-1.0ex]
\xline\verb~~\\[-1.0ex]
\xline\verb~~\\[-1.0ex]
\xline\verb~  private static abstract class TypedEncoder extends AttributedEncoder {~\\[-1.0ex]
\xline\verb~~\\[-1.0ex]
\xline\verb~    Element_type type ;~\\[-1.0ex]
\xline\verb~~\\[-1.0ex]
\xline\verb~    @Override protected void action(final Type x) {~\\[-1.0ex]
\xline\verb~      type = new Element_type() {~\\[-1.0ex]
\xline\verb~          @Override protected void initAttrs() {~\\[-1.0ex]
\xline\verb~            getAttr_xlink_href().setValue(x.get_href()) ;~\\[-1.0ex]
\xline\verb~          }~\\[-1.0ex]
\xline\verb~        } ;~\\[-1.0ex]
\xline\verb~    }~\\[-1.0ex]
\xline\verb~~\\[-1.0ex]
\xline\verb~  }~\\[-1.0ex]
\xline\verb~~\\[-1.0ex]
\xline\verb~~\\[-1.0ex]
\xline\verb~  private static class GraphEncoder extends TypedEncoder {~\\[-1.0ex]
\xline\verb~~\\[-1.0ex]
\xline\verb~    final Collection<Element_part> parts = new ArrayList<Element_part>() ;~\\[-1.0ex]
\xline\verb~~\\[-1.0ex]
\xline\verb~    Element_graph encode(final Graph x) {~\\[-1.0ex]
\xline\verb~      match(x) ;~\\[-1.0ex]
\xline\verb~      return new Element_graph(type,~\\[-1.0ex]
\xline\verb~                               attrs.toArray(ATTRS),~\\[-1.0ex]
\xline\verb~                               parts.toArray(PARTS)) {~\\[-1.0ex]
\xline\verb~        @Override protected void initAttrs() {~\\[-1.0ex]
\xline\verb~          getAttr_id().setValue(x.get_id()) ;~\\[-1.0ex]
\xline\verb~        }~\\[-1.0ex]
\xline\verb~      } ;~\\[-1.0ex]
\xline\verb~    }~\\[-1.0ex]
\xline\verb~~\\[-1.0ex]
\xline\verb~    @Override protected void action(Node x) {~\\[-1.0ex]
\xline\verb~      parts.add(new NodeEncoder().encode(x)) ;~\\[-1.0ex]
\xline\verb~    }~\\[-1.0ex]
\xline\verb~~\\[-1.0ex]
\xline\verb~    @Override protected void action(Edge x) {~\\[-1.0ex]
\xline\verb~      parts.add(new EdgeEncoder().encode(x)) ;~\\[-1.0ex]
\xline\verb~    }~\\[-1.0ex]
\xline\verb~~\\[-1.0ex]
\xline\verb~    @Override protected void action(Rel x) {~\\[-1.0ex]
\xline\verb~      parts.add(new RelEncoder().encode(x)) ;~\\[-1.0ex]
\xline\verb~    }~\\[-1.0ex]
\xline\verb~~\\[-1.0ex]
\xline\verb~  }~\\[-1.0ex]
\xline\verb~~\\[-1.0ex]
\xline\verb~  private static abstract class PartEncoder extends TypedEncoder {~\\[-1.0ex]
\xline\verb~~\\[-1.0ex]
\xline\verb~    final Collection<Element_graph> graphs = new ArrayList<Element_graph>() ;~\\[-1.0ex]
\xline\verb~~\\[-1.0ex]
\xline\verb~    @Override protected void action(Graph x) {~\\[-1.0ex]
\xline\verb~      graphs.add(new GraphEncoder().encode(x)) ;~\\[-1.0ex]
\xline\verb~    }~\\[-1.0ex]
\xline\verb~~\\[-1.0ex]
\xline\verb~  }~\\[-1.0ex]
\xline\verb~~\\[-1.0ex]
\xline\verb~  private static class NodeEncoder extends PartEncoder {~\\[-1.0ex]
\xline\verb~~\\[-1.0ex]
\xline\verb~    Element_node encode(final Node x) {~\\[-1.0ex]
\xline\verb~      match(x) ;~\\[-1.0ex]
\xline\verb~      return new Element_node(type,~\\[-1.0ex]
\xline\verb~                              attrs.toArray(ATTRS),~\\[-1.0ex]
\xline\verb~                              graphs.toArray(GRAPHS)) {~\\[-1.0ex]
\xline\verb~        @Override protected void initAttrs() {~\\[-1.0ex]
\xline\verb~          getAttr_id().setValue(x.get_id()) ;~\\[-1.0ex]
\xline\verb~        }~\\[-1.0ex]
\xline\verb~      } ;~\\[-1.0ex]
\xline\verb~    }~\\[-1.0ex]
\xline\verb~  }~\\[-1.0ex]
\xline\verb~~\\[-1.0ex]
\xline\verb~  private static abstract class EdgyEncoder extends PartEncoder {~\\[-1.0ex]
\xline\verb~~\\[-1.0ex]
\xline\verb~    java.lang.String id ;~\\[-1.0ex]
\xline\verb~    Attr_isdirected.Value isdirected ;~\\[-1.0ex]
\xline\verb~~\\[-1.0ex]
\xline\verb~    @Override protected void action(Edgy x) {~\\[-1.0ex]
\xline\verb~      super.action(x) ;~\\[-1.0ex]
\xline\verb~      id = x.get_id() ;~\\[-1.0ex]
\xline\verb~      if (x.get_isdirected() != null)~\\[-1.0ex]
\xline\verb~        isdirected = x.get_isdirected()~\\[-1.0ex]
\xline\verb~          ? Attr_isdirected.Value.Value_true~\\[-1.0ex]
\xline\verb~          : Attr_isdirected.Value.Value_false ;~\\[-1.0ex]
\xline\verb~    }~\\[-1.0ex]
\xline\verb~~\\[-1.0ex]
\xline\verb~  }~\\[-1.0ex]
\xline\verb~~\\[-1.0ex]
\xline\verb~  private static class EdgeEncoder extends EdgyEncoder {~\\[-1.0ex]
\xline\verb~~\\[-1.0ex]
\xline\verb~    Element_edge encode(final Edge x) {~\\[-1.0ex]
\xline\verb~      match(x) ;~\\[-1.0ex]
\xline\verb~      return new Element_edge(type,~\\[-1.0ex]
\xline\verb~                              attrs.toArray(ATTRS),~\\[-1.0ex]
\xline\verb~                              graphs.toArray(GRAPHS)) {~\\[-1.0ex]
\xline\verb~        @Override protected void initAttrs() {~\\[-1.0ex]
\xline\verb~          getAttr_id().setValue(id) ;~\\[-1.0ex]
\xline\verb~          getAttr_from().setValue(x.get_from().get_id()) ;~\\[-1.0ex]
\xline\verb~          getAttr_to().setValue(x.get_to().get_id()) ;~\\[-1.0ex]
\xline\verb~          if (x.get_fromorder() != null)~\\[-1.0ex]
\xline\verb~            getAttr_fromorder().setValue(x.get_fromorder().toString()) ;~\\[-1.0ex]
\xline\verb~          if (x.get_toorder() != null)~\\[-1.0ex]
\xline\verb~            getAttr_toorder().setValue(x.get_toorder().toString()) ;~\\[-1.0ex]
\xline\verb~          getAttr_isdirected().setValue(isdirected) ;~\\[-1.0ex]
\xline\verb~        }~\\[-1.0ex]
\xline\verb~      } ;~\\[-1.0ex]
\xline\verb~    }~\\[-1.0ex]
\xline\verb~~\\[-1.0ex]
\xline\verb~  }~\\[-1.0ex]
\xline\verb~~\\[-1.0ex]
\xline\verb~  private static class RelEncoder extends EdgyEncoder {~\\[-1.0ex]
\xline\verb~~\\[-1.0ex]
\xline\verb~    private Collection<Element_relend> relends = new ArrayList<Element_relend>() ;~\\[-1.0ex]
\xline\verb~~\\[-1.0ex]
\xline\verb~    Element_rel encode(final Rel x) {~\\[-1.0ex]
\xline\verb~      match(x) ;~\\[-1.0ex]
\xline\verb~      return new Element_rel(type,~\\[-1.0ex]
\xline\verb~                             attrs.toArray(ATTRS),~\\[-1.0ex]
\xline\verb~                             graphs.toArray(GRAPHS),~\\[-1.0ex]
\xline\verb~                             relends.toArray(RELENDS)) {~\\[-1.0ex]
\xline\verb~        @Override protected void initAttrs() {~\\[-1.0ex]
\xline\verb~          getAttr_id().setValue(id) ;~\\[-1.0ex]
\xline\verb~          getAttr_isdirected().setValue(isdirected) ;~\\[-1.0ex]
\xline\verb~        }~\\[-1.0ex]
\xline\verb~      } ;~\\[-1.0ex]
\xline\verb~    }~\\[-1.0ex]
\xline\verb~~\\[-1.0ex]
\xline\verb~    @Override protected void action(Relend x) {~\\[-1.0ex]
\xline\verb~      relends.add(new RelendEncoder().encode(x)) ;~\\[-1.0ex]
\xline\verb~    }~\\[-1.0ex]
\xline\verb~~\\[-1.0ex]
\xline\verb~  }~\\[-1.0ex]
\xline\verb~~\\[-1.0ex]
\xline\verb~  private static class RelendEncoder extends AttributedEncoder {~\\[-1.0ex]
\xline\verb~~\\[-1.0ex]
\xline\verb~    Element_relend encode(final Relend x) {~\\[-1.0ex]
\xline\verb~      match(x) ;~\\[-1.0ex]
\xline\verb~      return new Element_relend(attrs.toArray(ATTRS)) {~\\[-1.0ex]
\xline\verb~        @Override protected void initAttrs() {~\\[-1.0ex]
\xline\verb~          getAttr_target().setValue(x.get_target().get_id()) ;~\\[-1.0ex]
\xline\verb~          getAttr_role().setValue(x.get_role()) ;~\\[-1.0ex]
\xline\verb~          if (x.get_direction() != null)~\\[-1.0ex]
\xline\verb~            switch (x.get_direction()) {~\\[-1.0ex]
\xline\verb~            case in:~\\[-1.0ex]
\xline\verb~              getAttr_direction().setValue(Element_relend.Attr_direction.Value.Value_in) ; break ;~\\[-1.0ex]
\xline\verb~            case out:~\\[-1.0ex]
\xline\verb~              getAttr_direction().setValue(Element_relend.Attr_direction.Value.Value_out) ; break ;~\\[-1.0ex]
\xline\verb~            case none:~\\[-1.0ex]
\xline\verb~              getAttr_direction().setValue(Element_relend.Attr_direction.Value.Value_none) ; break ;~\\[-1.0ex]
\xline\verb~            }~\\[-1.0ex]
\xline\verb~          if (x.get_startorder() != null)~\\[-1.0ex]
\xline\verb~            getAttr_startorder().setValue(x.get_startorder().toString()) ;~\\[-1.0ex]
\xline\verb~          if (x.get_endorder() != null)~\\[-1.0ex]
\xline\verb~            getAttr_endorder().setValue(x.get_endorder().toString()) ;~\\[-1.0ex]
\xline\verb~        }~\\[-1.0ex]
\xline\verb~      } ;~\\[-1.0ex]
\xline\verb~    }~\\[-1.0ex]
\xline\verb~~\\[-1.0ex]
\xline\verb~  }~\\[-1.0ex]
\xline\verb~~\\[-1.0ex]
\xline\verb~  private static class AttrEncoder extends AttributedEncoder {~\\[-1.0ex]
\xline\verb~~\\[-1.0ex]
\xline\verb~    private Element_anyval val ;~\\[-1.0ex]
\xline\verb~~\\[-1.0ex]
\xline\verb~    Element_attr encode(final Attr x) {~\\[-1.0ex]
\xline\verb~      actionNoMatch(x) ;~\\[-1.0ex]
\xline\verb~      return new Element_attr(attrs.toArray(ATTRS), val) {~\\[-1.0ex]
\xline\verb~        @Override protected void initAttrs() {~\\[-1.0ex]
\xline\verb~          getAttr_id().setValue(x.get_id()) ;     ~\\[-1.0ex]
\xline\verb~          getAttr_name().setValue(x.get_name()) ;~\\[-1.0ex]
\xline\verb~          getAttr_kind().setValue(x.get_kind()) ;~\\[-1.0ex]
\xline\verb~        }~\\[-1.0ex]
\xline\verb~      } ;~\\[-1.0ex]
\xline\verb~    }~\\[-1.0ex]
\xline\verb~~\\[-1.0ex]
\xline\verb~    @Override protected void action(Val x) {~\\[-1.0ex]
\xline\verb~      val = new ValEncoder().encode(x) ;~\\[-1.0ex]
\xline\verb~    }~\\[-1.0ex]
\xline\verb~~\\[-1.0ex]
\xline\verb~  }~\\[-1.0ex]
\xline\verb~~\\[-1.0ex]
\xline\verb~  private static class ValEncoder extends SinglePass {~\\[-1.0ex]
\xline\verb~~\\[-1.0ex]
\xline\verb~    private Element_anyval val ;~\\[-1.0ex]
\xline\verb~~\\[-1.0ex]
\xline\verb~    Element_anyval encode(Val x) {~\\[-1.0ex]
\xline\verb~      match(x) ;~\\[-1.0ex]
\xline\verb~      return val ;~\\[-1.0ex]
\xline\verb~    }~\\[-1.0ex]
\xline\verb~~\\[-1.0ex]
\xline\verb~    @Override protected void action(final Locator x) {~\\[-1.0ex]
\xline\verb~      val = new Element_locator() {~\\[-1.0ex]
\xline\verb~          @Override protected void initAttrs() {~\\[-1.0ex]
\xline\verb~            getAttr_xlink_href().setValue(x.get_href()) ;~\\[-1.0ex]
\xline\verb~          }~\\[-1.0ex]
\xline\verb~        } ;~\\[-1.0ex]
\xline\verb~    }~\\[-1.0ex]
\xline\verb~~\\[-1.0ex]
\xline\verb~    @Override protected void action(Bool x) {~\\[-1.0ex]
\xline\verb~      val = new Element_bool(java.lang.String.valueOf(x.get_value())) ;~\\[-1.0ex]
\xline\verb~    }~\\[-1.0ex]
\xline\verb~~\\[-1.0ex]
\xline\verb~    @Override protected void action(Int x) {~\\[-1.0ex]
\xline\verb~      val = new Element_int(java.lang.String.valueOf(x.get_value())) ;~\\[-1.0ex]
\xline\verb~    }~\\[-1.0ex]
\xline\verb~~\\[-1.0ex]
\xline\verb~    @Override protected void action(Float x) {~\\[-1.0ex]
\xline\verb~      val = new Element_float(java.lang.String.valueOf(x.get_value())) ;~\\[-1.0ex]
\xline\verb~    }~\\[-1.0ex]
\xline\verb~~\\[-1.0ex]
\xline\verb~    @Override protected void action(String x) {~\\[-1.0ex]
\xline\verb~      val = new Element_string(x.get_value()) ;~\\[-1.0ex]
\xline\verb~    }~\\[-1.0ex]
\xline\verb~~\\[-1.0ex]
\xline\verb~    @Override protected void action(Enum x) {~\\[-1.0ex]
\xline\verb~      val = new Element_enum(x.get_value()) ;~\\[-1.0ex]
\xline\verb~    }~\\[-1.0ex]
\xline\verb~~\\[-1.0ex]
\xline\verb~    @Override protected void action(Seq x) {~\\[-1.0ex]
\xline\verb~      val = new Element_seq(subvals(x.get_elems())) ;~\\[-1.0ex]
\xline\verb~    }~\\[-1.0ex]
\xline\verb~~\\[-1.0ex]
\xline\verb~    @Override protected void action(Set x) {~\\[-1.0ex]
\xline\verb~      val = new Element_set(subvals(x.get_elems())) ;~\\[-1.0ex]
\xline\verb~    }~\\[-1.0ex]
\xline\verb~~\\[-1.0ex]
\xline\verb~    @Override protected void action(Bag x) {~\\[-1.0ex]
\xline\verb~      val = new Element_bag(subvals(x.get_elems())) ;~\\[-1.0ex]
\xline\verb~    }~\\[-1.0ex]
\xline\verb~~\\[-1.0ex]
\xline\verb~    @Override protected void action(Tup x) {~\\[-1.0ex]
\xline\verb~      val = new Element_tup(subvals(x.get_elems())) ;~\\[-1.0ex]
\xline\verb~    }~\\[-1.0ex]
\xline\verb~~\\[-1.0ex]
\xline\verb~    private Element_anyval[] subvals(Collection<Val> xs) {~\\[-1.0ex]
\xline\verb~      final Collection<Element_anyval> elems = new ArrayList<Element_anyval>(xs.size()) ;~\\[-1.0ex]
\xline\verb~      for (Val x : xs)~\\[-1.0ex]
\xline\verb~        elems.add(new ValEncoder().encode(x)) ;~\\[-1.0ex]
\xline\verb~      return elems.toArray(new Element_anyval[xs.size()]) ;~\\[-1.0ex]
\xline\verb~    }~\\[-1.0ex]
\xline\verb~~\\[-1.0ex]
\xline\verb~  }~\\[-1.0ex]
\xline\verb~~\\[-1.0ex]
\xline\verb~}~\\[-1.0ex]

\egroup

\section{The Firm Mode, as source to \umod{}}
\label{file_firm_umod}
\xlinecounterreset{}
\bgroup\footnotesize
\xline\verb~MODEL Firm =~\\[-1.0ex]
\xline\verb~~\\[-1.0ex]
\xline\verb~/** Firm Model version 0.1~\\[-1.0ex]
\xline\verb~   ~\\[-1.0ex]
\xline\verb~   This model is a more recent version (20110813). It is ~\\[-1.0ex]
\xline\verb~   based on the documentation contained in~\\[-1.0ex]
\xline\verb~   http://www.info.uni-karlsruhe.de/papers/firmdoc.ps.gz~\\[-1.0ex]
\xline\verb~   <p>~\\[-1.0ex]
\xline\verb~   (The more recent table in ~\\[-1.0ex]
\xline\verb~   http://pp.info.uni-karlsruhe.de/uploads/publikationen/braun11wir.pdf~\\[-1.0ex]
\xline\verb~   differs slightly, and is taken instead for some definitios not yet used in~\\[-1.0ex]
\xline\verb~   the testcase, e.g. "alloc" etc.)~\\[-1.0ex]
\xline\verb~   <p>~\\[-1.0ex]
\xline\verb~   The fundamental principles of the this Firm realization by an umod model:~\\[-1.0ex]
\xline\verb~   <p>~\\[-1.0ex]
\xline\verb~   Node classes are translated to umod model element classes / java classes.~\\[-1.0ex]
\xline\verb~   Edges are translated to references. ~\\[-1.0ex]
\xline\verb~   <p>~\\[-1.0ex]
\xline\verb~   The above-mentioned documentation introduces "sockets" for separating~\\[-1.0ex]
\xline\verb~   and qualifying edges. ~\\[-1.0ex]
\xline\verb~   <p>~\\[-1.0ex]
\xline\verb~   With the outgoing edges this is realized explicitly by~\\[-1.0ex]
\xline\verb~   mapping "sockets" to fields. Further sequential order on the edges is realized~\\[-1.0ex]
\xline\verb~   by using appropriate container types (SEQ, MAP, etc)~\\[-1.0ex]
\xline\verb~   <p>~\\[-1.0ex]
\xline\verb~   With the incoming edges the "sockets" are always given implicitly, ~\\[-1.0ex]
\xline\verb~   by the "starting socket" of the edge.~\\[-1.0ex]
\xline\verb~   <p>~\\[-1.0ex]
\xline\verb~   Whenever further order of incoming edges is required, e.g. when a~\\[-1.0ex]
\xline\verb~   "call" or "start" node produces more than one numeric data, or~\\[-1.0ex]
\xline\verb~   with the "phi" and "cond" node, the data is ~\\[-1.0ex]
\xline\verb~   treated as forming a  "Tuple" and explicit "Proj" nodes are inserted~\\[-1.0ex]
\xline\verb~   between the producing socket and the reading conumer.~\\[-1.0ex]
\xline\verb~ */~\\[-1.0ex]
\xline\verb~~\\[-1.0ex]
\xline\verb~EXT Location = eu.bandm.tools.message.Location~\\[-1.0ex]
\xline\verb~    <eu.bandm.tools.message.XMLDocumentIdentifier>~\\[-1.0ex]
\xline\verb~~\\[-1.0ex]
\xline\verb~~\\[-1.0ex]
\xline\verb~VISITOR 0 Matcher MULTIPHASE ;~\\[-1.0ex]
\xline\verb~VISITOR 0 SimpleVisitor ; ~\\[-1.0ex]
\xline\verb~VISITOR 0 Rewriter IS REWRITER ; ~\\[-1.0ex]
\xline\verb~VISITOR 0 CoRewriter IS COREWRITER ; ~\\[-1.0ex]
\xline\verb~~\\[-1.0ex]
\xline\verb~ENUM NumericType =   p, Iu, Is, Su, Ss, Bu, Bs, E, F, D, C, b,~\\[-1.0ex]
\xline\verb~                     NotYetComputed~\\[-1.0ex]
\xline\verb~~\\[-1.0ex]
\xline\verb~INTERFACE~\\[-1.0ex]
\xline\verb~  MemoryState~\\[-1.0ex]
\xline\verb~  Numeric~\\[-1.0ex]
\xline\verb~  ControlFlow~\\[-1.0ex]
\xline\verb~~\\[-1.0ex]
\xline\verb~TOPLEVEL CLASS~\\[-1.0ex]
\xline\verb~~\\[-1.0ex]
\xline\verb~FirmNode ~\\[-1.0ex]
\xline\verb~        location OPT Location           ! C 0/0 ; ~\\[-1.0ex]
\xline\verb~        gxlId OPT string ~\\[-1.0ex]
\xline\verb~~\\[-1.0ex]
\xline\verb~| Unknown IMPLEMENTS MemoryState, Numeric, ControlFlow~\\[-1.0ex]
\xline\verb~~\\[-1.0ex]
\xline\verb~| Bad IMPLEMENTS MemoryState, Numeric, ControlFlow~\\[-1.0ex]
\xline\verb~~\\[-1.0ex]
\xline\verb~<< JAVA public static final Bad BAD = new Bad(); $$~\\[-1.0ex]
\xline\verb~~\\[-1.0ex]
\xline\verb~| Block~\\[-1.0ex]
\xline\verb~        predecs  int->ControlFlow       !                       V 0/0 R ; // 1/0 R ; ~\\[-1.0ex]
\xline\verb~~\\[-1.0ex]
\xline\verb~| BlockNode~\\[-1.0ex]
\xline\verb~        block Block                     ! C 0/1                 V 0/0 ; ~\\[-1.0ex]
\xline\verb~~\\[-1.0ex]
\xline\verb~| | End~\\[-1.0ex]
\xline\verb~| | Cond ~\\[-1.0ex]
\xline\verb~        selector Numeric                ! C 0/2                 V 0/0 ;  ~\\[-1.0ex]
\xline\verb~        // int or boolean !!~\\[-1.0ex]
\xline\verb~~\\[-1.0ex]
\xline\verb~| | ControlFlowNode IMPLEMENTS ControlFlow ~\\[-1.0ex]
\xline\verb~| | MemoryNode IMPLEMENTS MemoryState~\\[-1.0ex]
\xline\verb~| | NumericNode IMPLEMENTS Numeric~\\[-1.0ex]
\xline\verb~~\\[-1.0ex]
\xline\verb~~\\[-1.0ex]
\xline\verb~EXTEND CLASS~\\[-1.0ex]
\xline\verb~~\\[-1.0ex]
\xline\verb~ControlFlowNode~\\[-1.0ex]
\xline\verb~| Start IMPLEMENTS MemoryState, Numeric~\\[-1.0ex]
\xline\verb~| Return ~\\[-1.0ex]
\xline\verb~        memstate MemoryState            ! C 0/2                 V 0/0; ~\\[-1.0ex]
\xline\verb~        results SEQ Numeric             !                       V 0/1 ; ~\\[-1.0ex]
\xline\verb~| Jmp~\\[-1.0ex]
\xline\verb~| Proj_X~\\[-1.0ex]
\xline\verb~        input Cond                      ! C 0/2                 V 0/0 ;~\\[-1.0ex]
\xline\verb~        selection int                   ! C 0/3                 ;~\\[-1.0ex]
\xline\verb~~\\[-1.0ex]
\xline\verb~MemoryNode~\\[-1.0ex]
\xline\verb~| NoMem~\\[-1.0ex]
\xline\verb~| Sync ~\\[-1.0ex]
\xline\verb~        predecs SET MemoryState         !                       V 0/0 ; ~\\[-1.0ex]
\xline\verb~| MemModification~\\[-1.0ex]
\xline\verb~    preState MemoryState                ! C 0/2                 V 0/1 ; ~\\[-1.0ex]
\xline\verb~    position Numeric                    ! C 0/3                 V 0/2 ; ~\\[-1.0ex]
\xline\verb~| | Store ~\\[-1.0ex]
\xline\verb~    valueNumeric Numeric                ! C 0/4                 V 0/3 ; ~\\[-1.0ex]
\xline\verb~| | Free~\\[-1.0ex]
\xline\verb~~\\[-1.0ex]
\xline\verb~~\\[-1.0ex]
\xline\verb~NumericNode~\\[-1.0ex]
\xline\verb~        type NumericType                ! C 0/2 ;~\\[-1.0ex]
\xline\verb~~\\[-1.0ex]
\xline\verb~| Phi~\\[-1.0ex]
\xline\verb~        alternatives int->Numeric       !                       V 0/0 R ; ~\\[-1.0ex]
\xline\verb~// must be in sync with "this.block.predecs" !~\\[-1.0ex]
\xline\verb~~\\[-1.0ex]
\xline\verb~~\\[-1.0ex]
\xline\verb~| Proj_N // de-compose a tupel ~\\[-1.0ex]
\xline\verb~        predec Numeric /*Tuple_N*/      ! C 0/3                 V 0/0 R ; ~\\[-1.0ex]
\xline\verb~        pos int                         ! C 0/4 ;~\\[-1.0ex]
\xline\verb~~\\[-1.0ex]
\xline\verb~| Tuple_N IMPLEMENTS Numeric~\\[-1.0ex]
\xline\verb~        combines SEQ Numeric            !                       V 0/0 ; ~\\[-1.0ex]
\xline\verb~| | Call IMPLEMENTS MemoryState~\\[-1.0ex]
\xline\verb~        preState MemoryState            ! C 0/3                 V 0/1 ; ~\\[-1.0ex]
\xline\verb~~\\[-1.0ex]
\xline\verb~| Nullary ~\\[-1.0ex]
\xline\verb~| | NumericConst~\\[-1.0ex]
\xline\verb~    unparsedValue string            ! C 0/3;~\\[-1.0ex]
\xline\verb~    intValue      OPT int ~\\[-1.0ex]
\xline\verb~    floatValue    OPT float ~\\[-1.0ex]
\xline\verb~    stringValue   OPT string~\\[-1.0ex]
\xline\verb~    charValue     OPT char~\\[-1.0ex]
\xline\verb~    booleanValue  OPT bool~\\[-1.0ex]
\xline\verb~~\\[-1.0ex]
\xline\verb~| | SymConst~\\[-1.0ex]
\xline\verb~    unparsedValue string            ! C 0/3;~\\[-1.0ex]
\xline\verb~~\\[-1.0ex]
\xline\verb~| Unary~\\[-1.0ex]
\xline\verb~    on  Numeric                 ! C 0/3         V 0/0 ;~\\[-1.0ex]
\xline\verb~| | Conv~\\[-1.0ex]
\xline\verb~| | Minus~\\[-1.0ex]
\xline\verb~| | Not~\\[-1.0ex]
\xline\verb~| | Rotl~\\[-1.0ex]
\xline\verb~| | Shl~\\[-1.0ex]
\xline\verb~| | Shr~\\[-1.0ex]
\xline\verb~| | Shrs~\\[-1.0ex]
\xline\verb~~\\[-1.0ex]
\xline\verb~| Binary ~\\[-1.0ex]
\xline\verb~    left Numeric                        ! C 0/3         V 0/0 ;~\\[-1.0ex]
\xline\verb~    right Numeric                       ! C 0/4         V 0/1 ;~\\[-1.0ex]
\xline\verb~| | Add~\\[-1.0ex]
\xline\verb~| | And~\\[-1.0ex]
\xline\verb~| | Div~\\[-1.0ex]
\xline\verb~| | Eor~\\[-1.0ex]
\xline\verb~| | Mod~\\[-1.0ex]
\xline\verb~| | Mul~\\[-1.0ex]
\xline\verb~| | Or~\\[-1.0ex]
\xline\verb~| | Sub~\\[-1.0ex]
\xline\verb~~\\[-1.0ex]
\xline\verb~| | Cmp~\\[-1.0ex]
\xline\verb~~\\[-1.0ex]
\xline\verb~| Ternary~\\[-1.0ex]
\xline\verb~  first  Numeric                        ! C 0/3         V 0/1 ; // boolean!~\\[-1.0ex]
\xline\verb~  second Numeric                        ! C 0/4         V 0/2 ;~\\[-1.0ex]
\xline\verb~  third  Numeric                        ! C 0/5         V 0/3 ;~\\[-1.0ex]
\xline\verb~| | Mux~\\[-1.0ex]
\xline\verb~~\\[-1.0ex]
\xline\verb~| MemOperations IMPLEMENTS MemoryState~\\[-1.0ex]
\xline\verb~    preState MemoryState                ! C 0/3         V 0/1 ; ~\\[-1.0ex]
\xline\verb~| | Alloc ~\\[-1.0ex]
\xline\verb~    size Numeric                        ! C 0/4         V 0/2 ; ~\\[-1.0ex]
\xline\verb~| | Load ~\\[-1.0ex]
\xline\verb~    position Numeric                    ! C 0/4         V 0/2 ; ~\\[-1.0ex]
\xline\verb~| | Sel ~\\[-1.0ex]
\xline\verb~    position Numeric                    ! C 0/4         V 0/2 ; ~\\[-1.0ex]
\xline\verb~~\\[-1.0ex]
\xline\verb~| BadNumeric~\\[-1.0ex]
\xline\verb~| UnknownNumeric~\\[-1.0ex]
\xline\verb~~\\[-1.0ex]
\xline\verb~~\\[-1.0ex]
\xline\verb~<<JAVA~\\[-1.0ex]
\xline\verb~  public static class VisitBlocksOnce extends SimpleVisitor {~\\[-1.0ex]
\xline\verb~    final protected java.util.Set<Block> visitedBlocks ~\\[-1.0ex]
\xline\verb~    = new java.util.HashSet<Block>();~\\[-1.0ex]
\xline\verb~    public java.util.Set<Block> visited(){~\\[-1.0ex]
\xline\verb~      return java.util.Collections.unmodifiableSet(visitedBlocks);~\\[-1.0ex]
\xline\verb~    }~\\[-1.0ex]
\xline\verb~    @Override public void action (final Block block){~\\[-1.0ex]
\xline\verb~      if (visitedBlocks.contains(block))~\\[-1.0ex]
\xline\verb~        return ; ~\\[-1.0ex]
\xline\verb~      visitedBlocks.add(block);~\\[-1.0ex]
\xline\verb~      super.action(block);~\\[-1.0ex]
\xline\verb~    }~\\[-1.0ex]
\xline\verb~  }~\\[-1.0ex]
\xline\verb~$$~\\[-1.0ex]
\xline\verb~~\\[-1.0ex]
\xline\verb~/* == NOT YET IMPLEMENTED:~\\[-1.0ex]
\xline\verb~ASM             B x variable   -->   variable   Inline assembler~\\[-1.0ex]
\xline\verb~==================*/~\\[-1.0ex]
\xline\verb~~\\[-1.0ex]
\xline\verb~END MODEL~\\[-1.0ex]

\egroup

\section{Decoding a GXL model into a Firm model}
\label{file_gxl2firm_java}
\xlinecounterreset{}
\bgroup\footnotesize
\xline\verb~package eu.bandm.ttc2011.case2.transformations ;~\\[-1.0ex]
\xline\verb~~\\[-1.0ex]
\xline\verb~import java.util.Map; ~\\[-1.0ex]
\xline\verb~import java.util.HashMap; ~\\[-1.0ex]
\xline\verb~import java.util.Set; ~\\[-1.0ex]
\xline\verb~import java.util.HashSet; ~\\[-1.0ex]
\xline\verb~~\\[-1.0ex]
\xline\verb~import eu.bandm.tools.message.MessageReceiver ; ~\\[-1.0ex]
\xline\verb~import eu.bandm.tools.message.MessageTee ; ~\\[-1.0ex]
\xline\verb~import eu.bandm.tools.message.MessageCounter  ; ~\\[-1.0ex]
\xline\verb~import eu.bandm.tools.message.SimpleMessage ;  ~\\[-1.0ex]
\xline\verb~import eu.bandm.tools.message.XMLDocumentIdentifier ; ~\\[-1.0ex]
\xline\verb~import eu.bandm.tools.message.Location;~\\[-1.0ex]
\xline\verb~~\\[-1.0ex]
\xline\verb~import eu.bandm.tools.ops.Multimap ; ~\\[-1.0ex]
\xline\verb~import eu.bandm.tools.ops.HashMultimap ; ~\\[-1.0ex]
\xline\verb~~\\[-1.0ex]
\xline\verb~import eu.bandm.tools.ops.Pattern ; ~\\[-1.0ex]
\xline\verb~import static eu.bandm.tools.ops.Pattern.eq ; ~\\[-1.0ex]
\xline\verb~import static eu.bandm.tools.ops.Pattern.any; ~\\[-1.0ex]
\xline\verb~import static eu.bandm.tools.ops.Pattern.Variable ; ~\\[-1.0ex]
\xline\verb~~\\[-1.0ex]
\xline\verb~~\\[-1.0ex]
\xline\verb~import eu.bandm.ttc2011.case2.gxlcodec.* ; ~\\[-1.0ex]
\xline\verb~import eu.bandm.ttc2011.case2.firm_01.* ; ~\\[-1.0ex]
\xline\verb~~\\[-1.0ex]
\xline\verb~import java.lang.String ; ~\\[-1.0ex]
\xline\verb~~\\[-1.0ex]
\xline\verb~~\\[-1.0ex]
\xline\verb~/** Translates between two umod-generated java models,~\\[-1.0ex]
\xline\verb~namely from the {@link eu.bandm.ttc2011.case2.gxlcodec.GxlModel gxl model}~\\[-1.0ex]
\xline\verb~into the {@link eu.bandm.ttc2011.case2.firm_01.Firm firm model}.~\\[-1.0ex]
\xline\verb~~\\[-1.0ex]
\xline\verb~~\\[-1.0ex]
\xline\verb~Please note the nomenclature:~\\[-1.0ex]
\xline\verb~<pre>~\\[-1.0ex]
\xline\verb~    MODEL Glx~\\[-1.0ex]
\xline\verb~      Node~\\[-1.0ex]
\xline\verb~      Edge~\\[-1.0ex]
\xline\verb~~\\[-1.0ex]
\xline\verb~    MODEL Firm~\\[-1.0ex]
\xline\verb~      FirmNode~\\[-1.0ex]
\xline\verb~      | Block~\\[-1.0ex]
\xline\verb~      | BlockNode // node "contained in a block"~\\[-1.0ex]
\xline\verb~</pre>~\\[-1.0ex]
\xline\verb~*/~\\[-1.0ex]
\xline\verb~~\\[-1.0ex]
\xline\verb~public class Gxl2Firm {~\\[-1.0ex]
\xline\verb~~\\[-1.0ex]
\xline\verb~  final Set <Node> all_gxl_nodes = new HashSet<Node>();~\\[-1.0ex]
\xline\verb~~\\[-1.0ex]
\xline\verb~  final Map <String, String> metanodeid2nodeclassname~\\[-1.0ex]
\xline\verb~    = new HashMap<String, String>();~\\[-1.0ex]
\xline\verb~  final Map <String, String> nodeclassname2metanodeid~\\[-1.0ex]
\xline\verb~    = new HashMap<String, String>();~\\[-1.0ex]
\xline\verb~~\\[-1.0ex]
\xline\verb~  final Map <Node, Map<Integer, Edge>> node2outgoing~\\[-1.0ex]
\xline\verb~    = new HashMap<Node, Map<Integer, Edge>>();~\\[-1.0ex]
\xline\verb~  ~\\[-1.0ex]
\xline\verb~  final Map <Node, String> node2nodetype = new HashMap<Node, String>();~\\[-1.0ex]
\xline\verb~  ~\\[-1.0ex]
\xline\verb~  final Map <Node, FirmNode> correspondent = new HashMap <Node, FirmNode>();~\\[-1.0ex]
\xline\verb~~\\[-1.0ex]
\xline\verb~  final Set <Node> undertranslation = new HashSet<Node>();~\\[-1.0ex]
\xline\verb~~\\[-1.0ex]
\xline\verb~  ~\\[-1.0ex]
\xline\verb~  protected void ERROR(Exception ex){~\\[-1.0ex]
\xline\verb~    msgT.receive(SimpleMessage.<XMLDocumentIdentifier>error(ex, "exception"));~\\[-1.0ex]
\xline\verb~  }~\\[-1.0ex]
\xline\verb~  protected void ERROR(Location<XMLDocumentIdentifier> loc, String ex){~\\[-1.0ex]
\xline\verb~    msgT.receive(SimpleMessage.<XMLDocumentIdentifier>error(loc, ex));~\\[-1.0ex]
\xline\verb~  }~\\[-1.0ex]
\xline\verb~  protected void ERROR(String ex){~\\[-1.0ex]
\xline\verb~    msgT.receive(SimpleMessage.<XMLDocumentIdentifier>error(ex));~\\[-1.0ex]
\xline\verb~  }~\\[-1.0ex]
\xline\verb~  protected void WARNING(Location<XMLDocumentIdentifier> loc, String ex){~\\[-1.0ex]
\xline\verb~    msgT.receive(SimpleMessage.<XMLDocumentIdentifier>warning(loc, ex));~\\[-1.0ex]
\xline\verb~  }~\\[-1.0ex]
\xline\verb~  ~\\[-1.0ex]
\xline\verb~~\\[-1.0ex]
\xline\verb~  protected MessageTee<SimpleMessage<XMLDocumentIdentifier>> msgT~\\[-1.0ex]
\xline\verb~    = new  MessageTee<SimpleMessage<XMLDocumentIdentifier>>();~\\[-1.0ex]
\xline\verb~  protected MessageCounter<SimpleMessage<XMLDocumentIdentifier>> msgC ~\\[-1.0ex]
\xline\verb~    = new MessageCounter<SimpleMessage<XMLDocumentIdentifier>>();~\\[-1.0ex]
\xline\verb~  {msgT.add(msgC);}~\\[-1.0ex]
\xline\verb~~\\[-1.0ex]
\xline\verb~  protected boolean severeErrors(){~\\[-1.0ex]
\xline\verb~    return msgC.getCriticalCount()>0;~\\[-1.0ex]
\xline\verb~  }~\\[-1.0ex]
\xline\verb~~\\[-1.0ex]
\xline\verb~  ~\\[-1.0ex]
\xline\verb~  protected Gxl gxl ; ~\\[-1.0ex]
\xline\verb~  protected Graph metamodel, objectmodel ; ~\\[-1.0ex]
\xline\verb~  Node startnode = null, glx_endnode = null ; ~\\[-1.0ex]
\xline\verb~~\\[-1.0ex]
\xline\verb~  End firm_endnode = null ; ~\\[-1.0ex]
\xline\verb~  ~\\[-1.0ex]
\xline\verb~  public static final String STRING_HREF_METAMODEL ~\\[-1.0ex]
\xline\verb~    = "http://www.gupro.de/GXL/gxl-1.0.gxl#gxl-1.0" ;~\\[-1.0ex]
\xline\verb~  public static final String STRING_HREF_OBJECTMODEL            = "#Firm" ;~\\[-1.0ex]
\xline\verb~  public static final String STRING_HREF_OBJECTMODEL_2~\\[-1.0ex]
\xline\verb~           = "#InstructionSelection"  ;~\\[-1.0ex]
\xline\verb~  public static final String STRING_NODENAME_START              = "Start" ;~\\[-1.0ex]
\xline\verb~  public static final String STRING_NODENAME_END                = "End" ;~\\[-1.0ex]
\xline\verb~  public static final String ATTRIBUTE_NAME_NAME         = "name" ;~\\[-1.0ex]
\xline\verb~  public static final String ATTRIBUTE_NAME_POSITION     = "position" ;~\\[-1.0ex]
\xline\verb~  public static final String ATTRIBUTE_NAME_VALUE        = "value" ;~\\[-1.0ex]
\xline\verb~~\\[-1.0ex]
\xline\verb~  public static final String STRING_BLOCKTYPE                   = "Block" ;~\\[-1.0ex]
\xline\verb~  public static final String STRING_STARTBLOCKTYPE              = "StartBlock" ;~\\[-1.0ex]
\xline\verb~  public static final String STRING_ENDBLOCKTYPE                = "EndBlock" ;~\\[-1.0ex]
\xline\verb~  public static final int ORDER_BLOCK_CONTAINMENT = -1 ; ~\\[-1.0ex]
\xline\verb~~\\[-1.0ex]
\xline\verb~  public static final String STRING_EDGETYPE_TRUE               = "#True" ;~\\[-1.0ex]
\xline\verb~  public static final String STRING_EDGETYPE_FALSE              = "#False" ;~\\[-1.0ex]
\xline\verb~~\\[-1.0ex]
\xline\verb~  // numeric parts of right side of a start node is a tuple as follows:~\\[-1.0ex]
\xline\verb~  public static final int POSITION_STACKFRAME_ARGUMENT = 0 ; ~\\[-1.0ex]
\xline\verb~  public static final int POSITION_HEAPPOINTER_ARGUMENT = 1 ; ~\\[-1.0ex]
\xline\verb~  public static final int POSITION_ARGUMENTS_ARGUMENT = 2 ; ~\\[-1.0ex]
\xline\verb~~\\[-1.0ex]
\xline\verb~~\\[-1.0ex]
\xline\verb~  protected Pattern<Attributed> hasAttr(String name, Pattern<? super Val> val) {~\\[-1.0ex]
\xline\verb~    return Attributed.get_attrs(Attr.get_name(eq(name))~\\[-1.0ex]
\xline\verb~                                .and(Attr.get_val(val)).somewhere()) ;~\\[-1.0ex]
\xline\verb~  }~\\[-1.0ex]
\xline\verb~  ~\\[-1.0ex]
\xline\verb~  ~\\[-1.0ex]
\xline\verb~  /** ASSUME the GXL model contains TWO graphs:~\\[-1.0ex]
\xline\verb~      <ol>~\\[-1.0ex]
\xline\verb~      <li> A first Graph representing the meta-model, i.e. describing "Firm" itself.~\\[-1.0ex]
\xline\verb~      <br>~\\[-1.0ex]
\xline\verb~      Its "type" field conatains the ~\\[-1.0ex]
\xline\verb~      href-value "{@code http://www.gupro.de/GXL/gxl-1.0.gxl#gxl-1.0}" <br>~\\[-1.0ex]
\xline\verb~      Its "id" contains "SCE_Firm". <br>~\\[-1.0ex]
\xline\verb~      Most of its nodes represent node classes of the object graph.<br>~\\[-1.0ex]
\xline\verb~      With these nodes, the child element {@code <Gxl:Attr name="name">}~\\[-1.0ex]
\xline\verb~      gives a hint which node class is meant.~\\[-1.0ex]
\xline\verb~      <li>~\\[-1.0ex]
\xline\verb~      As second graph in the gxl structure, there is the object model itself.~\\[-1.0ex]
\xline\verb~      </ol>~\\[-1.0ex]
\xline\verb~      Its "type" field contains a reference (local to this file!)~\\[-1.0ex]
\xline\verb~      to the node with id = "Firm" in the meta-model <br>~\\[-1.0ex]
\xline\verb~      <br>~\\[-1.0ex]
\xline\verb~      All firm nodes in the are represented as gxl nodes in the second~\\[-1.0ex]
\xline\verb~      graph. Theur "type" attribute~\\[-1.0ex]
\xline\verb~      is a href (local to this file!) to a node in the meta model.~\\[-1.0ex]
\xline\verb~      <br>~\\[-1.0ex]
\xline\verb~      All edges go between nodes (gxl would permit others!). ~\\[-1.0ex]
\xline\verb~      <br>(NB The order~\\[-1.0ex]
\xline\verb~      of the outgoing edges is NOT given by the gxl-defined field "endorder",~\\[-1.0ex]
\xline\verb~      but by a dedicated gxl-attribute with name {@link ATTRIBUTE_NAME_VALUE}!)~\\[-1.0ex]
\xline\verb~      <br>~\\[-1.0ex]
\xline\verb~      This method is the central entry method. <br>~\\[-1.0ex]
\xline\verb~      <br>~\\[-1.0ex]
\xline\verb~      It decodes the "model" graph contained in the Gxl into a Firm object.~\\[-1.0ex]
\xline\verb~      <ol>~\\[-1.0ex]
\xline\verb~      <li> It calls "new Collector().match()" to separate both models~\\[-1.0ex]
\xline\verb~      and to collect all edges between nodes.~\\[-1.0ex]
\xline\verb~      <li> It calls "find_start_and_end()" to identify the start and the end node ~\\[-1.0ex]
\xline\verb~      in the gxl model.~\\[-1.0ex]
\xline\verb~      <li> It calls "convert()" on the end node to produce the "firm" end node.~\\[-1.0ex]
\xline\verb~      </ol>~\\[-1.0ex]
\xline\verb~      Afterwards the user can inquire for ~\\[-1.0ex]
\xline\verb~      <ol>~\\[-1.0ex]
\xline\verb~     <li> the "endnode" of the Firm model,~\\[-1.0ex]
\xline\verb~     <li> the meta-model as a Gxl graph, ~\\[-1.0ex]
\xline\verb~     <li> and the map from node class names to identifiers of the meta model.~\\[-1.0ex]
\xline\verb~     </ol>~\\[-1.0ex]
\xline\verb~     The last two are required for subsequent re-encoding of a tranformed Firm model.~\\[-1.0ex]
\xline\verb~  */~\\[-1.0ex]
\xline\verb~  public void decode_meta_and_object~\\[-1.0ex]
\xline\verb~    (final MessageReceiver<SimpleMessage<XMLDocumentIdentifier>>msg,~\\[-1.0ex]
\xline\verb~     final Gxl gxl){~\\[-1.0ex]
\xline\verb~    this.msgT.add(msg);~\\[-1.0ex]
\xline\verb~    this.gxl = gxl ; ~\\[-1.0ex]
\xline\verb~    metamodel=objectmodel=null ; ~\\[-1.0ex]
\xline\verb~    new Collector().match(gxl);~\\[-1.0ex]
\xline\verb~    if (all_gxl_nodes.isEmpty())~\\[-1.0ex]
\xline\verb~      ERROR("no model nodes found in file "~\\[-1.0ex]
\xline\verb~            +"(perhaps due to a wrong <graph><type xlink:href='#???'> entry ?)");~\\[-1.0ex]
\xline\verb~    find_start_and_end();~\\[-1.0ex]
\xline\verb~    if (glx_endnode==null)~\\[-1.0ex]
\xline\verb~      ERROR("no endnode in graph");~\\[-1.0ex]
\xline\verb~    if (startnode==null)~\\[-1.0ex]
\xline\verb~      ERROR("no startnode in graph");~\\[-1.0ex]
\xline\verb~    if (severeErrors())~\\[-1.0ex]
\xline\verb~      return; ~\\[-1.0ex]
\xline\verb~    undertranslation.add(glx_endnode);~\\[-1.0ex]
\xline\verb~    firm_endnode = (End)convert(glx_endnode);~\\[-1.0ex]
\xline\verb~    return  ;~\\[-1.0ex]
\xline\verb~  }~\\[-1.0ex]
\xline\verb~  ~\\[-1.0ex]
\xline\verb~  public End get_resulted_endnode(){~\\[-1.0ex]
\xline\verb~    return firm_endnode;~\\[-1.0ex]
\xline\verb~  }~\\[-1.0ex]
\xline\verb~  public Graph get_resulted_metamodel(){~\\[-1.0ex]
\xline\verb~    return metamodel;~\\[-1.0ex]
\xline\verb~  }~\\[-1.0ex]
\xline\verb~  public Map<String,String> get_nodeClassName2metaNodeId(){~\\[-1.0ex]
\xline\verb~    return java.util.Collections.unmodifiableMap(nodeclassname2metanodeid);~\\[-1.0ex]
\xline\verb~  }~\\[-1.0ex]
\xline\verb~~\\[-1.0ex]
\xline\verb~  protected Map<Integer, Edge> get_node2outgoing (final Node from){~\\[-1.0ex]
\xline\verb~    Map<Integer,Edge> map = node2outgoing.get(from);~\\[-1.0ex]
\xline\verb~    if (map==null){~\\[-1.0ex]
\xline\verb~      map = new HashMap<Integer,Edge>();~\\[-1.0ex]
\xline\verb~      node2outgoing.put(from, map);~\\[-1.0ex]
\xline\verb~    }~\\[-1.0ex]
\xline\verb~    return map ;~\\[-1.0ex]
\xline\verb~  }~\\[-1.0ex]
\xline\verb~~\\[-1.0ex]
\xline\verb~  // ==========================================================================~\\[-1.0ex]
\xline\verb~~\\[-1.0ex]
\xline\verb~  /** Analyses the gxl files as presented by the authors of the task.~\\[-1.0ex]
\xline\verb~      <br/>~\\[-1.0ex]
\xline\verb~      1) Memorizes the meta-model graph as a whole, for subsequent re-encoding.~\\[-1.0ex]
\xline\verb~      <br/>~\\[-1.0ex]
\xline\verb~      2) makes a map from meta-model node ids to firm class names, also for re-encoding~\\[-1.0ex]
\xline\verb~      <br/>~\\[-1.0ex]
\xline\verb~      3) identifies the model graph, and then <br/>~\\[-1.0ex]
\xline\verb~      <br/>~\\[-1.0ex]
\xline\verb~         - collects all of its nodes into {@link #all_gxl_nodes} <br/>and memorizes~\\[-1.0ex]
\xline\verb~         - collects all edges into {@link #ndoe2outgoing} : Node -> int -> Node <br/>~\\[-1.0ex]
\xline\verb~  */~\\[-1.0ex]
\xline\verb~~\\[-1.0ex]
\xline\verb~  public class Collector  extends SinglePass {~\\[-1.0ex]
\xline\verb~~\\[-1.0ex]
\xline\verb~    boolean metamode = true ; ~\\[-1.0ex]
\xline\verb~    String metanodeid ; ~\\[-1.0ex]
\xline\verb~    ~\\[-1.0ex]
\xline\verb~    @Override public void action (final Graph g){~\\[-1.0ex]
\xline\verb~      final Type t = g.get_type();~\\[-1.0ex]
\xline\verb~      if (t==null){~\\[-1.0ex]
\xline\verb~        ERROR("Type == null not forseen.");~\\[-1.0ex]
\xline\verb~        return; ~\\[-1.0ex]
\xline\verb~      }~\\[-1.0ex]
\xline\verb~      final String href = t.get_href();~\\[-1.0ex]
\xline\verb~      if (href.equals(STRING_HREF_METAMODEL))~\\[-1.0ex]
\xline\verb~        if (metamodel!=null){~\\[-1.0ex]
\xline\verb~          ERROR("More than one metamodel?");~\\[-1.0ex]
\xline\verb~          return; ~\\[-1.0ex]
\xline\verb~        }~\\[-1.0ex]
\xline\verb~        else{~\\[-1.0ex]
\xline\verb~          metamodel=g ; ~\\[-1.0ex]
\xline\verb~          metamode=true ; ~\\[-1.0ex]
\xline\verb~        }~\\[-1.0ex]
\xline\verb~      else if (href.equals(STRING_HREF_OBJECTMODEL)~\\[-1.0ex]
\xline\verb~               || href.equals(STRING_HREF_OBJECTMODEL_2))~\\[-1.0ex]
\xline\verb~        if (objectmodel!=null){~\\[-1.0ex]
\xline\verb~          ERROR("More than one objectmodel?");~\\[-1.0ex]
\xline\verb~          return; ~\\[-1.0ex]
\xline\verb~        }~\\[-1.0ex]
\xline\verb~        else{~\\[-1.0ex]
\xline\verb~          objectmodel=g ; ~\\[-1.0ex]
\xline\verb~          metamode=false;~\\[-1.0ex]
\xline\verb~        }~\\[-1.0ex]
\xline\verb~      super.action(g);~\\[-1.0ex]
\xline\verb~    }~\\[-1.0ex]
\xline\verb~    ~\\[-1.0ex]
\xline\verb~    @Override public void action(final Node n){~\\[-1.0ex]
\xline\verb~      if (metamode){~\\[-1.0ex]
\xline\verb~        final Variable<String> var = Pattern.<String>variable();~\\[-1.0ex]
\xline\verb~        if(hasAttr(ATTRIBUTE_NAME_NAME, ~\\[-1.0ex]
\xline\verb~                   eu.bandm.ttc2011.case2.gxlcodec.String.cast~\\[-1.0ex]
\xline\verb~                   (eu.bandm.ttc2011.case2.gxlcodec.String.get_value(var))~\\[-1.0ex]
\xline\verb~                   .or(any)).match(n)){~\\[-1.0ex]
\xline\verb~          final String s = var.getValue();~\\[-1.0ex]
\xline\verb~          if (s==null)~\\[-1.0ex]
\xline\verb~            ERROR(n.get_location(),~\\[-1.0ex]
\xline\verb~                    "name attribute of a meta-node must carry a string value");~\\[-1.0ex]
\xline\verb~          else{~\\[-1.0ex]
\xline\verb~            final String metanodeid = n.get_id();~\\[-1.0ex]
\xline\verb~            metanodeid2nodeclassname.put(metanodeid, s);~\\[-1.0ex]
\xline\verb~            nodeclassname2metanodeid.put(s, metanodeid);~\\[-1.0ex]
\xline\verb~          }~\\[-1.0ex]
\xline\verb~        }~\\[-1.0ex]
\xline\verb~        else~\\[-1.0ex]
\xline\verb~          ;~\\[-1.0ex]
\xline\verb~          //      WARNING(n.get_location(), ~\\[-1.0ex]
\xline\verb~          //              "meta-node should have an attribute named \"name\"");~\\[-1.0ex]
\xline\verb~      }~\\[-1.0ex]
\xline\verb~      else~\\[-1.0ex]
\xline\verb~        all_gxl_nodes.add(n);~\\[-1.0ex]
\xline\verb~    }~\\[-1.0ex]
\xline\verb~    ~\\[-1.0ex]
\xline\verb~~\\[-1.0ex]
\xline\verb~~\\[-1.0ex]
\xline\verb~    @Override public void action(final Edge e){~\\[-1.0ex]
\xline\verb~      if (metamode)~\\[-1.0ex]
\xline\verb~        return ; ~\\[-1.0ex]
\xline\verb~      /** NO, the examples included in the task use <attr name="position"><int> ~\\[-1.0ex]
\xline\verb~      Integer fromOrder = e.get_fromorder();~\\[-1.0ex]
\xline\verb~       */~\\[-1.0ex]
\xline\verb~      Integer fromOrder = findFromOrder(e);~\\[-1.0ex]
\xline\verb~      ~\\[-1.0ex]
\xline\verb~      if (fromOrder==null){~\\[-1.0ex]
\xline\verb~        ERROR(e.get_location(), "from order requrired for all edges");~\\[-1.0ex]
\xline\verb~        return ; ~\\[-1.0ex]
\xline\verb~      }~\\[-1.0ex]
\xline\verb~      if (!(e.get_from() instanceof Node)){~\\[-1.0ex]
\xline\verb~        ERROR(e.get_location(), "start of each edge muat be of node type");~\\[-1.0ex]
\xline\verb~        return ; ~\\[-1.0ex]
\xline\verb~      }~\\[-1.0ex]
\xline\verb~      final Node from = (Node)e.get_from();~\\[-1.0ex]
\xline\verb~      final Map<Integer,Edge> map = get_node2outgoing(from); ~\\[-1.0ex]
\xline\verb~      final Edge oldEdge = map.get(fromOrder);~\\[-1.0ex]
\xline\verb~      if (oldEdge!=null){~\\[-1.0ex]
\xline\verb~        ERROR (e.get_location(),~\\[-1.0ex]
\xline\verb~               "duplicate use of order number "+fromOrder+" for this node and "~\\[-1.0ex]
\xline\verb~               +" previously for "+oldEdge.get_location());~\\[-1.0ex]
\xline\verb~        return ;~\\[-1.0ex]
\xline\verb~      } ~\\[-1.0ex]
\xline\verb~      map.put(fromOrder, e);~\\[-1.0ex]
\xline\verb~    }~\\[-1.0ex]
\xline\verb~    ~\\[-1.0ex]
\xline\verb~    ~\\[-1.0ex]
\xline\verb~    protected Integer findFromOrder(Edge e){~\\[-1.0ex]
\xline\verb~      Variable<Integer> i = Pattern.<Integer>variable();~\\[-1.0ex]
\xline\verb~      /*Edge.get_attrs(Attr.get_id(eq(ATTRIBUTE_NAME_POSITION))~\\[-1.0ex]
\xline\verb~                           .and(Attr.get_val(Int.cast(Int.get_value(i)))~\\[-1.0ex]
\xline\verb~                           ).somewhere()*/~\\[-1.0ex]
\xline\verb~      hasAttr(ATTRIBUTE_NAME_POSITION,~\\[-1.0ex]
\xline\verb~              Int.cast(Int.get_value(i))~\\[-1.0ex]
\xline\verb~              ).match(e);~\\[-1.0ex]
\xline\verb~      return i.getValue();~\\[-1.0ex]
\xline\verb~    }~\\[-1.0ex]
\xline\verb~~\\[-1.0ex]
\xline\verb~  }//class Collector~\\[-1.0ex]
\xline\verb~  ~\\[-1.0ex]
\xline\verb~  ~\\[-1.0ex]
\xline\verb~  // ==========================================================================~\\[-1.0ex]
\xline\verb~  ~\\[-1.0ex]
\xline\verb~  protected void find_start_and_end (/*GLO IN all_gxl_nodes,~\\[-1.0ex]
\xline\verb~                                       GLO OUT startnode, endnode, node2nodeclass */){~\\[-1.0ex]
\xline\verb~    for (final Node node : all_gxl_nodes){~\\[-1.0ex]
\xline\verb~      final Type t = node.get_type();~\\[-1.0ex]
\xline\verb~      if (t==null){~\\[-1.0ex]
\xline\verb~        ERROR(node.get_location(), "empty node type attribute?");~\\[-1.0ex]
\xline\verb~        return ; ~\\[-1.0ex]
\xline\verb~      }~\\[-1.0ex]
\xline\verb~      final String s = t.get_href();~\\[-1.0ex]
\xline\verb~      final String s1 = s.substring(1);~\\[-1.0ex]
\xline\verb~      final String nodeclass = metanodeid2nodeclassname.get(s1);~\\[-1.0ex]
\xline\verb~      if (nodeclass==null){~\\[-1.0ex]
\xline\verb~        ERROR(node.get_location(), "unknown node type \""+s+"\" ?");~\\[-1.0ex]
\xline\verb~        return ; ~\\[-1.0ex]
\xline\verb~      }~\\[-1.0ex]
\xline\verb~      node2nodetype.put(node, nodeclass);~\\[-1.0ex]
\xline\verb~      if (nodeclass.equals(STRING_NODENAME_END))~\\[-1.0ex]
\xline\verb~        if (glx_endnode != null)~\\[-1.0ex]
\xline\verb~          ERROR(node.get_location(), "A second end node found");~\\[-1.0ex]
\xline\verb~        else ~\\[-1.0ex]
\xline\verb~          glx_endnode = node;~\\[-1.0ex]
\xline\verb~      else if (nodeclass.equals(STRING_NODENAME_START))~\\[-1.0ex]
\xline\verb~        if (startnode != null)~\\[-1.0ex]
\xline\verb~          ERROR(node.get_location(), "A second start node found");~\\[-1.0ex]
\xline\verb~        else ~\\[-1.0ex]
\xline\verb~          startnode = node;~\\[-1.0ex]
\xline\verb~    }~\\[-1.0ex]
\xline\verb~  }~\\[-1.0ex]
\xline\verb~~\\[-1.0ex]
\xline\verb~  // --------------------------------------------------------------------------~\\[-1.0ex]
\xline\verb~~\\[-1.0ex]
\xline\verb~  protected Map<Start, Proj_N> allArguments = new HashMap<Start, Proj_N>();~\\[-1.0ex]
\xline\verb~  ~\\[-1.0ex]
\xline\verb~  protected Proj_N getAllArgs (final Start start){~\\[-1.0ex]
\xline\verb~    if (allArguments.containsKey(start))~\\[-1.0ex]
\xline\verb~      return (allArguments.get(start));~\\[-1.0ex]
\xline\verb~    final Proj_N proj = new Proj_N(start.get_location(), start.get_block(),~\\[-1.0ex]
\xline\verb~                                   NumericType.NotYetComputed,~\\[-1.0ex]
\xline\verb~                                   start, POSITION_ARGUMENTS_ARGUMENT);~\\[-1.0ex]
\xline\verb~    allArguments.put(start,proj);~\\[-1.0ex]
\xline\verb~    return proj;~\\[-1.0ex]
\xline\verb~  }~\\[-1.0ex]
\xline\verb~~\\[-1.0ex]
\xline\verb~~\\[-1.0ex]
\xline\verb~  protected String noEdgeFound_errormsg (final Node node,~\\[-1.0ex]
\xline\verb~                                         final int position,~\\[-1.0ex]
\xline\verb~                                         final Class cls){~\\[-1.0ex]
\xline\verb~    return "Expecting an edge starting at position "+position~\\[-1.0ex]
\xline\verb~      +" of node "+node.get_id()+" of type "+node2nodetype.get(node)~\\[-1.0ex]
\xline\verb~      +" pointing to an encoding of a "+cls.getSimpleName();~\\[-1.0ex]
\xline\verb~  }~\\[-1.0ex]
\xline\verb~~\\[-1.0ex]
\xline\verb~~\\[-1.0ex]
\xline\verb~~\\[-1.0ex]
\xline\verb~  /** Follow the Gxl.Edge object of the Gxl.Node at the given position~\\[-1.0ex]
\xline\verb~      and deliver the un-decoded Gxl.Node.~\\[-1.0ex]
\xline\verb~      Generate error message if no such Node can be found and strict=true.~\\[-1.0ex]
\xline\verb~      Only called from {@link #convert_target(Node,int,Class,boolean)}.~\\[-1.0ex]
\xline\verb~  */~\\[-1.0ex]
\xline\verb~  protected Node find_targetencoding (final Node node,~\\[-1.0ex]
\xline\verb~                                      final int position,~\\[-1.0ex]
\xline\verb~                                      final Class cls, // DIAGNOSIS only!~\\[-1.0ex]
\xline\verb~                                      final boolean strict){~\\[-1.0ex]
\xline\verb~    final Edge edge  = get_node2outgoing(node).get(position);~\\[-1.0ex]
\xline\verb~    if (edge==null){~\\[-1.0ex]
\xline\verb~      if (strict)~\\[-1.0ex]
\xline\verb~        ERROR(node.get_location(), "No edge at all at this position. "~\\[-1.0ex]
\xline\verb~              +noEdgeFound_errormsg(node, position, cls));~\\[-1.0ex]
\xline\verb~      return null ; ~\\[-1.0ex]
\xline\verb~    }~\\[-1.0ex]
\xline\verb~    final Part target = edge.get_to();~\\[-1.0ex]
\xline\verb~    if (!(target instanceof Node)){~\\[-1.0ex]
\xline\verb~      ERROR(node.get_location(), "Edge points to something not a graph node. "~\\[-1.0ex]
\xline\verb~            +noEdgeFound_errormsg(node, position, cls));~\\[-1.0ex]
\xline\verb~      return null ; ~\\[-1.0ex]
\xline\verb~    }~\\[-1.0ex]
\xline\verb~    return (Node) target ; ~\\[-1.0ex]
\xline\verb~  }~\\[-1.0ex]
\xline\verb~~\\[-1.0ex]
\xline\verb~~\\[-1.0ex]
\xline\verb~  /** Follow the Gxl.Edge object of the Gxl.Node at the given position,~\\[-1.0ex]
\xline\verb~      create the corresponding Firm object by calling {@link #convert(Node)},~\\[-1.0ex]
\xline\verb~      and ensure that this result is of type T.~\\[-1.0ex]
\xline\verb~  */~\\[-1.0ex]
\xline\verb~  protected <T> T convert_target ~\\[-1.0ex]
\xline\verb~  //protected <T extends FirmNode> T convert_target ~\\[-1.0ex]
\xline\verb~                       (final Node node,~\\[-1.0ex]
\xline\verb~                        final int position,~\\[-1.0ex]
\xline\verb~                        final Class<T> cls, ~\\[-1.0ex]
\xline\verb~                        final boolean strict){~\\[-1.0ex]
\xline\verb~    final Node targetnode = find_targetencoding(node, position, cls, strict);~\\[-1.0ex]
\xline\verb~    if (targetnode==null) // Error messages are already generated.~\\[-1.0ex]
\xline\verb~      return null;~\\[-1.0ex]
\xline\verb~    final FirmNode oldtranslation = correspondent.get(targetnode);~\\[-1.0ex]
\xline\verb~    if (oldtranslation!=null){~\\[-1.0ex]
\xline\verb~      if (!(cls.isInstance(oldtranslation))){~\\[-1.0ex]
\xline\verb~        ERROR(targetnode.get_location(), ~\\[-1.0ex]
\xline\verb~              "translation result is of type "~\\[-1.0ex]
\xline\verb~              +oldtranslation.getClass().getSimpleName()~\\[-1.0ex]
\xline\verb~              +"but now "+cls.getSimpleName()+" is expected? "~\\[-1.0ex]
\xline\verb~              +noEdgeFound_errormsg(node,position,cls));~\\[-1.0ex]
\xline\verb~        return null ; ~\\[-1.0ex]
\xline\verb~      }~\\[-1.0ex]
\xline\verb~      else~\\[-1.0ex]
\xline\verb~        return cls.cast(oldtranslation);~\\[-1.0ex]
\xline\verb~    }~\\[-1.0ex]
\xline\verb~    if (undertranslation.contains(targetnode)){~\\[-1.0ex]
\xline\verb~      /**/System.err.println("BACK PATCH not yet implemented for LOOP: "~\\[-1.0ex]
\xline\verb~                             +noEdgeFound_errormsg(node, position, cls));~\\[-1.0ex]
\xline\verb~      return null ; ~\\[-1.0ex]
\xline\verb~    }~\\[-1.0ex]
\xline\verb~    undertranslation.add(targetnode);~\\[-1.0ex]
\xline\verb~    final FirmNode newTranslation = convert(targetnode);~\\[-1.0ex]
\xline\verb~    if (newTranslation!=null)~\\[-1.0ex]
\xline\verb~      if (!(cls.isInstance(newTranslation))){~\\[-1.0ex]
\xline\verb~        ERROR(targetnode.get_location(),~\\[-1.0ex]
\xline\verb~              "Translation result is of class "~\\[-1.0ex]
\xline\verb~              +newTranslation.getClass().getSimpleName()~\\[-1.0ex]
\xline\verb~              +", but should be of "+cls.getSimpleName()+" ."~\\[-1.0ex]
\xline\verb~            +noEdgeFound_errormsg(node, position, cls));~\\[-1.0ex]
\xline\verb~        return null ; ~\\[-1.0ex]
\xline\verb~      }~\\[-1.0ex]
\xline\verb~    correspondent.put(targetnode, newTranslation);~\\[-1.0ex]
\xline\verb~    return cls.cast(newTranslation);~\\[-1.0ex]
\xline\verb~  }~\\[-1.0ex]
\xline\verb~~\\[-1.0ex]
\xline\verb~  // --------------------------------------------------------------------~\\[-1.0ex]
\xline\verb~~\\[-1.0ex]
\xline\verb~  /** Comprehend all three block types ("StartBlock", "EndBlock", etc.) ~\\[-1.0ex]
\xline\verb~      into one single model class.~\\[-1.0ex]
\xline\verb~  */~\\[-1.0ex]
\xline\verb~  protected boolean typeIsBlocktype (final String type){~\\[-1.0ex]
\xline\verb~    return type.equals(STRING_BLOCKTYPE)||type.equals(STRING_STARTBLOCKTYPE)~\\[-1.0ex]
\xline\verb~      ||type.equals(STRING_ENDBLOCKTYPE) ;~\\[-1.0ex]
\xline\verb~  }~\\[-1.0ex]
\xline\verb~    ~\\[-1.0ex]
\xline\verb~~\\[-1.0ex]
\xline\verb~  Location<XMLDocumentIdentifier> sourceLocation ; ~\\[-1.0ex]
\xline\verb~~\\[-1.0ex]
\xline\verb~  /** ASSUME Only two callers are (1) frequently convert_target(), ~\\[-1.0ex]
\xline\verb~      and (2) only once top-level, the external call for the end-node.~\\[-1.0ex]
\xline\verb~      Only called if node is not yet translated.~\\[-1.0ex]
\xline\verb~   */~\\[-1.0ex]
\xline\verb~  protected FirmNode convert (final Node node /* GLO ... */){~\\[-1.0ex]
\xline\verb~~\\[-1.0ex]
\xline\verb~    final String type = node2nodetype.get(node);~\\[-1.0ex]
\xline\verb~    if (typeIsBlocktype(type))~\\[-1.0ex]
\xline\verb~      return convert_block(node);~\\[-1.0ex]
\xline\verb~~\\[-1.0ex]
\xline\verb~    final Block block ~\\[-1.0ex]
\xline\verb~      = convert_target (node, ORDER_BLOCK_CONTAINMENT, Block.class, true);~\\[-1.0ex]
\xline\verb~    if (block==null) // Error messages are already emitted!~\\[-1.0ex]
\xline\verb~      return null ; ~\\[-1.0ex]
\xline\verb~~\\[-1.0ex]
\xline\verb~    sourceLocation = node.get_location();~\\[-1.0ex]
\xline\verb~~\\[-1.0ex]
\xline\verb~    if (type.equals("Return")){~\\[-1.0ex]
\xline\verb~      final MemoryState ms = convert_target(node, 0, MemoryState.class, true/*???*/);~\\[-1.0ex]
\xline\verb~      final Return ret = new Return(sourceLocation, block, ms);~\\[-1.0ex]
\xline\verb~      for (Integer i : node2outgoing.get(node).keySet()){~\\[-1.0ex]
\xline\verb~        if (i<=0) continue;~\\[-1.0ex]
\xline\verb~        ret.get_results().add(convert_target(node, i, Numeric.class, true));~\\[-1.0ex]
\xline\verb~      }~\\[-1.0ex]
\xline\verb~      return ret ; ~\\[-1.0ex]
\xline\verb~    }~\\[-1.0ex]
\xline\verb~    if (type.equals("Start")){~\\[-1.0ex]
\xline\verb~      return new Start(sourceLocation, block);~\\[-1.0ex]
\xline\verb~    }~\\[-1.0ex]
\xline\verb~    if (type.equals("Jmp")){~\\[-1.0ex]
\xline\verb~      return new Jmp(sourceLocation, block);~\\[-1.0ex]
\xline\verb~    }~\\[-1.0ex]
\xline\verb~    if (type.equals("Phi")){~\\[-1.0ex]
\xline\verb~      final Phi phi = new Phi(sourceLocation, block,NumericType.NotYetComputed);~\\[-1.0ex]
\xline\verb~      for (Integer i : get_node2outgoing(node).keySet()){~\\[-1.0ex]
\xline\verb~        if (i==-1) continue;~\\[-1.0ex]
\xline\verb~        phi.get_alternatives().put(i,convert_target(node, i, Numeric.class, true));~\\[-1.0ex]
\xline\verb~      }~\\[-1.0ex]
\xline\verb~      return phi ;~\\[-1.0ex]
\xline\verb~    }~\\[-1.0ex]
\xline\verb~    if (type.equals("Cond")){~\\[-1.0ex]
\xline\verb~      final Numeric selector = convert_target(node, 0, Numeric.class, true);~\\[-1.0ex]
\xline\verb~      return new Cond(sourceLocation, block, selector);~\\[-1.0ex]
\xline\verb~    }~\\[-1.0ex]
\xline\verb~    if (type.equals("End"))~\\[-1.0ex]
\xline\verb~      return new End(sourceLocation, block);~\\[-1.0ex]
\xline\verb~~\\[-1.0ex]
\xline\verb~    /*~\\[-1.0ex]
\xline\verb~    if (type.equals("Store")){~\\[-1.0ex]
\xline\verb~      return new Start(sourceLocation, block);~\\[-1.0ex]
\xline\verb~    }~\\[-1.0ex]
\xline\verb~    */~\\[-1.0ex]
\xline\verb~    if (type.equals("Argument")){~\\[-1.0ex]
\xline\verb~      // "#Argument" is not documented, but used in testdata/testcase.gxl~\\[-1.0ex]
\xline\verb~      // we realize it as a projection out of a projection.~\\[-1.0ex]
\xline\verb~      final Start start = convert_target(node, 0, Start.class, true);~\\[-1.0ex]
\xline\verb~      // the Gxl Attribute named "position" gives the index of the argument~\\[-1.0ex]
\xline\verb~      // into the stack image with "Start"~\\[-1.0ex]
\xline\verb~~\\[-1.0ex]
\xline\verb~      Variable<Integer> i = Pattern.<Integer>variable();~\\[-1.0ex]
\xline\verb~      if(!hasAttr(ATTRIBUTE_NAME_POSITION,~\\[-1.0ex]
\xline\verb~                  Int.cast(Int.get_value(i))~\\[-1.0ex]
\xline\verb~                  ).match(node)){~\\[-1.0ex]
\xline\verb~        ERROR(sourceLocation, "position attribute missing for Argument");~\\[-1.0ex]
\xline\verb~        return Firm.BAD ; ~\\[-1.0ex]
\xline\verb~      }~\\[-1.0ex]
\xline\verb~      final Start start0 = convert_target(node, 0, Start.class, true);~\\[-1.0ex]
\xline\verb~      return new Proj_N(sourceLocation, block, NumericType.NotYetComputed,~\\[-1.0ex]
\xline\verb~                        getAllArgs(start0), i.getValue());~\\[-1.0ex]
\xline\verb~    }~\\[-1.0ex]
\xline\verb~    if (is_numeric_nullary(type))~\\[-1.0ex]
\xline\verb~      return convert_numeric_nullary(block, type, node);~\\[-1.0ex]
\xline\verb~    if (is_numeric_unary(type))~\\[-1.0ex]
\xline\verb~      return convert_numeric_unary(block, type, node);~\\[-1.0ex]
\xline\verb~    if (is_numeric_binary(type))~\\[-1.0ex]
\xline\verb~      return convert_numeric_binary(block, type, node);~\\[-1.0ex]
\xline\verb~    if (is_numeric_ternary(type))~\\[-1.0ex]
\xline\verb~      return convert_numeric_ternary(block, type, node);~\\[-1.0ex]
\xline\verb~    throw new Error("No rule for gxl node  "+node.get_id()+" of type "+type);~\\[-1.0ex]
\xline\verb~    //    return null ; ~\\[-1.0ex]
\xline\verb~  }~\\[-1.0ex]
\xline\verb~~\\[-1.0ex]
\xline\verb~  protected boolean is_numeric_nullary(String t){~\\[-1.0ex]
\xline\verb~    return t.equals("Const") ; ~\\[-1.0ex]
\xline\verb~  }~\\[-1.0ex]
\xline\verb~~\\[-1.0ex]
\xline\verb~  // ---------------------------------------------------------------------~\\[-1.0ex]
\xline\verb~~\\[-1.0ex]
\xline\verb~  protected class DecodeConstValue extends SinglePass{~\\[-1.0ex]
\xline\verb~    public NumericType type ; ~\\[-1.0ex]
\xline\verb~    public int intValue ; ~\\[-1.0ex]
\xline\verb~    public double floatValue ; ~\\[-1.0ex]
\xline\verb~~\\[-1.0ex]
\xline\verb~    @Override protected void action(Attr a){~\\[-1.0ex]
\xline\verb~      if (!(a.get_name().equals(ATTRIBUTE_NAME_VALUE)))~\\[-1.0ex]
\xline\verb~        return;~\\[-1.0ex]
\xline\verb~      match(a.get_val());~\\[-1.0ex]
\xline\verb~    }~\\[-1.0ex]
\xline\verb~    @Override protected void action(Locator i){~\\[-1.0ex]
\xline\verb~      ERROR(i.get_location(), "constants of lcator type should not appear");~\\[-1.0ex]
\xline\verb~    }~\\[-1.0ex]
\xline\verb~    @Override protected void action(Int i){~\\[-1.0ex]
\xline\verb~      intValue = i.get_value();~\\[-1.0ex]
\xline\verb~      type = NumericType.Is ; // PROVIS !~\\[-1.0ex]
\xline\verb~    }~\\[-1.0ex]
\xline\verb~    @Override protected void action(Bool i){~\\[-1.0ex]
\xline\verb~      ERROR(i.get_location(), "constants of boolean type should not appear");~\\[-1.0ex]
\xline\verb~    }~\\[-1.0ex]
\xline\verb~    @Override protected void action(eu.bandm.ttc2011.case2.gxlcodec.Float i){~\\[-1.0ex]
\xline\verb~      floatValue = i.get_value();~\\[-1.0ex]
\xline\verb~      type = NumericType.F ; // PROVIS !~\\[-1.0ex]
\xline\verb~    }~\\[-1.0ex]
\xline\verb~    @Override protected void action(eu.bandm.ttc2011.case2.gxlcodec.String i){~\\[-1.0ex]
\xline\verb~      ERROR(i.get_location(), "constants of string type should not appear");~\\[-1.0ex]
\xline\verb~    }~\\[-1.0ex]
\xline\verb~    @Override protected void action(eu.bandm.ttc2011.case2.gxlcodec.Enum i){~\\[-1.0ex]
\xline\verb~      ERROR(i.get_location(), "constants of enum type should not appear");~\\[-1.0ex]
\xline\verb~    }~\\[-1.0ex]
\xline\verb~    @Override protected void action(Aggregate i){~\\[-1.0ex]
\xline\verb~      ERROR(i.get_location(), "constants of aggregate type should not appear");~\\[-1.0ex]
\xline\verb~    }~\\[-1.0ex]
\xline\verb~  }// class DecodeConstValue~\\[-1.0ex]
\xline\verb~~\\[-1.0ex]
\xline\verb~  protected NumericNode convert_numeric_nullary ~\\[-1.0ex]
\xline\verb~    (final Block b, final String type, final Node node){~\\[-1.0ex]
\xline\verb~    final DecodeConstValue d = new DecodeConstValue();~\\[-1.0ex]
\xline\verb~    d.match(node);~\\[-1.0ex]
\xline\verb~    switch(d.type){~\\[-1.0ex]
\xline\verb~      case Is: ~\\[-1.0ex]
\xline\verb~        NumericConst intConst = new NumericConst (sourceLocation, b, d.type, ~\\[-1.0ex]
\xline\verb~                                                  ""+d.intValue);~\\[-1.0ex]
\xline\verb~        intConst.set_intValue(d.intValue);~\\[-1.0ex]
\xline\verb~        return intConst ; ~\\[-1.0ex]
\xline\verb~    case F: ~\\[-1.0ex]
\xline\verb~      NumericConst fConst = new NumericConst (sourceLocation, b, d.type, ~\\[-1.0ex]
\xline\verb~                                              ""+d.floatValue);~\\[-1.0ex]
\xline\verb~      fConst.set_floatValue(d.floatValue);~\\[-1.0ex]
\xline\verb~      return fConst ; ~\\[-1.0ex]
\xline\verb~    }~\\[-1.0ex]
\xline\verb~    return null ; // ERROR already generated.~\\[-1.0ex]
\xline\verb~  }~\\[-1.0ex]
\xline\verb~  ~\\[-1.0ex]
\xline\verb~  protected boolean is_numeric_unary(String t){~\\[-1.0ex]
\xline\verb~    return t.equals("Conv")|| t.equals("Minus")|| t.equals("Not")~\\[-1.0ex]
\xline\verb~      || t.equals("Rotl")|| t.equals("Shl")~\\[-1.0ex]
\xline\verb~      || t.equals("Shr")|| t.equals("Shrs");~\\[-1.0ex]
\xline\verb~  }~\\[-1.0ex]
\xline\verb~~\\[-1.0ex]
\xline\verb~  protected NumericNode convert_numeric_unary ~\\[-1.0ex]
\xline\verb~    (final Block b, final String type, final Node node){~\\[-1.0ex]
\xline\verb~    final Numeric left  = convert_target(node,0, Numeric.class, true);~\\[-1.0ex]
\xline\verb~    if (type.equals("Conv"))~\\[-1.0ex]
\xline\verb~      return new Conv(sourceLocation, b, NumericType.NotYetComputed, left);~\\[-1.0ex]
\xline\verb~    else if (type.equals("Minus"))~\\[-1.0ex]
\xline\verb~      return new Minus(sourceLocation, b, NumericType.NotYetComputed, left);~\\[-1.0ex]
\xline\verb~    else if (type.equals("Not"))~\\[-1.0ex]
\xline\verb~      return new Not(sourceLocation, b, NumericType.NotYetComputed, left);~\\[-1.0ex]
\xline\verb~    else if (type.equals("Rotl"))~\\[-1.0ex]
\xline\verb~      return new Rotl(sourceLocation, b, NumericType.NotYetComputed, left);~\\[-1.0ex]
\xline\verb~    else if (type.equals("Shl"))~\\[-1.0ex]
\xline\verb~      return new Shl(sourceLocation, b, NumericType.NotYetComputed, left);~\\[-1.0ex]
\xline\verb~    else if (type.equals("Shr"))~\\[-1.0ex]
\xline\verb~      return new Shr(sourceLocation, b, NumericType.NotYetComputed, left);~\\[-1.0ex]
\xline\verb~    else // if (type.equals("Shrs"))~\\[-1.0ex]
\xline\verb~      return new Shrs(sourceLocation, b, NumericType.NotYetComputed, left);~\\[-1.0ex]
\xline\verb~  }~\\[-1.0ex]
\xline\verb~~\\[-1.0ex]
\xline\verb~  protected boolean is_numeric_binary(String t){~\\[-1.0ex]
\xline\verb~    return t.equals("Add")|| t.equals("And")|| t.equals("Cmp")|| t.equals("Div")~\\[-1.0ex]
\xline\verb~      || t.equals("Eor")|| t.equals("Mod")~\\[-1.0ex]
\xline\verb~      || t.equals("Mul")|| t.equals("Or") ~\\[-1.0ex]
\xline\verb~      || t.equals("Sub")  ;~\\[-1.0ex]
\xline\verb~  }~\\[-1.0ex]
\xline\verb~~\\[-1.0ex]
\xline\verb~  protected NumericNode convert_numeric_binary ~\\[-1.0ex]
\xline\verb~    (final Block b, final String type, ~\\[-1.0ex]
\xline\verb~     final Node node){~\\[-1.0ex]
\xline\verb~    final Numeric left = convert_target(node, 0, Numeric.class, true);~\\[-1.0ex]
\xline\verb~    final Numeric right = convert_target(node, 1, Numeric.class, true);~\\[-1.0ex]
\xline\verb~    if (type.equals("Add"))~\\[-1.0ex]
\xline\verb~      return new Add(sourceLocation, b, NumericType.NotYetComputed, left, right);~\\[-1.0ex]
\xline\verb~    else if (type.equals("And"))~\\[-1.0ex]
\xline\verb~      return new And(sourceLocation, b, NumericType.NotYetComputed, left, right);~\\[-1.0ex]
\xline\verb~    else if (type.equals("Cmp"))~\\[-1.0ex]
\xline\verb~      return new Cmp(sourceLocation, b, NumericType.NotYetComputed, left, right);~\\[-1.0ex]
\xline\verb~    else if (type.equals("Div"))~\\[-1.0ex]
\xline\verb~      return new Div(sourceLocation, b, NumericType.NotYetComputed, left, right);~\\[-1.0ex]
\xline\verb~    else if (type.equals("Eor"))~\\[-1.0ex]
\xline\verb~      return new Eor(sourceLocation, b, NumericType.NotYetComputed, left, right);~\\[-1.0ex]
\xline\verb~    else if (type.equals("Mod"))~\\[-1.0ex]
\xline\verb~      return new Mod(sourceLocation, b, NumericType.NotYetComputed, left, right);~\\[-1.0ex]
\xline\verb~    else if (type.equals("Mul"))~\\[-1.0ex]
\xline\verb~      return new Mul(sourceLocation, b, NumericType.NotYetComputed, left, right);~\\[-1.0ex]
\xline\verb~    else if (type.equals("Or"))~\\[-1.0ex]
\xline\verb~      return new Or(sourceLocation, b, NumericType.NotYetComputed, left, right);~\\[-1.0ex]
\xline\verb~    else // if (type.equals("Sub"))~\\[-1.0ex]
\xline\verb~      return new Sub(sourceLocation, b, NumericType.NotYetComputed, left, right);~\\[-1.0ex]
\xline\verb~  }~\\[-1.0ex]
\xline\verb~~\\[-1.0ex]
\xline\verb~  protected boolean is_numeric_ternary(String t){~\\[-1.0ex]
\xline\verb~    return t.equals("Mux") ; ~\\[-1.0ex]
\xline\verb~  }~\\[-1.0ex]
\xline\verb~~\\[-1.0ex]
\xline\verb~  protected NumericNode convert_numeric_ternary ~\\[-1.0ex]
\xline\verb~    (final Block b, final String type, ~\\[-1.0ex]
\xline\verb~     final Node node){~\\[-1.0ex]
\xline\verb~    final Numeric selector = convert_target(node, 0, Numeric.class, true);~\\[-1.0ex]
\xline\verb~    final Numeric left = convert_target(node, 1, Numeric.class, true);~\\[-1.0ex]
\xline\verb~    final Numeric right = convert_target(node, 2, Numeric.class, true);~\\[-1.0ex]
\xline\verb~    // if (type.equals("Mux"))~\\[-1.0ex]
\xline\verb~    return new Mux(sourceLocation, b, NumericType.NotYetComputed, ~\\[-1.0ex]
\xline\verb~                   selector, left, right);~\\[-1.0ex]
\xline\verb~  }~\\[-1.0ex]
\xline\verb~~\\[-1.0ex]
\xline\verb~  protected Block convert_block(final Node node){~\\[-1.0ex]
\xline\verb~    final Block block = new Block(node.get_location());~\\[-1.0ex]
\xline\verb~    if (node.get_id()!=null)~\\[-1.0ex]
\xline\verb~      block.set_gxlId(node.get_id());~\\[-1.0ex]
\xline\verb~~\\[-1.0ex]
\xline\verb~    final Map<Integer, Edge> outgoing = get_node2outgoing(node);~\\[-1.0ex]
\xline\verb~    for (Integer i : outgoing.keySet()){~\\[-1.0ex]
\xline\verb~      final Type edgetype = outgoing.get(i).get_type();~\\[-1.0ex]
\xline\verb~      boolean isfalse = false, istrue = false ; ~\\[-1.0ex]
\xline\verb~      ControlFlow cf = null ; ~\\[-1.0ex]
\xline\verb~      if (edgetype!=null)~\\[-1.0ex]
\xline\verb~        // ATTENTION here also "numeric conditions / switch(){}" ~\\[-1.0ex]
\xline\verb~        //   has to be decoded into a Proj_X node~\\[-1.0ex]
\xline\verb~        if (edgetype.get_href().equals(STRING_EDGETYPE_FALSE))~\\[-1.0ex]
\xline\verb~          isfalse=true ; ~\\[-1.0ex]
\xline\verb~        else if (edgetype.get_href().equals(STRING_EDGETYPE_TRUE))~\\[-1.0ex]
\xline\verb~          istrue=true ; ~\\[-1.0ex]
\xline\verb~      if (isfalse||istrue){~\\[-1.0ex]
\xline\verb~        final Cond cond = convert_target(node, i, Cond.class, true);~\\[-1.0ex]
\xline\verb~        cf = new Proj_X (cond.get_location(),~\\[-1.0ex]
\xline\verb~                         cond.get_block(),~\\[-1.0ex]
\xline\verb~                         cond, ~\\[-1.0ex]
\xline\verb~                         istrue ? 1 : 0);~\\[-1.0ex]
\xline\verb~      }~\\[-1.0ex]
\xline\verb~      else~\\[-1.0ex]
\xline\verb~        cf = convert_target(node, i, ControlFlow.class, true);~\\[-1.0ex]
\xline\verb~      if (cf!=null)~\\[-1.0ex]
\xline\verb~        block.get_predecs().put (i, cf);~\\[-1.0ex]
\xline\verb~      // else error case, has already been signaled above.~\\[-1.0ex]
\xline\verb~    }~\\[-1.0ex]
\xline\verb~    return block ;~\\[-1.0ex]
\xline\verb~  }~\\[-1.0ex]
\xline\verb~~\\[-1.0ex]
\xline\verb~}~\\[-1.0ex]
\xline\verb~// eof~\\[-1.0ex]

\egroup

\section{Encoding the Firm model into a GXL model}
\label{file_firm2gxl_java}
\xlinecounterreset{}
\bgroup\footnotesize
\xline\verb~package eu.bandm.ttc2011.case2.transformations ;~\\[-1.0ex]
\xline\verb~~\\[-1.0ex]
\xline\verb~import java.util.Map ; ~\\[-1.0ex]
\xline\verb~import java.util.HashMap ; ~\\[-1.0ex]
\xline\verb~import java.util.Set ; ~\\[-1.0ex]
\xline\verb~import java.util.HashSet ; ~\\[-1.0ex]
\xline\verb~~\\[-1.0ex]
\xline\verb~import eu.bandm.ttc2011.case2.gxlcodec.* ; ~\\[-1.0ex]
\xline\verb~import eu.bandm.ttc2011.case2.firm_01.* ; ~\\[-1.0ex]
\xline\verb~~\\[-1.0ex]
\xline\verb~import java.lang.String ; ~\\[-1.0ex]
\xline\verb~~\\[-1.0ex]
\xline\verb~import static eu.bandm.ttc2011.case2.transformations~\\[-1.0ex]
\xline\verb~  .Gxl2Firm.ATTRIBUTE_NAME_POSITION ;~\\[-1.0ex]
\xline\verb~import static eu.bandm.ttc2011.case2.transformations~\\[-1.0ex]
\xline\verb~  .Gxl2Firm.ORDER_BLOCK_CONTAINMENT ;~\\[-1.0ex]
\xline\verb~import static eu.bandm.ttc2011.case2.transformations~\\[-1.0ex]
\xline\verb~  .Gxl2Firm.ATTRIBUTE_NAME_VALUE ; ~\\[-1.0ex]
\xline\verb~~\\[-1.0ex]
\xline\verb~~\\[-1.0ex]
\xline\verb~public class Firm2Gxl extends SimpleVisitor  {~\\[-1.0ex]
\xline\verb~~\\[-1.0ex]
\xline\verb~  protected Gxl result ; ~\\[-1.0ex]
\xline\verb~  protected Graph metamodel ;~\\[-1.0ex]
\xline\verb~  protected End endnode ; ~\\[-1.0ex]
\xline\verb~  protected Graph model ; ~\\[-1.0ex]
\xline\verb~  protected Map<String, String> nodeType2id ; ~\\[-1.0ex]
\xline\verb~  protected Map<FirmNode, Node> node2node = new HashMap<FirmNode, Node>();~\\[-1.0ex]
\xline\verb~~\\[-1.0ex]
\xline\verb~~\\[-1.0ex]
\xline\verb~  public Gxl translate (Graph metamodel, Map<String,String> nodeType2id, ~\\[-1.0ex]
\xline\verb~                        String graphId, End endnode){~\\[-1.0ex]
\xline\verb~    this.metamodel=metamodel;~\\[-1.0ex]
\xline\verb~    this.nodeType2id=nodeType2id;~\\[-1.0ex]
\xline\verb~    this.endnode = endnode ;~\\[-1.0ex]
\xline\verb~    result = new Gxl();~\\[-1.0ex]
\xline\verb~    result.get_graphs().add(metamodel);~\\[-1.0ex]
\xline\verb~    model = new Graph(graphId);~\\[-1.0ex]
\xline\verb~    result.get_graphs().add(model);~\\[-1.0ex]
\xline\verb~    new Writeout_nodesAndBlocks().match (endnode);~\\[-1.0ex]
\xline\verb~    new Writeout_edges().match (endnode);~\\[-1.0ex]
\xline\verb~    return result ; ~\\[-1.0ex]
\xline\verb~  }~\\[-1.0ex]
\xline\verb~~\\[-1.0ex]
\xline\verb~  protected class Writeout_nodesAndBlocks extends SimpleVisitor{~\\[-1.0ex]
\xline\verb~~\\[-1.0ex]
\xline\verb~    protected void write_node(final FirmNode n, String type){~\\[-1.0ex]
\xline\verb~      if (node2node.containsKey(n))~\\[-1.0ex]
\xline\verb~        return ; ~\\[-1.0ex]
\xline\verb~      final String newId = "node"+String.valueOf(10000+node2node.size());~\\[-1.0ex]
\xline\verb~      final Node node = new Node(newId);~\\[-1.0ex]
\xline\verb~      node.set_type(new Type("#"+nodeType2id.get(type)));~\\[-1.0ex]
\xline\verb~      node2node.put(n, node);~\\[-1.0ex]
\xline\verb~      model.get_parts().add(node);~\\[-1.0ex]
\xline\verb~    }~\\[-1.0ex]
\xline\verb~~\\[-1.0ex]
\xline\verb~~\\[-1.0ex]
\xline\verb~    @Override protected void action (final Block n){~\\[-1.0ex]
\xline\verb~      if (node2node.containsKey(n))~\\[-1.0ex]
\xline\verb~        return ; ~\\[-1.0ex]
\xline\verb~      write_node(n, "Block");~\\[-1.0ex]
\xline\verb~      super.action(n);~\\[-1.0ex]
\xline\verb~    }~\\[-1.0ex]
\xline\verb~~\\[-1.0ex]
\xline\verb~    @Override protected void action (final Start n){~\\[-1.0ex]
\xline\verb~      if (node2node.containsKey(n))~\\[-1.0ex]
\xline\verb~        return ; ~\\[-1.0ex]
\xline\verb~      write_node(n, "Start");~\\[-1.0ex]
\xline\verb~      super.action(n);~\\[-1.0ex]
\xline\verb~    }~\\[-1.0ex]
\xline\verb~    @Override protected void action (final Return n){~\\[-1.0ex]
\xline\verb~      if (node2node.containsKey(n))~\\[-1.0ex]
\xline\verb~        return ; ~\\[-1.0ex]
\xline\verb~      write_node(n, "Return");~\\[-1.0ex]
\xline\verb~      super.action(n);~\\[-1.0ex]
\xline\verb~    }~\\[-1.0ex]
\xline\verb~    @Override protected void action (final End n){~\\[-1.0ex]
\xline\verb~      if (node2node.containsKey(n))~\\[-1.0ex]
\xline\verb~        return ; ~\\[-1.0ex]
\xline\verb~      write_node(n, "End");~\\[-1.0ex]
\xline\verb~      super.action(n);~\\[-1.0ex]
\xline\verb~    }~\\[-1.0ex]
\xline\verb~    @Override protected void action (final Jmp n){~\\[-1.0ex]
\xline\verb~      if (node2node.containsKey(n))~\\[-1.0ex]
\xline\verb~        return ; ~\\[-1.0ex]
\xline\verb~      write_node(n, "Jmp");~\\[-1.0ex]
\xline\verb~      super.action(n);~\\[-1.0ex]
\xline\verb~    }~\\[-1.0ex]
\xline\verb~    @Override protected void action (final Cond n){~\\[-1.0ex]
\xline\verb~      if (node2node.containsKey(n))~\\[-1.0ex]
\xline\verb~        return ; ~\\[-1.0ex]
\xline\verb~      write_node(n, "Cond");~\\[-1.0ex]
\xline\verb~      super.action(n);~\\[-1.0ex]
\xline\verb~    }~\\[-1.0ex]
\xline\verb~    /* do NOT generate a node in gxl model!~\\[-1.0ex]
\xline\verb~    @Override protected void action (final Proj_X n){~\\[-1.0ex]
\xline\verb~      super.action(n);~\\[-1.0ex]
\xline\verb~    }~\\[-1.0ex]
\xline\verb~    */~\\[-1.0ex]
\xline\verb~    @Override protected void action (final Sync n){~\\[-1.0ex]
\xline\verb~      if (node2node.containsKey(n))~\\[-1.0ex]
\xline\verb~        return ; ~\\[-1.0ex]
\xline\verb~      write_node(n, "Sync");~\\[-1.0ex]
\xline\verb~      super.action(n);~\\[-1.0ex]
\xline\verb~    }~\\[-1.0ex]
\xline\verb~    @Override protected void action (final Phi n){~\\[-1.0ex]
\xline\verb~      if (node2node.containsKey(n))~\\[-1.0ex]
\xline\verb~        return ; ~\\[-1.0ex]
\xline\verb~      write_node(n, "Phi");~\\[-1.0ex]
\xline\verb~      super.action(n);~\\[-1.0ex]
\xline\verb~    }~\\[-1.0ex]
\xline\verb~    @Override protected void action (final NumericConst n){~\\[-1.0ex]
\xline\verb~      if (node2node.containsKey(n))~\\[-1.0ex]
\xline\verb~        return ; ~\\[-1.0ex]
\xline\verb~      write_node(n, "Const");~\\[-1.0ex]
\xline\verb~      eu.bandm.ttc2011.case2.gxlcodec.String s ~\\[-1.0ex]
\xline\verb~        = new eu.bandm.ttc2011.case2.gxlcodec.String(n.get_unparsedValue());~\\[-1.0ex]
\xline\verb~      Attr attr = new Attr(ATTRIBUTE_NAME_VALUE, s);~\\[-1.0ex]
\xline\verb~      node2node.get(n).get_attrs().add(attr);~\\[-1.0ex]
\xline\verb~      super.action(n);~\\[-1.0ex]
\xline\verb~    }~\\[-1.0ex]
\xline\verb~    @Override protected void action (final Conv n){~\\[-1.0ex]
\xline\verb~      if (node2node.containsKey(n))~\\[-1.0ex]
\xline\verb~        return ; ~\\[-1.0ex]
\xline\verb~      write_node(n, "Conv");~\\[-1.0ex]
\xline\verb~      super.action(n);~\\[-1.0ex]
\xline\verb~    }~\\[-1.0ex]
\xline\verb~    @Override protected void action (final Minus n){~\\[-1.0ex]
\xline\verb~      if (node2node.containsKey(n))~\\[-1.0ex]
\xline\verb~        return ; ~\\[-1.0ex]
\xline\verb~      write_node(n, "Minus");~\\[-1.0ex]
\xline\verb~      super.action(n);~\\[-1.0ex]
\xline\verb~    }~\\[-1.0ex]
\xline\verb~    @Override protected void action (final Not n){~\\[-1.0ex]
\xline\verb~      if (node2node.containsKey(n))~\\[-1.0ex]
\xline\verb~        return ; ~\\[-1.0ex]
\xline\verb~      write_node(n, "Not");~\\[-1.0ex]
\xline\verb~      super.action(n);~\\[-1.0ex]
\xline\verb~    }~\\[-1.0ex]
\xline\verb~    @Override protected void action (final Rotl n){~\\[-1.0ex]
\xline\verb~      if (node2node.containsKey(n))~\\[-1.0ex]
\xline\verb~        return ; ~\\[-1.0ex]
\xline\verb~      write_node(n, "Rotl");~\\[-1.0ex]
\xline\verb~      super.action(n);~\\[-1.0ex]
\xline\verb~    }~\\[-1.0ex]
\xline\verb~    @Override protected void action (final Shl n){~\\[-1.0ex]
\xline\verb~      if (node2node.containsKey(n))~\\[-1.0ex]
\xline\verb~        return ; ~\\[-1.0ex]
\xline\verb~      write_node(n, "Shl");~\\[-1.0ex]
\xline\verb~      super.action(n);~\\[-1.0ex]
\xline\verb~    }~\\[-1.0ex]
\xline\verb~    @Override protected void action (final Shr n){~\\[-1.0ex]
\xline\verb~      if (node2node.containsKey(n))~\\[-1.0ex]
\xline\verb~        return ; ~\\[-1.0ex]
\xline\verb~      write_node(n, "Shr");~\\[-1.0ex]
\xline\verb~      super.action(n);~\\[-1.0ex]
\xline\verb~    }~\\[-1.0ex]
\xline\verb~    @Override protected void action (final Shrs n){~\\[-1.0ex]
\xline\verb~      if (node2node.containsKey(n))~\\[-1.0ex]
\xline\verb~        return ; ~\\[-1.0ex]
\xline\verb~      write_node(n, "Shrs");~\\[-1.0ex]
\xline\verb~      super.action(n);~\\[-1.0ex]
\xline\verb~    }~\\[-1.0ex]
\xline\verb~    @Override protected void action (final Add n){~\\[-1.0ex]
\xline\verb~      if (node2node.containsKey(n))~\\[-1.0ex]
\xline\verb~        return ; ~\\[-1.0ex]
\xline\verb~      write_node(n, "Add");~\\[-1.0ex]
\xline\verb~      super.action(n);~\\[-1.0ex]
\xline\verb~    }~\\[-1.0ex]
\xline\verb~    @Override protected void action (final And n){~\\[-1.0ex]
\xline\verb~      if (node2node.containsKey(n))~\\[-1.0ex]
\xline\verb~        return ; ~\\[-1.0ex]
\xline\verb~      write_node(n, "And");~\\[-1.0ex]
\xline\verb~      super.action(n);~\\[-1.0ex]
\xline\verb~    }~\\[-1.0ex]
\xline\verb~    @Override protected void action (final Div n){~\\[-1.0ex]
\xline\verb~      if (node2node.containsKey(n))~\\[-1.0ex]
\xline\verb~        return ; ~\\[-1.0ex]
\xline\verb~      write_node(n, "Div");~\\[-1.0ex]
\xline\verb~      super.action(n);~\\[-1.0ex]
\xline\verb~    }~\\[-1.0ex]
\xline\verb~    @Override protected void action (final Eor n){~\\[-1.0ex]
\xline\verb~      if (node2node.containsKey(n))~\\[-1.0ex]
\xline\verb~        return ; ~\\[-1.0ex]
\xline\verb~      write_node(n, "Eor");~\\[-1.0ex]
\xline\verb~      super.action(n);~\\[-1.0ex]
\xline\verb~    }~\\[-1.0ex]
\xline\verb~    @Override protected void action (final Mul n){~\\[-1.0ex]
\xline\verb~      if (node2node.containsKey(n))~\\[-1.0ex]
\xline\verb~        return ; ~\\[-1.0ex]
\xline\verb~      write_node(n, "Mul");~\\[-1.0ex]
\xline\verb~      super.action(n);~\\[-1.0ex]
\xline\verb~    }~\\[-1.0ex]
\xline\verb~    @Override protected void action (final Or n){~\\[-1.0ex]
\xline\verb~      if (node2node.containsKey(n))~\\[-1.0ex]
\xline\verb~        return ; ~\\[-1.0ex]
\xline\verb~      write_node(n, "Or");~\\[-1.0ex]
\xline\verb~      super.action(n);~\\[-1.0ex]
\xline\verb~    }~\\[-1.0ex]
\xline\verb~    @Override protected void action (final Sub n){~\\[-1.0ex]
\xline\verb~      if (node2node.containsKey(n))~\\[-1.0ex]
\xline\verb~        return ; ~\\[-1.0ex]
\xline\verb~      write_node(n, "Sub");~\\[-1.0ex]
\xline\verb~      super.action(n);~\\[-1.0ex]
\xline\verb~    }~\\[-1.0ex]
\xline\verb~    @Override protected void action (final Cmp n){~\\[-1.0ex]
\xline\verb~      if (node2node.containsKey(n))~\\[-1.0ex]
\xline\verb~        return ; ~\\[-1.0ex]
\xline\verb~      write_node(n, "Cmp");~\\[-1.0ex]
\xline\verb~      super.action(n);~\\[-1.0ex]
\xline\verb~    }~\\[-1.0ex]
\xline\verb~    @Override protected void action (final Mux n){~\\[-1.0ex]
\xline\verb~      if (node2node.containsKey(n))~\\[-1.0ex]
\xline\verb~        return ; ~\\[-1.0ex]
\xline\verb~      write_node(n, "Mux");~\\[-1.0ex]
\xline\verb~      super.action(n);~\\[-1.0ex]
\xline\verb~    }~\\[-1.0ex]
\xline\verb~  }// class Writeout_nodesAndBlocks~\\[-1.0ex]
\xline\verb~~\\[-1.0ex]
\xline\verb~  // ===========================================================================~\\[-1.0ex]
\xline\verb~~\\[-1.0ex]
\xline\verb~  protected class Writeout_edges extends SimpleVisitor{~\\[-1.0ex]
\xline\verb~~\\[-1.0ex]
\xline\verb~    final java.util.Set<FirmNode> visited = new HashSet<FirmNode>();~\\[-1.0ex]
\xline\verb~~\\[-1.0ex]
\xline\verb~    protected boolean done (final FirmNode node){~\\[-1.0ex]
\xline\verb~      if (visited.contains(node))~\\[-1.0ex]
\xline\verb~        return true ; ~\\[-1.0ex]
\xline\verb~      visited.add(node);~\\[-1.0ex]
\xline\verb~      return false ;~\\[-1.0ex]
\xline\verb~    }~\\[-1.0ex]
\xline\verb~    protected Edge write_edge(final FirmNode from, final FirmNode to, final int pos){~\\[-1.0ex]
\xline\verb~      final Edge edge = new Edge(node2node.get(from), node2node.get(to));~\\[-1.0ex]
\xline\verb~      final Int i = new Int(pos);~\\[-1.0ex]
\xline\verb~      final Attr attr = new Attr(ATTRIBUTE_NAME_POSITION,i);~\\[-1.0ex]
\xline\verb~      edge.get_attrs().add(attr);~\\[-1.0ex]
\xline\verb~      model.get_parts().add(edge);~\\[-1.0ex]
\xline\verb~      return edge ; ~\\[-1.0ex]
\xline\verb~    }~\\[-1.0ex]
\xline\verb~~\\[-1.0ex]
\xline\verb~    protected void write_containing(final BlockNode from){~\\[-1.0ex]
\xline\verb~      write_edge(from, from.get_block(), ORDER_BLOCK_CONTAINMENT);~\\[-1.0ex]
\xline\verb~    }~\\[-1.0ex]
\xline\verb~~\\[-1.0ex]
\xline\verb~    @Override protected void action (final Block n){~\\[-1.0ex]
\xline\verb~      if (done(n)) return ;~\\[-1.0ex]
\xline\verb~      for (Integer i : n.get_predecs().keySet()){~\\[-1.0ex]
\xline\verb~        final ControlFlow target = n.get_predecs().get(i);~\\[-1.0ex]
\xline\verb~        if (target instanceof Proj_X){~\\[-1.0ex]
\xline\verb~          final Proj_X proj = (Proj_X)target;~\\[-1.0ex]
\xline\verb~          final Edge e = write_edge(n, proj.get_input(), i);~\\[-1.0ex]
\xline\verb~          e.set_type(new Type(proj.get_selection()==0 ~\\[-1.0ex]
\xline\verb~                              ? Gxl2Firm.STRING_EDGETYPE_FALSE~\\[-1.0ex]
\xline\verb~                              : Gxl2Firm.STRING_EDGETYPE_TRUE));~\\[-1.0ex]
\xline\verb~        }~\\[-1.0ex]
\xline\verb~        else~\\[-1.0ex]
\xline\verb~          write_edge(n, (FirmNode)target, i);~\\[-1.0ex]
\xline\verb~      }~\\[-1.0ex]
\xline\verb~      super.action(n);~\\[-1.0ex]
\xline\verb~    }~\\[-1.0ex]
\xline\verb~~\\[-1.0ex]
\xline\verb~    @Override protected void action (final Start n){~\\[-1.0ex]
\xline\verb~      if (done(n)) return ;~\\[-1.0ex]
\xline\verb~      write_containing(n);~\\[-1.0ex]
\xline\verb~    }~\\[-1.0ex]
\xline\verb~    @Override protected void action (final Return n){~\\[-1.0ex]
\xline\verb~      if (done(n)) return ;~\\[-1.0ex]
\xline\verb~      write_containing(n);~\\[-1.0ex]
\xline\verb~      write_edge(n, (BlockNode)n.get_memstate(), 0);~\\[-1.0ex]
\xline\verb~      int i = 1 ; ~\\[-1.0ex]
\xline\verb~      for (Numeric num : n.get_results())~\\[-1.0ex]
\xline\verb~        write_edge(n, (FirmNode)num, i++);~\\[-1.0ex]
\xline\verb~      super.action(n);~\\[-1.0ex]
\xline\verb~    }~\\[-1.0ex]
\xline\verb~~\\[-1.0ex]
\xline\verb~    @Override protected void action (final End n){~\\[-1.0ex]
\xline\verb~      if (done(n)) return ;~\\[-1.0ex]
\xline\verb~      write_containing(n);~\\[-1.0ex]
\xline\verb~      super.action(n);~\\[-1.0ex]
\xline\verb~    }~\\[-1.0ex]
\xline\verb~    @Override protected void action (final Jmp n){~\\[-1.0ex]
\xline\verb~      if (done(n)) return ;~\\[-1.0ex]
\xline\verb~      write_containing(n);~\\[-1.0ex]
\xline\verb~      super.action(n);~\\[-1.0ex]
\xline\verb~    }~\\[-1.0ex]
\xline\verb~    @Override protected void action (final Cond n){~\\[-1.0ex]
\xline\verb~      if (done(n)) return ;~\\[-1.0ex]
\xline\verb~      write_containing(n);~\\[-1.0ex]
\xline\verb~      write_edge(n, (NumericNode)n.get_selector(), 0);~\\[-1.0ex]
\xline\verb~      super.action(n);~\\[-1.0ex]
\xline\verb~    }~\\[-1.0ex]
\xline\verb~    @Override protected void action (final Sync n){~\\[-1.0ex]
\xline\verb~      if (done(n)) return ;~\\[-1.0ex]
\xline\verb~      write_containing(n);~\\[-1.0ex]
\xline\verb~      int i = 0 ; ~\\[-1.0ex]
\xline\verb~      for (MemoryState mem : n.get_predecs())~\\[-1.0ex]
\xline\verb~        write_edge(n, (FirmNode)mem, i++);~\\[-1.0ex]
\xline\verb~      super.action(n);~\\[-1.0ex]
\xline\verb~    }~\\[-1.0ex]
\xline\verb~    @Override protected void action (final Phi n){~\\[-1.0ex]
\xline\verb~      if (done(n)) return ;~\\[-1.0ex]
\xline\verb~      write_containing(n);~\\[-1.0ex]
\xline\verb~      for (Integer i : n.get_alternatives().keySet())~\\[-1.0ex]
\xline\verb~        write_edge(n, (BlockNode)n.get_alternatives().get(i), i);~\\[-1.0ex]
\xline\verb~      super.action(n);~\\[-1.0ex]
\xline\verb~    }~\\[-1.0ex]
\xline\verb~    @Override protected void action (final NumericConst n){~\\[-1.0ex]
\xline\verb~      if (done(n)) return ;~\\[-1.0ex]
\xline\verb~      write_containing(n);~\\[-1.0ex]
\xline\verb~      super.action(n);~\\[-1.0ex]
\xline\verb~    }~\\[-1.0ex]
\xline\verb~    @Override protected void action (final Unary n){~\\[-1.0ex]
\xline\verb~      if (done(n)) return ;~\\[-1.0ex]
\xline\verb~      write_containing(n);~\\[-1.0ex]
\xline\verb~      write_edge(n, (NumericNode)n.get_on(), 0);~\\[-1.0ex]
\xline\verb~      super.action(n);~\\[-1.0ex]
\xline\verb~    }~\\[-1.0ex]
\xline\verb~    @Override protected void action (final Binary n){~\\[-1.0ex]
\xline\verb~      if (done(n)) return ;~\\[-1.0ex]
\xline\verb~      write_containing(n);~\\[-1.0ex]
\xline\verb~      write_edge(n, (NumericNode)n.get_left(), 0);~\\[-1.0ex]
\xline\verb~      write_edge(n, (NumericNode)n.get_right(), 1);~\\[-1.0ex]
\xline\verb~      super.action(n);~\\[-1.0ex]
\xline\verb~    }~\\[-1.0ex]
\xline\verb~    @Override protected void action (final Ternary n){~\\[-1.0ex]
\xline\verb~      if (done(n)) return ;~\\[-1.0ex]
\xline\verb~      write_containing(n);~\\[-1.0ex]
\xline\verb~      write_edge(n, (NumericNode)n.get_first(), 0);~\\[-1.0ex]
\xline\verb~      write_edge(n, (NumericNode)n.get_second(), 1);~\\[-1.0ex]
\xline\verb~      write_edge(n, (NumericNode)n.get_third(), 2);~\\[-1.0ex]
\xline\verb~      super.action(n);~\\[-1.0ex]
\xline\verb~    }~\\[-1.0ex]
\xline\verb~~\\[-1.0ex]
\xline\verb~  }// class Writeout_edges~\\[-1.0ex]
\xline\verb~    ~\\[-1.0ex]
\xline\verb~}~\\[-1.0ex]
\xline\verb~~\\[-1.0ex]
\xline\verb~~\\[-1.0ex]
\xline\verb~// eof~\\[-1.0ex]

\egroup

\section{Consistency checks on the Firm model}
\label{file_checker_java}
\xlinecounterreset{}
\bgroup\footnotesize
\xline\verb~package eu.bandm.ttc2011.case2.transformations ;~\\[-1.0ex]
\xline\verb~~\\[-1.0ex]
\xline\verb~import java.util.Map ; ~\\[-1.0ex]
\xline\verb~import java.util.HashMap ; ~\\[-1.0ex]
\xline\verb~import java.util.Set; ~\\[-1.0ex]
\xline\verb~import java.util.HashSet ; ~\\[-1.0ex]
\xline\verb~~\\[-1.0ex]
\xline\verb~import eu.bandm.tools.message.SimpleMessage ;~\\[-1.0ex]
\xline\verb~import eu.bandm.tools.message.Location ;~\\[-1.0ex]
\xline\verb~import eu.bandm.tools.message.Locatable ;~\\[-1.0ex]
\xline\verb~import eu.bandm.tools.message.XMLDocumentIdentifier ;~\\[-1.0ex]
\xline\verb~~\\[-1.0ex]
\xline\verb~import eu.bandm.tools.message.MessageReceiver ; ~\\[-1.0ex]
\xline\verb~import eu.bandm.tools.message.MessageCounter ; ~\\[-1.0ex]
\xline\verb~import eu.bandm.tools.message.MessageTee ; ~\\[-1.0ex]
\xline\verb~~\\[-1.0ex]
\xline\verb~import eu.bandm.ttc2011.case2.firm_01.* ; ~\\[-1.0ex]
\xline\verb~~\\[-1.0ex]
\xline\verb~~\\[-1.0ex]
\xline\verb~/** ~\\[-1.0ex]
\xline\verb~Performs basic consistenty checks on a firm model.~\\[-1.0ex]
\xline\verb~<p>~\\[-1.0ex]
\xline\verb~Every firm model is represented by its "endnode", from which all nodes are~\\[-1.0ex]
\xline\verb~reachable.~\\[-1.0ex]
\xline\verb~<p>~\\[-1.0ex]
\xline\verb~In detail it checks~\\[-1.0ex]
\xline\verb~<ul>~\\[-1.0ex]
\xline\verb~<li> every block (except the end block) ~\\[-1.0ex]
\xline\verb~contains exactly one(1) control flow node ("jump" or "cond"; which represents~\\[-1.0ex]
\xline\verb~what to happen after the complete execution of the block's code)~\\[-1.0ex]
\xline\verb~<li>~\\[-1.0ex]
\xline\verb~the block which contains the endnode ("end block") does NOT have a~\\[-1.0ex]
\xline\verb~control flow node~\\[-1.0ex]
\xline\verb~<li> every phi node has as many inputs (outgoing edges)~\\[-1.0ex]
\xline\verb~ as the block it is contained in.~\\[-1.0ex]
\xline\verb~<li> the inputs (outgoing edges) of both are numbered consecutively starting with zero.~\\[-1.0ex]
\xline\verb~</ul>~\\[-1.0ex]
\xline\verb~<p>~\\[-1.0ex]
\xline\verb~Additionally some transient cache values are updated to reflect the results,~\\[-1.0ex]
\xline\verb~but they are not maintained in the following code!~\\[-1.0ex]
\xline\verb~<p>~\\[-1.0ex]
\xline\verb~Some properties have already been checked when the Gxl model has~\\[-1.0ex]
\xline\verb~been translated, see {@link Gxl2Firm}. These are:~\\[-1.0ex]
\xline\verb~<ul>~\\[-1.0ex]
\xline\verb~<li>~\\[-1.0ex]
\xline\verb~That there is only one "end" and only one "start" node.~\\[-1.0ex]
\xline\verb~<li>~\\[-1.0ex]
\xline\verb~That every "BlockNode" is related to a "Block".~\\[-1.0ex]
\xline\verb~<li>~\\[-1.0ex]
\xline\verb~That the numeric nodes have all of their numeric arguments, starting at ~\\[-1.0ex]
\xline\verb~position number 0(zero), consecutively.~\\[-1.0ex]
\xline\verb~</ul>~\\[-1.0ex]
\xline\verb~~\\[-1.0ex]
\xline\verb~~\\[-1.0ex]
\xline\verb~*/~\\[-1.0ex]
\xline\verb~~\\[-1.0ex]
\xline\verb~public class Checker extends Firm.VisitBlocksOnce  {~\\[-1.0ex]
\xline\verb~~\\[-1.0ex]
\xline\verb~  final MessageTee<SimpleMessage<XMLDocumentIdentifier>> msg~\\[-1.0ex]
\xline\verb~    = new  MessageTee<SimpleMessage<XMLDocumentIdentifier>>();~\\[-1.0ex]
\xline\verb~  final MessageCounter<SimpleMessage<XMLDocumentIdentifier>> msgC ~\\[-1.0ex]
\xline\verb~    = new MessageCounter<SimpleMessage<XMLDocumentIdentifier>>();~\\[-1.0ex]
\xline\verb~  {msg.add(msgC);}~\\[-1.0ex]
\xline\verb~~\\[-1.0ex]
\xline\verb~  protected void ERROR (Location<XMLDocumentIdentifier> loc, String txt){~\\[-1.0ex]
\xline\verb~    msg.receive(SimpleMessage.error(loc, txt));~\\[-1.0ex]
\xline\verb~  }~\\[-1.0ex]
\xline\verb~  ~\\[-1.0ex]
\xline\verb~~\\[-1.0ex]
\xline\verb~  public boolean check ~\\[-1.0ex]
\xline\verb~    ( final End endnode,~\\[-1.0ex]
\xline\verb~      final MessageReceiver<SimpleMessage<XMLDocumentIdentifier>> msg){~\\[-1.0ex]
\xline\verb~    this.msg.add(msg) ; ~\\[-1.0ex]
\xline\verb~    match(endnode);~\\[-1.0ex]
\xline\verb~    final Block endblock = endnode.get_block();~\\[-1.0ex]
\xline\verb~    if (finalNode.containsKey(endblock))~\\[-1.0ex]
\xline\verb~      ERROR(finalNode.get(endnode).get_location(),~\\[-1.0ex]
\xline\verb~            "The block containing the end node does contain a jump/cond node");~\\[-1.0ex]
\xline\verb~    if (visitedBlocks.size()!=finalNode.size()+1)~\\[-1.0ex]
\xline\verb~      for (Block b : visitedBlocks)~\\[-1.0ex]
\xline\verb~        if (!(b==endblock || finalNode.containsKey(b)))~\\[-1.0ex]
\xline\verb~          ERROR(b.get_location(), "this block does not contain a jump/cond node");~\\[-1.0ex]
\xline\verb~    return msgC.getCriticalCount()==0;~\\[-1.0ex]
\xline\verb~  }~\\[-1.0ex]
\xline\verb~~\\[-1.0ex]
\xline\verb~~\\[-1.0ex]
\xline\verb~  public static boolean isIntegerType (NumericType t){~\\[-1.0ex]
\xline\verb~    switch (t){~\\[-1.0ex]
\xline\verb~      case Iu: case  Is: case  Su: case  Ss: case  Bu: case  Bs: ~\\[-1.0ex]
\xline\verb~        return  true ;~\\[-1.0ex]
\xline\verb~    default: return false ;~\\[-1.0ex]
\xline\verb~    }~\\[-1.0ex]
\xline\verb~  }~\\[-1.0ex]
\xline\verb~~\\[-1.0ex]
\xline\verb~  public static boolean isValidCondSelector (Numeric n){~\\[-1.0ex]
\xline\verb~    final NumericType t = ((NumericNode)n).get_type();~\\[-1.0ex]
\xline\verb~    return (t==NumericType.b) || isIntegerType(t);~\\[-1.0ex]
\xline\verb~  }~\\[-1.0ex]
\xline\verb~~\\[-1.0ex]
\xline\verb~~\\[-1.0ex]
\xline\verb~  protected Map<Block, BlockNode> finalNode~\\[-1.0ex]
\xline\verb~    = new HashMap<Block, BlockNode>();~\\[-1.0ex]
\xline\verb~~\\[-1.0ex]
\xline\verb~  /** Checks whether "set" contains only numbers from zero to pred(set.size()),~\\[-1.0ex]
\xline\verb~      ergo all these numbers.~\\[-1.0ex]
\xline\verb~   */~\\[-1.0ex]
\xline\verb~  protected boolean checkIndexSet (final Set<Integer> set){~\\[-1.0ex]
\xline\verb~    final int count = set.size();~\\[-1.0ex]
\xline\verb~    for (final Integer i:set)~\\[-1.0ex]
\xline\verb~      if (i<0||i>=count)~\\[-1.0ex]
\xline\verb~        return false ;~\\[-1.0ex]
\xline\verb~    return true ; ~\\[-1.0ex]
\xline\verb~  }~\\[-1.0ex]
\xline\verb~~\\[-1.0ex]
\xline\verb~~\\[-1.0ex]
\xline\verb~  @Override public void action (final Jmp cf){~\\[-1.0ex]
\xline\verb~    memoControlFlow(cf);~\\[-1.0ex]
\xline\verb~    super.action(cf);~\\[-1.0ex]
\xline\verb~  }~\\[-1.0ex]
\xline\verb~  @Override public void action (final Cond cond){~\\[-1.0ex]
\xline\verb~    memoControlFlow(cond);~\\[-1.0ex]
\xline\verb~    super.action(cond);~\\[-1.0ex]
\xline\verb~  }~\\[-1.0ex]
\xline\verb~  @Override public void action (final Start start){~\\[-1.0ex]
\xline\verb~    memoControlFlow(start);~\\[-1.0ex]
\xline\verb~    super.action(start);~\\[-1.0ex]
\xline\verb~  }~\\[-1.0ex]
\xline\verb~  @Override public void action (final Return r){~\\[-1.0ex]
\xline\verb~    memoControlFlow(r);~\\[-1.0ex]
\xline\verb~    super.action(r);~\\[-1.0ex]
\xline\verb~  }~\\[-1.0ex]
\xline\verb~~\\[-1.0ex]
\xline\verb~  /** Checks whether every Block contains only one control flow, ~\\[-1.0ex]
\xline\verb~      and enters this into the transient map "finalNode"~\\[-1.0ex]
\xline\verb~  */~\\[-1.0ex]
\xline\verb~    //  @Override public void action (final ControlFlowNode cf){~\\[-1.0ex]
\xline\verb~  protected void memoControlFlow (final BlockNode cf){~\\[-1.0ex]
\xline\verb~    final Block block = cf.get_block();~\\[-1.0ex]
\xline\verb~    if (finalNode.containsKey(block))~\\[-1.0ex]
\xline\verb~      if (finalNode.get(block)!=cf)~\\[-1.0ex]
\xline\verb~        ERROR(block.get_location(), "Block contains more than one control flow, "~\\[-1.0ex]
\xline\verb~            +"namely defined at "+finalNode.get(block).get_location()~\\[-1.0ex]
\xline\verb~            +" and defined at "+cf.get_location());~\\[-1.0ex]
\xline\verb~    finalNode.put(block,cf);~\\[-1.0ex]
\xline\verb~  }~\\[-1.0ex]
\xline\verb~~\\[-1.0ex]
\xline\verb~  /** Checks whether the indices of the predecs of the phi node~\\[-1.0ex]
\xline\verb~      are the same as that of the containing block.~\\[-1.0ex]
\xline\verb~  */~\\[-1.0ex]
\xline\verb~  @Override public void action (final Phi phi){~\\[-1.0ex]
\xline\verb~    final Block b = phi.get_block();~\\[-1.0ex]
\xline\verb~    final int count = phi.get_alternatives().size();~\\[-1.0ex]
\xline\verb~    if (b.get_predecs().size()!=count){~\\[-1.0ex]
\xline\verb~      ERROR(phi.get_location(), ~\\[-1.0ex]
\xline\verb~            "phi node has not the same count of outgoing edges(/input) "~\\[-1.0ex]
\xline\verb~            +"as containing block.");~\\[-1.0ex]
\xline\verb~    }~\\[-1.0ex]
\xline\verb~    else if (!checkIndexSet(phi.get_alternatives().keySet()))~\\[-1.0ex]
\xline\verb~      ERROR(phi.get_location(), ~\\[-1.0ex]
\xline\verb~            "phi node alternatives not numbered from zero, consecutively.");~\\[-1.0ex]
\xline\verb~    super.action(phi);~\\[-1.0ex]
\xline\verb~  }~\\[-1.0ex]
\xline\verb~~\\[-1.0ex]
\xline\verb~~\\[-1.0ex]
\xline\verb~  /** Checks whether the indices of the predecs of the block are ~\\[-1.0ex]
\xline\verb~      starting from zero and consecutive.~\\[-1.0ex]
\xline\verb~  */~\\[-1.0ex]
\xline\verb~  @Override public void action (final Block b){~\\[-1.0ex]
\xline\verb~    if (!checkIndexSet(b.get_predecs().keySet()))~\\[-1.0ex]
\xline\verb~      ERROR(b.get_location(), ~\\[-1.0ex]
\xline\verb~            "block predecessors (/outgoing edges) "~\\[-1.0ex]
\xline\verb~            +"not numbered from zero, consecutively.");~\\[-1.0ex]
\xline\verb~    super.action(b);~\\[-1.0ex]
\xline\verb~  }~\\[-1.0ex]
\xline\verb~~\\[-1.0ex]
\xline\verb~}~\\[-1.0ex]
\xline\verb~~\\[-1.0ex]
\xline\verb~~\\[-1.0ex]
\xline\verb~~\\[-1.0ex]
\xline\verb~~\\[-1.0ex]
\xline\verb~// eof~\\[-1.0ex]

\egroup

\section{Constant Folding}
\label{file_constantfolding_java}
\xlinecounterreset{}
\bgroup\footnotesize
\xline\verb~package eu.bandm.ttc2011.case2.transformations ;~\\[-1.0ex]
\xline\verb~~\\[-1.0ex]
\xline\verb~import java.util.Set ;~\\[-1.0ex]
\xline\verb~import java.util.HashSet ;~\\[-1.0ex]
\xline\verb~import java.util.Map ;~\\[-1.0ex]
\xline\verb~import java.util.HashMap ;~\\[-1.0ex]
\xline\verb~~\\[-1.0ex]
\xline\verb~import eu.bandm.tools.message.XMLDocumentIdentifier; ~\\[-1.0ex]
\xline\verb~import eu.bandm.tools.message.Location; ~\\[-1.0ex]
\xline\verb~import eu.bandm.tools.message.SimpleMessage ; ~\\[-1.0ex]
\xline\verb~import eu.bandm.tools.message.MessageReceiver ; ~\\[-1.0ex]
\xline\verb~import eu.bandm.tools.message.MessageCounter ; ~\\[-1.0ex]
\xline\verb~import eu.bandm.tools.message.MessageTee ; ~\\[-1.0ex]
\xline\verb~~\\[-1.0ex]
\xline\verb~import eu.bandm.tools.umod.runtime.CheckedMap_RD;~\\[-1.0ex]
\xline\verb~~\\[-1.0ex]
\xline\verb~import static eu.bandm.tools.ops.Collections.the ; ~\\[-1.0ex]
\xline\verb~~\\[-1.0ex]
\xline\verb~import eu.bandm.ttc2011.case2.firm_01.* ; ~\\[-1.0ex]
\xline\verb~~\\[-1.0ex]
\xline\verb~public class ConstantFolding {~\\[-1.0ex]
\xline\verb~~\\[-1.0ex]
\xline\verb~  final MessageTee<SimpleMessage<XMLDocumentIdentifier>> msg~\\[-1.0ex]
\xline\verb~    = new  MessageTee<SimpleMessage<XMLDocumentIdentifier>>();~\\[-1.0ex]
\xline\verb~  final MessageCounter<SimpleMessage<XMLDocumentIdentifier>> msgC ~\\[-1.0ex]
\xline\verb~    = new MessageCounter<SimpleMessage<XMLDocumentIdentifier>>();~\\[-1.0ex]
\xline\verb~  {msg.add(msgC);}~\\[-1.0ex]
\xline\verb~~\\[-1.0ex]
\xline\verb~  protected void ERROR (Location<XMLDocumentIdentifier> loc, String txt){~\\[-1.0ex]
\xline\verb~    msg.receive(SimpleMessage.error(loc, txt));~\\[-1.0ex]
\xline\verb~  }~\\[-1.0ex]
\xline\verb~  ~\\[-1.0ex]
\xline\verb~  public ConstantFolding ~\\[-1.0ex]
\xline\verb~    (final MessageReceiver<SimpleMessage<XMLDocumentIdentifier>> msgExt){~\\[-1.0ex]
\xline\verb~    msg.add(msgExt);~\\[-1.0ex]
\xline\verb~  }~\\[-1.0ex]
\xline\verb~~\\[-1.0ex]
\xline\verb~  public End rewrite (final End endnode, final int visualizationMode){~\\[-1.0ex]
\xline\verb~    final End e1 = new ConstPropagator().rewrite_typed(endnode);~\\[-1.0ex]
\xline\verb~    //    return e1;~\\[-1.0ex]
\xline\verb~    new Visual().visualize(e1, "after constant expression propagation", ~\\[-1.0ex]
\xline\verb~                           visualizationMode);~\\[-1.0ex]
\xline\verb~~\\[-1.0ex]
\xline\verb~    final End e2 = new CoRewriter().rewrite_typed(e1);~\\[-1.0ex]
\xline\verb~    new DeadBlockEliminator().checkAlive(e2.get_block());~\\[-1.0ex]
\xline\verb~    new Visual().visualize(e2, "after dead block elimination", ~\\[-1.0ex]
\xline\verb~                           visualizationMode);~\\[-1.0ex]
\xline\verb~~\\[-1.0ex]
\xline\verb~    final End e3 = new DeadPhiEliminator().rewrite_typed(e2);~\\[-1.0ex]
\xline\verb~    // FIXME required for "const.gxl", REICHT NICHT !!~\\[-1.0ex]
\xline\verb~    final End e4 = new ConstPropagator().rewrite_typed(e3);~\\[-1.0ex]
\xline\verb~    return e4 ; ~\\[-1.0ex]
\xline\verb~  }~\\[-1.0ex]
\xline\verb~~\\[-1.0ex]
\xline\verb~  // =============================================================~\\[-1.0ex]
\xline\verb~~\\[-1.0ex]
\xline\verb~  protected class ConstPropagator extends Rewriter {~\\[-1.0ex]
\xline\verb~~\\[-1.0ex]
\xline\verb~    /** Replaces a unary numeric operation by a constant, iff operand is constant.~\\[-1.0ex]
\xline\verb~    */~\\[-1.0ex]
\xline\verb~    @Override public void rewriteFields(final Unary clone){~\\[-1.0ex]
\xline\verb~      rewriteFields((NumericNode)clone);~\\[-1.0ex]
\xline\verb~      final Numeric on = rewrite_typed(clone.get_on());~\\[-1.0ex]
\xline\verb~      if (on instanceof NumericConst) {~\\[-1.0ex]
\xline\verb~        final NumericConst nc ~\\[-1.0ex]
\xline\verb~          = new NumericConst(clone.get_location(),~\\[-1.0ex]
\xline\verb~                             clone.get_block(),~\\[-1.0ex]
\xline\verb~                             ((NumericConst)on).get_type(),~\\[-1.0ex]
\xline\verb~                             clone.get_type()+"("~\\[-1.0ex]
\xline\verb~                             +((NumericConst)on).get_unparsedValue()~\\[-1.0ex]
\xline\verb~                             +")"~\\[-1.0ex]
\xline\verb~                             );~\\[-1.0ex]
\xline\verb~        // for demo purpose only, instead of effective value calculation:~\\[-1.0ex]
\xline\verb~        nc.set_intValue(new Integer(0));~\\[-1.0ex]
\xline\verb~        substitute (nc);~\\[-1.0ex]
\xline\verb~      }~\\[-1.0ex]
\xline\verb~      else if (clone.set_on(on))~\\[-1.0ex]
\xline\verb~        substitute(clone);~\\[-1.0ex]
\xline\verb~    }~\\[-1.0ex]
\xline\verb~~\\[-1.0ex]
\xline\verb~~\\[-1.0ex]
\xline\verb~    /** Replaces a binary numeric operation by a constant, iff both~\\[-1.0ex]
\xline\verb~        operands are constants.~\\[-1.0ex]
\xline\verb~    */~\\[-1.0ex]
\xline\verb~    @Override public void rewriteFields(final Binary clone){~\\[-1.0ex]
\xline\verb~      rewriteFields((NumericNode)clone);~\\[-1.0ex]
\xline\verb~      final Numeric left = rewrite_typed(clone.get_left());~\\[-1.0ex]
\xline\verb~      final Numeric right = rewrite_typed(clone.get_right());~\\[-1.0ex]
\xline\verb~      if ((left instanceof NumericConst) && (right instanceof NumericConst)){~\\[-1.0ex]
\xline\verb~        final NumericConst nc ~\\[-1.0ex]
\xline\verb~          = new NumericConst(clone.get_location(),~\\[-1.0ex]
\xline\verb~                             clone.get_block(),~\\[-1.0ex]
\xline\verb~                             ((NumericConst)left).get_type(),~\\[-1.0ex]
\xline\verb~                             clone.get_type()+"("~\\[-1.0ex]
\xline\verb~                             +((NumericConst)left).get_unparsedValue()~\\[-1.0ex]
\xline\verb~                             +","+((NumericConst)right).get_unparsedValue()+")"~\\[-1.0ex]
\xline\verb~                             );~\\[-1.0ex]
\xline\verb~        // for demo purpose only, instead of effective value calculation:~\\[-1.0ex]
\xline\verb~        nc.set_intValue(new Integer(0));~\\[-1.0ex]
\xline\verb~        substitute (nc);~\\[-1.0ex]
\xline\verb~      }~\\[-1.0ex]
\xline\verb~      else {~\\[-1.0ex]
\xline\verb~        if (clone.set_left(left))~\\[-1.0ex]
\xline\verb~          substitute(clone);~\\[-1.0ex]
\xline\verb~        if (clone.set_right(right))~\\[-1.0ex]
\xline\verb~          substitute(clone);~\\[-1.0ex]
\xline\verb~      }~\\[-1.0ex]
\xline\verb~    }~\\[-1.0ex]
\xline\verb~~\\[-1.0ex]
\xline\verb~~\\[-1.0ex]
\xline\verb~    // out of place for demo purpose, since no explicit type checker provided~\\[-1.0ex]
\xline\verb~    protected Set<Cond> checked = new HashSet<Cond>();~\\[-1.0ex]
\xline\verb~~\\[-1.0ex]
\xline\verb~    /** Replaces a Proj_X  by an unconditional Jump or eliminates it, iff ~\\[-1.0ex]
\xline\verb~        the input to its Cond is a const.~\\[-1.0ex]
\xline\verb~    */~\\[-1.0ex]
\xline\verb~    @Override public void action (final Proj_X proj){~\\[-1.0ex]
\xline\verb~      super.action(proj);~\\[-1.0ex]
\xline\verb~      final Proj_X rewritten = (Proj_X)getResult() ; ~\\[-1.0ex]
\xline\verb~      final Cond cond = rewritten.get_input();~\\[-1.0ex]
\xline\verb~      final NumericNode selector = (NumericNode) cond.get_selector();~\\[-1.0ex]
\xline\verb~      if (selector instanceof NumericConst){~\\[-1.0ex]
\xline\verb~        if (Checker.isValidCondSelector(selector)){~\\[-1.0ex]
\xline\verb~          final int constSelection = ((NumericConst)selector).get_intValue();~\\[-1.0ex]
\xline\verb~          if (proj.get_selection()==constSelection){~\\[-1.0ex]
\xline\verb~            substitute(new Jmp(proj.get_location(), rewritten.get_block()));~\\[-1.0ex]
\xline\verb~          }~\\[-1.0ex]
\xline\verb~          else~\\[-1.0ex]
\xline\verb~            substitute_empty(); ~\\[-1.0ex]
\xline\verb~        }~\\[-1.0ex]
\xline\verb~        else if (!checked.contains(cond)){~\\[-1.0ex]
\xline\verb~          ERROR(proj.get_location(),~\\[-1.0ex]
\xline\verb~                "type check error, argument to cond must be of integer type");~\\[-1.0ex]
\xline\verb~          // thie error should of course have been found earlier, ~\\[-1.0ex]
\xline\verb~          // during a dedicated type checker's run.~\\[-1.0ex]
\xline\verb~          checked.add(cond);~\\[-1.0ex]
\xline\verb~        }~\\[-1.0ex]
\xline\verb~      }~\\[-1.0ex]
\xline\verb~    }~\\[-1.0ex]
\xline\verb~~\\[-1.0ex]
\xline\verb~    /** Since every loop in the model must pass through a "Block" model element,~\\[-1.0ex]
\xline\verb~        it can be cut be cloning the Block and memorizing *in advance*.~\\[-1.0ex]
\xline\verb~    */~\\[-1.0ex]
\xline\verb~    @Override public void action(final Block block){~\\[-1.0ex]
\xline\verb~      final Block clone = breakLoop(block);~\\[-1.0ex]
\xline\verb~      if (clone==null)~\\[-1.0ex]
\xline\verb~        return ; ~\\[-1.0ex]
\xline\verb~      rewriteFields (clone);~\\[-1.0ex]
\xline\verb~      original=block; ~\\[-1.0ex]
\xline\verb~      substitute(clone);~\\[-1.0ex]
\xline\verb~    }~\\[-1.0ex]
\xline\verb~~\\[-1.0ex]
\xline\verb~  }//class ConstPropagator~\\[-1.0ex]
\xline\verb~~\\[-1.0ex]
\xline\verb~~\\[-1.0ex]
\xline\verb~  // =============================================================~\\[-1.0ex]
\xline\verb~~\\[-1.0ex]
\xline\verb~  /** does visit ONLY blocks and Jmp/Proj_X/End~\\[-1.0ex]
\xline\verb~   */~\\[-1.0ex]
\xline\verb~  protected class DeadBlockEliminator {~\\[-1.0ex]
\xline\verb~    protected Set<Block> visited = new HashSet<Block>();~\\[-1.0ex]
\xline\verb~~\\[-1.0ex]
\xline\verb~    public boolean checkAlive (final Block block){~\\[-1.0ex]
\xline\verb~      if (visited.contains(block))~\\[-1.0ex]
\xline\verb~        return true ;~\\[-1.0ex]
\xline\verb~      final CheckedMap_RD<Integer, ControlFlow> tmp ~\\[-1.0ex]
\xline\verb~        = new CheckedMap_RD<Integer, ControlFlow>();~\\[-1.0ex]
\xline\verb~      for (Map.Entry<Integer, ControlFlow>me : block.get_predecs().entrySet()){~\\[-1.0ex]
\xline\verb~        ControlFlowNode cf = (ControlFlowNode)me.getValue();~\\[-1.0ex]
\xline\verb~        if ( (cf instanceof Start) || checkAlive(cf.get_block()) )~\\[-1.0ex]
\xline\verb~          tmp.put(me.getKey(), cf);~\\[-1.0ex]
\xline\verb~      }~\\[-1.0ex]
\xline\verb~      block.set_predecs(tmp);~\\[-1.0ex]
\xline\verb~      return  !tmp.isEmpty();~\\[-1.0ex]
\xline\verb~    }~\\[-1.0ex]
\xline\verb~~\\[-1.0ex]
\xline\verb~  } // class DeadBlockEliminator~\\[-1.0ex]
\xline\verb~~\\[-1.0ex]
\xline\verb~  // =============================================================~\\[-1.0ex]
\xline\verb~~\\[-1.0ex]
\xline\verb~  public class DeadPhiEliminator extends Rewriter{~\\[-1.0ex]
\xline\verb~~\\[-1.0ex]
\xline\verb~    // block loop prevention  MISSING~\\[-1.0ex]
\xline\verb~~\\[-1.0ex]
\xline\verb~    /** replace a Phi with only one input with this input~\\[-1.0ex]
\xline\verb~     */~\\[-1.0ex]
\xline\verb~    @Override public void action(final Phi phi){~\\[-1.0ex]
\xline\verb~      super.action(phi);~\\[-1.0ex]
\xline\verb~      final Phi rewritten = (Phi)getResult();~\\[-1.0ex]
\xline\verb~      final Set<Integer> keys = rewritten.get_block().get_predecs().keySet() ; ~\\[-1.0ex]
\xline\verb~      rewritten.get_alternatives().keySet().retainAll(keys);~\\[-1.0ex]
\xline\verb~      original = phi;~\\[-1.0ex]
\xline\verb~      switch(keys.size()){~\\[-1.0ex]
\xline\verb~      case 0: substitute (new Bad()); return ; ~\\[-1.0ex]
\xline\verb~      case 1: substitute (phi.get_alternatives().get(the(keys)));~\\[-1.0ex]
\xline\verb~        return ; ~\\[-1.0ex]
\xline\verb~      }~\\[-1.0ex]
\xline\verb~      substitute(rewritten);~\\[-1.0ex]
\xline\verb~    }~\\[-1.0ex]
\xline\verb~  }//class DeadPhiEliminator~\\[-1.0ex]
\xline\verb~~\\[-1.0ex]
\xline\verb~}~\\[-1.0ex]
\xline\verb~// eof~\\[-1.0ex]

\egroup



\end{document}